\documentclass[aps,prb,twocolumn,superscriptaddress,showpacs,longbibliography,nofootinbib]{revtex4-1}
\usepackage{amsmath}
\usepackage{graphicx}
\usepackage{amsfonts}
\usepackage{verbatim}
\usepackage{amssymb}
\usepackage{color}
\usepackage{dsfont}
\usepackage{amsmath} 
\usepackage{hyperref}
\usepackage{physics}
\usepackage{soul}
\hypersetup{colorlinks = true, urlcolor = blue, linkcolor = blue, citecolor = blue}
\usepackage[normalem]{ulem}

\newcommand{\bpm}{\begin{pmatrix}}
\newcommand{\epm}{\end{pmatrix}}

\newcommand{\cmtc}{Condensed Matter Theory Center and Joint Quantum Institute, Department of Physics, University of Maryland, College Park, Maryland 20742-4111, USA}
\newcommand{\wvu}{Department of Physics and Astronomy, West Virginia University, Morgantown, WV 26506, USA}

\begin{document}

\title{Spectral properties, topological patches, and effective phase diagrams of finite disordered Majorana nanowires}
\author{Sankar Das Sarma}
\affiliation{\cmtc}
\author{ Jay D. Sau}
\affiliation{\cmtc}
\author{ Tudor D. Stanescu}
\affiliation{\wvu}
\date{\today}

\begin{abstract}
We consider theoretically the physics of bulk topological superconductivity accompanied by boundary non-Abelian Majorana zero modes in semiconductor-superconductor (SM-SC) hybrid systems consisting of finite wires in the presence of correlated disorder arising from random charged impurities.  We find the system to manifest a highly complex behavior due to the subtle interplay between finite wire length and finite disorder, leading to copious low-energy in-gap states throughout the wire and considerably complicating the interpretation of tunneling spectroscopic transport measurements used extensively to search for Majorana modes.  The presence of disorder-induced low-energy states may lead to the non-existence of end Majorana zero modes even when tunneling spectroscopy manifests zero bias conductance peaks in local tunneling and signatures of bulk gap closing/reopening in the nonlocal transport.  In short wires within the intermediate disorder regime, apparent topology may manifest in small ranges ("patches") of parameter values, which may or may not survive the long wire limit depending on various details. Because of the dominance of disorder-induced in-gap states, the system may even occasionally have an appropriate topological invariant without manifesting isolated end Majorana zero modes.  We discuss our findings in the context of a recent breakthrough experiment from Microsoft reporting the simultaneous observations of zero bias conductance peaks in local tunneling and gap opening in nonlocal transport within small patches of parameter space.  Based on our analysis, we believe that the disorder strength to SC gap ratio must decrease further for the definitive realization of non-Abelian Majorana zero modes in SM-SC devices.
\end{abstract}

\maketitle

\section{Introduction}\label{S1}

Recent theoretical work has emphasized the key role of disorder in Majorana experiments, establishing that many features of the experimental observations, which are often construed to be signatures of topological Majorana zero modes, may actually be arising entirely from disorder effects in the trivial phase~\cite{das2023search}.  Since the presence of disorder is inevitable in experimental samples~\cite{das2023search,PanPhysRevB.101.024506,PanPhysRevResearch.2.013377,AhnPhysRevMaterials.5.124602,WoodsPhysRevApplied.16.054053,PanPhysRevB.103.195158}, it is crucial to understand all aspects of disorder in the Majorana physics.  Much of the recent theoretical work has focused on  the tunnel conductance spectroscopy in disordered systems, with the goal to critically analyze Majorana experiments reporting zero bias conductance peaks as evidence for the topological phase and leading to the conclusion that disorder can occasionally generate such zero bias peaks in the trivial phase, mimicking Majorana zero modes.  The current community consensus is that neither the topological phase nor the Majorana zero mode has likely been observed experimentally, with essentially all features of the observed tunneling spectroscopy in the published experimental literature in semiconductor nanowires being explicable as the appearance of disorder-induced 'ugly' conductance peaks arising from low energy trivial Andreev bound states~\cite{das2023search}.

Very recently, a notable experimental preprint~\cite{aghaee2022inas} from Microsoft has appeared reporting the observation of a topological gap in InAs-Al semiconductor-superconductor hybrid structures using a combination of local and nonlocal tunneling conductance data, as well as a detailed theoretical protocol controlling data collection and analysis. The extracted topological gap ($\sim 25~\mu$eV) is  relatively small~\cite{aghaee2022inas} in comparison with the typical values of the transport broadening ($\sim1~$meV) associated with the unintentional disorder present in the starting 2D InAs material used in the experiment.  
It should be noted that the "topological gap" here is extracted from the anti-symmetrized non-local conductance~\cite{rosdahl2018andreev}, and is a length-dependent 
property to characterize the finite-size devices. In the thermodynamic limit, the non-local conductance vanishes because of Anderson localization, except 
in the vicinity of the topological quantum phase transition (TQPT)~\cite{Akhmerov2011Quantized} where the non-local conductance is characterized as 
an Andreev rectifier~\cite{rosdahl2018andreev}. The current experiments~\cite{aghaee2022inas} take advantage of the finite size of the system  to use the transport gap as a characterization of disordered devices.
Measuring or directly estimating the disorder strength  in the ``active region,'' where a topological superconducting state is supposed to emerge, is difficult and therefore, the actual disorder in the samples is unknown, except for the starting mobility of the basic material being low ($\sim 50,000cm^2/V s$), which implies considerable disorder.

Inspired by this experiment and motivated by the ubiquitous presence of disorder in Majorana devices and by the intrinsic difficulty of estimating its strength, in the current work we carry out an in-depth theoretical analysis of the underlying Majorana physics as a function of disorder, focusing (in the first part of the work) on the low-energy spectral properties of the system  and (in the second part) on its transport properties, which have been the primary object of most theoretical (and essentially all experimental) papers on the subject.  
As much as possible, we use the best estimates for realistic disorder in our calculations, and use parameters close to the ones provided in Ref.~\onlinecite{aghaee2022inas}, although no attempt is made for any direct comparison with the experimental data.
Our goal is to develop a comprehensive picture of the low energy spectral features in disordered nanowires as functions of disorder, Zeeman field, chemical potential, and wire length, so that we gain an understanding of different disorder-induced regimes and of the corresponding low-energy properties of the hybrid system, which may or may not be ``visible'' in transport measurements.  We find that disorder is associated with some essential aspects of the low energy physics that remain invisible to both local and nonlocal transport measurements, which complicates a straightforward analysis of the experimental data.  In particular, in certain regimes one should be careful interpreting a transport gap as being the topological gap because it may actually by associated with finite size effects. There are cases when increasing the size of the system will not enhance the protection of the Majorana bound states that may emerge at the ends of the wire. Distinguishing this type of disorder-induced regime from the regime that enables the robust topological protection of the Majorana end modes is of critical theoretical and practical importance.
 
In the second part of the work, we combine both spectral and transport calculations, along with direct estimates of the relevant topological invariants, to comment on the effective  ``topological phase'' of the system, both as a function of disorder and as a function of wire length.  One important finding, as alluded to above, is that the manifestation of a gap in the nonlocal conductance may not necessarily imply that the system hosts non-Abelian Majorana zero modes, as gap features may arise from disorder-induced low energy bound sates with long localization length in finite systems.  Thus, although the use of full tunneling transport spectroscopy measuring both local and nonlocal conductance simultaneously from both wire ends is an experimental breakthrough of major import, we must still be careful in discerning disorder effects in finite (relatively short) wires producing misleading signals of topology in experiments.  What is necessary is the observation of the putative topological features (e.g., a gap) over a relatively large parameter space, e.g., over a large range of the applied magnetic field.  The observation of concrete topological features over a large parameter regime seems to be necessary for the unambiguous manifestation of a topological phase with stable Majorana zero modes.  This necessitates systems with larger values of gap to disorder ratios than are available today, but the Microsoft experiment has paved the way for future success. The thrust of this work is calculating  the low energy spectral (and transport) features of disordered Majorana nanowires realized in semiconductor-superconductor hybrid structures, in both long and short wire limits,  so that a clear general picture emerges  for understanding future experiments and their implications.  As such, we focus in the first part of the paper on the density of states calculations, including both total integrated density of states and spatially resolved local density of states,  as  functions of disorder, Zeeman field, chemical potential, and wire length.  Our main goal is theoretical transparency of the matters of principle, rather than  
simulations of specific experimental conditions, hence, we use a simple effective model for the Majorana nanowire and realistic correlated long-range disorder,  without incorporating any unnecessary complications that require unknown adjustable parameters.
The disorder also affects the trivial SC below TQPT by producing subgap Andreev bound states, but does not destroy the trivial SC gap at low enough magnetic field values. The region around the pristine TQPT gets increasingly dominated by low energy subgap states with increasing disorder, which leads to the eventual suppression of the TQPT itself for very strong disorder.
Key  additional aspects of the second part are obtaining the appropriate transport topological invariant and the Majorana localization (or coherence) length to ascertain the topological property and the integrity of the end Majorana zero modes in the nanowire along with the  transport and spectral properties.  The two parts use very similar, but not always identical parameters, ensuring that the two parts are accessible independently without referring to each other and explicitly showing that our main results are not a consequence of some parameter fine tuning.  We use realistic parameter vales for the currently accessible InAs-Al SM-SC hybrid nanowire systems based on estimates given in Ref.~\cite{aghaee2022inas}.

The remainder of this paper is organized as follows.  In Sec. \ref{S_Background}, we provide a background for the role of disorder in topological superconductors, emphasizing the thermodynamic limit in contrast with the finite disordered nanowires being considered in the current work.  We describe the theoretical model and our approach in Sec. \ref{S2}.  In Sec. \ref{S3} we provide detailed numerical results and analysis for the spectral properties of disordered nanowires in SM-SC systems, calculating and discussing both the total density of states (DOS) and the spatially resolved local density of states (LDOS) for  ideal wires (Sec. \ref{S3_1}) and both long disordered wires (Sec. \ref{S3_2}) and short disordered wires (Sec. \ref{S3_3}).  In Sec. \ref{S4}, we discuss our results  for the topological patches, presenting local and nonlocal conductance along with topological invariants for small regimes (``patches'') in parameter space manifesting aspects of topological properties and describing our inferred topological phase diagrams in the presence of disorder.Sections V A, B, C, D respectively describe the topological invariant,  the topological phase, the topological patches, and a summary of the  implications of the results of Sec. \ref{S4} for the current experimental samples.   We conclude in Sec. \ref{S6} with a detailed discussion of our results in the context of the Microsoft experiment, focusing on the role of disorder on the topological properties of finite nanowires in SM-SC hybrid systems.

\section{Background} \label{S_Background}

Since the key physics suppressing topology in Majorana nanowires is disorder, it is useful to provide a brief review of what is already known about disorder effects in SM-SC hybrid structures in the context of Majorana physics.  This background puts the current work in a proper perspective.

The importance of disorder for one-dimensional spinless p-wave SC (i.e. which is the effective topological phase in the Majorana nanowires in our SM-SC hybrid systems) was pointed out a long time ago~\cite{motrunich2001griffiths}, when it was established that strong disorder destroys the topological SC by causing a quantum phase transition to a trivial localized phase in the thermodynamic limit. This happens when the disorder is strong enough to cause the mean free path or the localization length (they are the same in 1D systems) to become shorter than the SC coherence length.  Since the SC coherence length (the localization length) increases (decreases) with decreasing SC  gap (increasing disorder), the importance of the effective dimensionless ratio of the topological SC gap to the disorder strength as a crucial controlling parameter has been known for a long time. Here SC gap refers to the spectral gap in the absence of disorder. A disordered time-reversal broken superconductor has a vanishing spectral gap in the thermodynamic limit~\cite{motrunich2001griffiths}.  In the thermodynamic limit, there are just two phases: a trivial Anderson localized phase for strong disorder  (i.e. small values of  gap to disorder ratio) and a topological SC phase for weak disorder (i.e. large values of gap to disorder ratio) separated by a quantum critical point.  It was later shown that the topological SC remains stable to weak disorder in SM-SC nanowires, even in the presence of interaction, thus guaranteeing the existence of a topological phase as long as the disorder is not too strong.~\cite{Lobos2012Interplay}  In addition, disorder creates an effective  Griffiths phase with the production of  low energy subgap localized states in the topological phase, leading eventually to the suppression of the topological SC for strong enough disorder~\cite{motrunich2001griffiths}. This early work on disorder effects on topological SC was later extended to the SM-SC nanowire systems by several groups, but the key physics discussed in the current work was not revealed or discussed in any of these early publications~\cite{Adagideli2014Effects,Liu2012Zero,Takei2013Soft,Brouwer2011Topological,RiederPhysRevB.88.060509,Bagrets2012Class,Pikulin2012Zero,Sau2013Density,Sau2012Experimental,PanRMTPhysRevB.106.115413}.

How do these disorder-induced thermodynamic quantum phase considerations of 1D topological SC systems affect our system of finite disordered nanowires in the SM-SC hybrid structures? The pristine Majorana nanowire has a Zeeman field (or chemical potential) driven topological quantum phase transition (TQPT)  from a trvial (spinful s-wave) SC phase to a topological (spinless p-wave) SC phase at a critical field.  This field-driven TQPT survives weak disorder, but  disappears in the strong disorder regime because the topological SC itself is suppressed for strong disorder. 
The disorder also affects the trivial SC below TQPT by producing subgap Andreev bound states, but does not destroy the trivial SC gap at low enough magnetic field values. However, the region around the pristine TQPT gets increasingly dominated by low energy subgap states with increasing disorder with the eventual suppression of the TQPT itself for very strong disorder.
 Thus, disorder produces subgap states, and strong disorder  eventually destroys the TQPT and the topological SC phase in the nanowire.  There are, however, serious issues arising from the finite wire length, which complicate the physics as we discuss in the current work.  

 The TQPT is ill-defined even for the disorder-free pristine system in ``short'' nanowires with the length being smaller than the SC coherence length.  In such, short wires, the end MZMs overlap producing Majorana oscillations with no non-Abelian zero modes localized at the wire ends independent of whether a bulk gap closes/opens or not.~\cite{DasSarma2012Splitting} Disorder in finite wires therefore is highly problematic since disorder induced low energy states may hybridize with the overlap-induced Majorana split modes, complicating any simple physical picture based on the long-wire pristine case.  A serious additional complication is that the crossover length from ``short'' to ``long'' limit is quantitatively unknown and cannot be directly estimated experimentally in disordered nanowires.  The crossover obviously depends on the disorder strength  as well as the basic wire parameters such as SC gap and Fermi velocity.  It is clear that all early experiments in Majorana nanowires, before the Microsoft experiment, are most certainly in the short wire limit as the wire length is typically $\sim 1$ micron in all of these early experiments, which misinterpreted the observation of disorder induced low energy subgap Andreev bound states as Majorana signatures.  Whether the current wire length ($\sim 3$ microns) used in the Microsoft experiment is long enough is unknown at this point.
 The existence of the end localized MZMs in the topological phase is compromised seriously by disorder, particularly in ``short'' wires, since the MZM wavefunctions may overlap.  Thus, there are several complications to consider with increasing disorder in finite SM-SC nanowires: Destruction of the topological SC and the TQPT as well as the role of the low-energy subgap localized states induced by disorder affecting the end MZMs.  The only clean result is that the TQPT and the topological SC survive weak disorder for long wires, and hence a large gap to disorder ratio ($>1$) guarantees the existence of MZMs in long wires (which are non-Abelian if the wire is long enough).  Unfortunately, it is unclear the extent to which the current samples, even in the breakthrough Microsoft experiment, are in the weak disorder (and/or long wire) regime since the actual effective disorder in the nanowire SM-SC samples is unknown from independent in situ measurements.  
 
 Based on our analysis, we believe that the current samples in the Microsoft experiment are in an ``intermediate'' disorder regime, which, depending on the precise conditions, may be either the weak or the strong disorder situation.  This is reflected in the reported operational ``topological'' gaps being rather small\cite{aghaee2022inas} and existing over rather narrow regimes of system parameters (e.g., magnetic field and gate voltage). Thus, the reported topological regimes are ``topological patches'' existing with small SC gaps over narrow regions of the parameter space in the magnetic field and gate voltage.  Our whole study is focused on understanding the complex nature of this intermediate patchy effective phase where the copious existence of disorder-induced low energy subgap states in finite wires considerably complicates the interpretation of various topological signatures, which are defined specifically for the pristine (and infinite) system, and ultimately undermines the stability of the Majorana end modes that may emerge. Our focus here is on the spectral properties, signatures of bulk gap, and the topological phase diagram, and not on the local tunneling induced zero bias conductance peaks (ZBCPs) which we have studied extensively elsewhere (see Ref.~\onlinecite{das2023search}  and references therein). It is now well-established that disorder may occasionally induce trivial ZBCPs (arising from Andreev bound states) mimicking MZM induced ZBCPs in disordered samples (and all earlier experiments have most likely seen these trivial ZBCPs), and this is not discussed much in this work.

We also mention that the effective disorder in SM-SC nanowires is Coulomb disorder arising from unintentional random charged impurities invariably present in the system.  Thus, the disorder must be characterized by both a strength and a correlation length as emphasized in our earlier work.~\cite{WoodsPhysRevApplied.16.054053,StanescuPhysRevB.106.085429}  The physics of Coulomb disorder is fundamentally different from the short-range on-site disorder often used in studying Anderson localization, but is crucial to understanding Majorana nanowires in SM-SC systems.  

\section{Theoretical model and approach}\label{S2}

In an ideal (i.e., clean and infinitely long) Majorana wire, the topological quantum phase transition (TQPT) driven by the Zeeman field, $\Gamma$, (or the chemical potential, $\mu$) is characterized by the closing and reopening of the (bulk) quasiparticle gap at critical points $\Gamma = \Gamma_c(\mu)$. In the topological phase, finite Majorana wires host a pair of mid-gap Majorana zero modes (MZMs) localized at the two ends of the system. The MZMs are separated by a finite energy gap from all quasiparticle excitations.  What becomes of this picture in a finite system in the presence of disorder? We address this key question by analyzing the numerical solution of a tight-binding model of the wire in the presence of disorder. There are several important aspects that we need to consider. First, finite size effects generate energy gaps that ``mask'' the vanishing of the quasiparticle gap (at the TQPT and elsewhere) and induce Majorana splitting oscillations (due to the overlap of the wave functions corresponding to the two Majorana end modes), making it difficult to clearly identify the topological phase (which, strictly speaking, is only defined in the thermodynamic limit). On the other hand, we have a finite numerical resolution associated with a finite broadening of the spectral features. Hence, to address the finite size problem, we consider systems that are long-enough, so that the induced energy gaps are below our numerical energy resolution. In the calculations, we typically consider systems of length $L= 40~\mu$m (i.e., at least one order of magnitude longer than the hybrid systems investigated experimentally), but a few results are checked using longer wires.  Short systems ($L= 3~\mu$m) are also considered, to make the connection with the experimental conditions. Second, the difference between ``bulk states'' and ``edge states'' can become ambiguous in the presence of disorder. To eliminate this ambiguity, we remove the edges and consider a disordered ``Majorana ring''. Of course, there are no end MZMs in the ring configuration. To connect with experimentally relevant conditions, we also investigate the role of open boundary conditions, in particular the emergence of MZMs in the presence of disorder. 

Based on the above considerations, we model the semiconductor (SM) component of the Majorana wire using a single-orbital one-dimensional (1D) tight-binding model that is numerically-accessible and contains a relatively small number of parameters. Numerical accessibility is required due to the relatively large size of the system, while the small number of system parameters simplifies the analysis of the low-energy physics. We note that enriching the model, e.g., including multi-orbital physics,\cite{Lutchyn2011Search,Stanescu2011Majorana} amounts to enlarging the parameter space; our model corresponds to all possible ``additional system parameters'' having trivial values. The 1D tight-binding model is described by the (second quantized) Hamiltonian 
\begin{eqnarray}
H\! &=& \! \sum_{i=1}^{N}\sum_{\sigma}\! \left\{ \!-t\left( \hat{c}_{{i}\sigma}^\dagger \hat{c}_{{i+1}\sigma}^{~} \!+ \!h.c.\! \right) +[V_{dis}(i) \!- \!\mu] \hat{c}_{{i}\sigma}^\dagger \hat{c}_{{i}\sigma}^{~}\!\!\right\} \label{Eq1} \\
&+& \! \sum_{i=1}^{N}\!\left\{ \Gamma\left( \hat{c}_{{i}\uparrow}^\dagger \hat{c}_{{i}\downarrow}^{~} \!+ \!h.c.\right) \!+\! \frac{\alpha}{2} \! \left( \hat{c}_{{i}\uparrow}^\dagger \hat{c}_{{i+1}\downarrow}^{~} \!-\! \hat{c}_{{i}\downarrow}^\dagger \hat{c}_{{i+1}\uparrow}^{~} \!+ \!h.c.\right)\! \right\}\! , \nonumber
\end{eqnarray}
 where $i$ labels the sites of a 1D lattice with lattice constant $a$,  $\hat{c}_{{i}\sigma}^\dagger$ ($\hat{c}_{{ i}\sigma}^{~}$) is the creation (annihilation) operator for an electron with spin $\sigma$ located at site $i$, and $V_{dis}(i)$ is a (position-dependent) disorder potential. The other model parameters are: $t$ -- nearest-neighbor hopping amplitude, $\mu$ -- chemical potential, $\Gamma$ -- (half) Zeeman splitting, and $\alpha$ -- Rashba spin-orbit coupling. To describe a Majorana ring, we identify the sites $i=N+1$ and $i=1$ in Eq. (\ref{Eq1}); ``standard'' open boundary conditions are obtained by eliminating the hopping and spin-orbit coupling contributions associated with the pair of sites $(N, 1)$. 
 Note that Eq. (\ref{Eq1}) corresponds to the ``standard'' tight-binding model for hybrid SM-SC Majorana structures \cite{Sau2010Generic,Sau2010NonAbelian,Lutchyn2010Majorana,Oreg2010Helical}, which has been studied extensively for more than 10 years, both theoretically and experimentally \cite{das2023search}.

\begin{figure}[t]
\begin{center}
\includegraphics[width=0.48\textwidth]{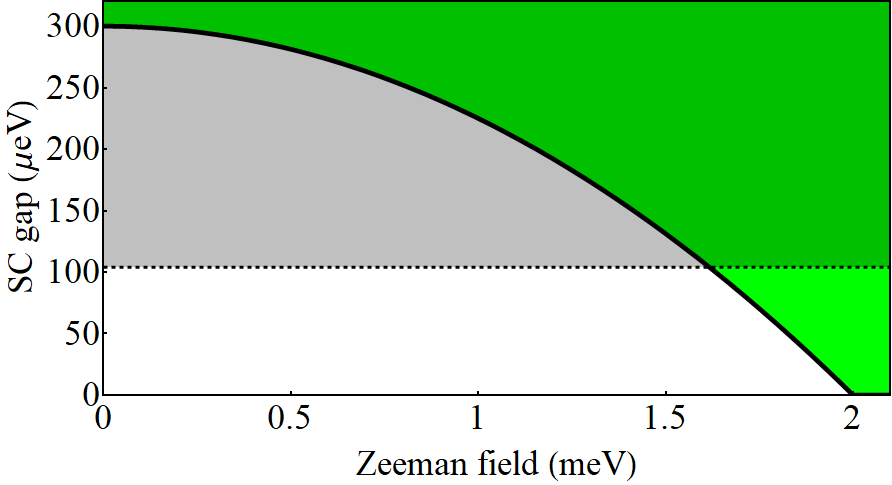}
\end{center}
\vspace{-3mm}
\caption{Parent SC gap as function of the applied Zeeman field for a system with $\Delta_0=0.3~$meV and $\Gamma_0=2~$meV [see Eq. (\ref{Eq2})]. The green shaded region corresponds to the SC quasiparticle spectrum. The properties of the hybrid SM-SC system will be investigated within the low-energy regime, $E < 100~\mu$eV  (highlighted area).}
\label{FIG1}
\vspace{-1mm}
\end{figure}

The SM wire is proximity coupled to a thin superconducting film that induces a finite pairing potential. In the presence of finite Zeeman splitting generated by a magnetic field parallel to the wire, the superconducting gap of the parent SC is suppressed and eventually vanishes when the applied field exceeds a certain critical value $B_0$. We model this behavior by assuming a field-dependent SC gap
\begin{equation}
\Delta(\Gamma) = \Delta_0\left[1-\left(\frac{\Gamma}{\Gamma_0}\right)^2\right], \label{Eq2}
\end{equation} 
where $\Delta_0$ is the parent SC gap at $B=0$ and $\Gamma_0 = \frac{1}{2} g \mu_B B_0$ is the critical Zeeman field in the SM nanowire ($\mu_B$ being the Bohr magneton and $g$ the gyromagnetic ratio of the semiconductor). The field dependence of the parent SC gap is illustrated in Fig. \ref{FIG1}. Note that the magnitude of the zero-field gap, $\Delta_0=0.3~$meV, is comparable to the values expected for Al thin films; the critical Zeeman field,   $\Gamma_0=2~$meV, corresponds to $g\approx 17$ for $B_0 \approx 4~$T, which probably overestimates the values characterizing Al-based structures by a factor $2-3$.  We note that these specific effective parameter values, which, in practice, are likely to vary from sample to sample even for nominally identical systems,  are not crucial for our general considerations. However, an important point is that there exists a finite critical field $B_0$, i.e.,  it is not experimentally possible to increase the applied Zeeman field arbitrarily to access the topological phase because at a high-enough field the bulk Al gap is quenched and all SC gaps vanish. The existence of the critical field $B_0$, where the bulk SC gap collapses (thus suppressing all topological physics in the SM-SC system), is a property of the system beyond Majorana physics that imposes an important constraint on the experimental situation.

Upon ``integrating out'' the SC degrees of freedom, the proximity effect induced in the SM wire is approximately described by the onsite self-energy contribution 
\begin{equation}
\Sigma_{SC}(\omega, \Gamma) =- \gamma ~\!\frac{\omega ~\!\sigma_0 \tau_0 +\Delta(\Gamma)~\! \sigma_y\tau_y}{\sqrt{\Delta(\Gamma)^2-\omega^2}},  \label{Eq3}
\end{equation}
where $\sigma_\mu$ and $\tau_\mu$ are Pauli matrices associated with the spin and particle-hole degrees of freedom, respectively, and $\gamma$ is the effective SM-SC coupling. In the numerical calculations we use the value $\gamma = 0.21~$meV, which, for $\Delta_0=0.3~$meV,  corresponds to an induced gap (at zero magnetic field) of about $0.125~$meV. 
 Eq. (\ref{Eq3}) is expected to represent a good approximation of the self-energy generated by a disordered thin film in the limit of weak SM-SC coupling. We emphasize that disorder in the SC represents a necessary ingredient for obtaining a robust proximity effect (because of the Fermi surface mismatch effect at the SM-SC interface) \cite{Stanescu2022Proximity}. In the presence of disorder (in the SC), the self-energy $\Sigma_{SC}$, which is proportional to the Green's function of the superconductor at the interface, becomes quasi-local. Furthermore, the effects due to ``induced disorder'' are negligible in the weak coupling limit, $\gamma < \Delta(\Gamma)$. 
Note, however, that for Zeeman field values approaching the critical field $\Gamma_0$ the system crosses into the strong coupling regime, $\gamma > \Delta(\Gamma)$, where the effects of ``induced disorder'' are expected to become significant \cite{Cole2016Proximity,Stanescu2022Proximity}. Hence, our approximation overestimates the stability of the  topological phase by neglecting the effects of ``induced disorder'', which are due to disorder necessarily present in the parent SC.  

\begin{figure}[t]
\begin{center}
\includegraphics[width=0.48\textwidth]{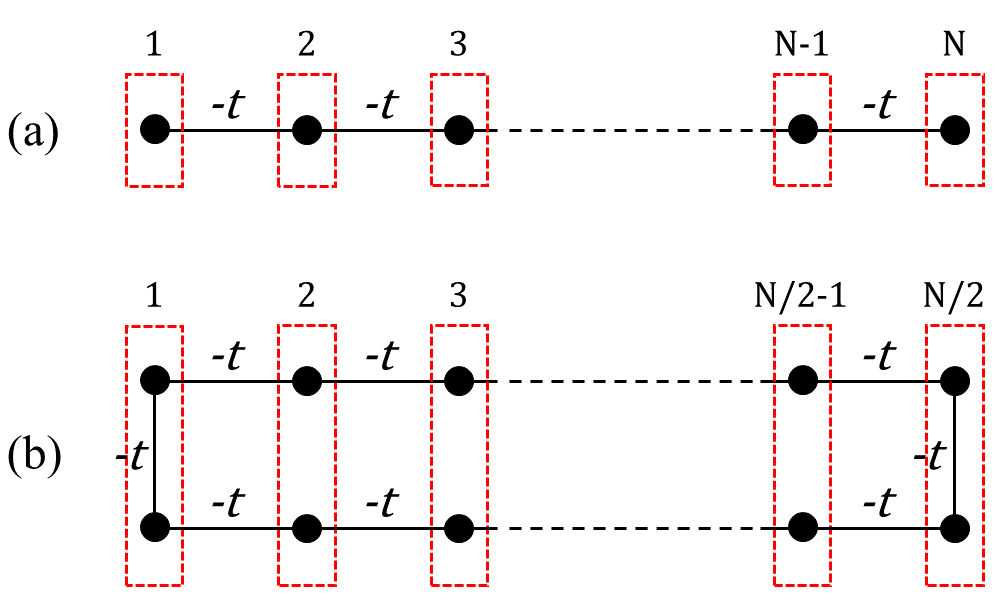}
\end{center}
\vspace{-3mm}
\caption{Schematic representation of the slicing used in the recursive Green's function method for (a) a system with open boundary conditions and (b) a Majorana ring.  For simplicity, we assume that the number $N$ of lattice sites is even.
In addition to the hopping $t$, nearest-neighbor slices are coupled through spin-orbit contributions [see Eqs. (\ref{Eq10}) and (\ref{Eq11})]. In (b) there are intra-slice hoping and spin-orbit contributions for slices $i=1$ and $i=N/2$ [Eq. (\ref{Eq14})].}
\label{FIG2}
\vspace{-1mm}
\end{figure}

The low-energy properties of the hybrid SM-SC system are described by the effective semiconductor Green’s function
\begin{equation}
G_{SM}(\omega) = \left[\omega~\! I_{4N} - H_{BdG} - \Sigma_{SC}\otimes I_N\right]^{-1}, \label{Eq4}
\end{equation}
where $I_M$ is an $M\times M$ identity matrix,  $\Sigma_{SC}$ is the $4\times4$ self-energy matrix given by Eq. (\ref{Eq3}), and $H_{BdG}$ is a $4N\times 4N$ matrix representing the contribution  to the (first quantized) Bogoliubov-de Gennes (BdG) Hamiltonian given by the semiconductor, i.e., corresponding to the Hamiltonian in Eq. (\ref{Eq1}). The quantities of interest that we focus on are the density of states (DOS),  $\rho$, and the  local density of states (LDOS), $\rho_L$, given by 
\begin{eqnarray}
\rho(\omega) &=& -\frac{1}{\pi}{\rm Im} {\rm Tr} [G_{SM}(\omega +i\eta)],              \label{Eq5} \\
\rho_L(\omega, i) &=& -\frac{1}{\pi}{\rm Im} {\rm Tr}_L [G_{SM}(\omega +i\eta)]_{ii},              \label{Eq6}
\end{eqnarray} 
where ${\rm Tr}$ is the trace over position ($i\in {\{1,2, \dots,N\}}$), spin, and the particle-hole degree of freedom, while ${\rm Tr}_L$ is the ``local'' trace over
spin and particle-hole variables. The parameter $\eta>0$ provides a (small) finite broadening of the spectral features and determines the energy resolution of our numerical solution. Typical values used in the calculations are $\eta\sim 0.5-2~\mu$eV.  Note that both $\rho$ and $\rho_L$ only involve the onsite Green's function, i.e., the diagonal matrix elements  $[G_{SM}(\omega +i\eta)]_{ii}$. While calculating the onsite Green's function can be done by simply performing the matrix inversion in Eq. (\ref{Eq4}), this brute force approach becomes numerically inefficient for large systems, particularly if one has to explore large regions of the parameter space. To address this technical challenge, we use a recursive Green's function method \cite{MacKinnon1985Calculation,Lewenkopf2013Recursive} that  involves $3N$ inversions of $4\times 4$ matrices for a system with open boundary conditions or $3N/2$ inversions of $8\times 8$ matrices for a Majorana ring. 
 More specifically, we divide the system into nearest-neighbor-coupled slices, as shown in Fig. \ref{FIG2}, and define the auxiliary left ($G^L$) and right ($G^R$) Green's functions with on-slice values $G_i^{L(R)}$ given by 
\begin{eqnarray}
G_i^{L(R)} &=& \left[g^{-1}(i) - \Sigma_i^{L(R)}\right]^{-1},   \label{Eq7} \\
 \Sigma_i^{L} &=& {\cal T}^T G_{i-1}^{L}{\cal T},                     \label{Eq8} \\
 \Sigma_i^{R} &=& {\cal T} G_{i+1}^{R}{\cal T}^T.                   \label{Eq9}
\end{eqnarray}
For a system with open boundary conditions (i.e., a ``standard'' Majorana wire), the inter-slice coupling, ${\cal T}$, is given by the $4\times 4$ matrix 
\begin{equation}
{\cal T} = -t \sigma_0 \tau_z +\frac{i\alpha}{2} \sigma_y \tau_z. \label{Eq10}
\end{equation}
For the Majorana ring, we introduce a new set of Pauli matrices, $\lambda_\mu$, associated with the two sites within each slice [see Fig. \ref{FIG2}(b)]. The corresponding  inter-slice coupling is the $8\times 8$ matrix 
\begin{equation}
{\cal T} = -t \sigma_0 \tau_z \lambda_0+\frac{i\alpha}{2} \sigma_y \tau_z \lambda_z. \label{Eq11}
\end{equation}
The local inverse Green's function $g^{-1}(i)$ can be separated into  position-independent and position-dependent contributions. For the Majorana wire, the local inverse Green's function  is a $4\times 4$ matrix of  the form
\begin{equation}
g^{-1}(i) = g_0^{-1}(\omega) - V_{dis}(i)~\! \sigma_0\tau_z, 
\end{equation}
with a position-independent contribution given by
\begin{eqnarray}
 g_0^{-1}(\omega) &=& \omega\left(1+\frac{\gamma}{\sqrt{\Delta(\Gamma)^2 - \omega^2}}\right)\sigma_0\tau_0 -(\epsilon_0-\mu)\sigma_0\tau_z \nonumber \\
 & -& \Gamma \sigma_x\tau_z + \frac{\gamma\Delta(\Gamma)}{\sqrt{\Delta(\Gamma)^2-\omega^2}} \sigma_y\tau_y,
\end{eqnarray}
where the constant $\epsilon_0 = 2t\cos[\pi/(N+1)]$ was included for convenience, to define the chemical potential with respect to the bottom of the SM band. 
For the Majorana ring, the position-independent contribution becomes $ g_0^{-1}(\omega) \rightarrow  g_0^{-1}(\omega) \lambda_0$, while the position-dependent contribution takes the form
\begin{eqnarray}
&~&-\frac{1}{2}V_{dis}(i)~\! \sigma_0\tau_z (\lambda_0 \!+\! \lambda_z) -\frac{1}{2}V_{dis}(N\!+\!1\!-\!i)~\! \sigma_0\tau_z (\lambda_0 \!-\! \lambda_z) \nonumber \\
&~&- t \sigma_0 \tau_z \lambda_x (\delta_{i,1} +\delta_{i,N/2}) +\frac{\alpha}{2} \sigma_y \tau_z \lambda_y (\delta_{i,1} - \delta_{i,N/2}), \label{Eq14}
\end{eqnarray}
where $\delta_{i,j}$ is the Kronecker delta. 

The left Green's function and the corresponding self-energy are determined recursively using Eqs. (\ref{Eq7}) and (\ref{Eq8}), starting with the slice $i=1$ and $\Sigma_1^L=0$ (where $0$ represents the appropriate zero matrix). Similarly, we calculate the right Green's function and $\Sigma_i^R$ using Eqs. (\ref{Eq7}) and (\ref{Eq9}),
starting with $i=N$,   $\Sigma_N^R=0$ (for a system with open boundary conditions) or  $i=N/2$,   $\Sigma_{N/2}^R=0$ (for a Majorana ring). Finally, we calculate the Green's function of interest (to be used for determining the DOS and LDOS) using the expression
\begin{equation}
[G_{SM}]_{ii} =  \left[g^{-1}(i) - \Sigma_i^{L}- \Sigma_i^{R}\right]^{-1},
\end{equation}    
with $\Sigma_i^{L}$ and  $\Sigma_i^{R}$ determined recursively, as described above. 

In the numerical calculations, unless explicitly stated otherwise,  we use the following values of the model parameters: lattice constant $a=10~$nm; number of lattice sites $N=4000$ (which corresponds to a system length $L= Na=40~\mu$m); nearest neighbor hopping $t=12.7~$meV (corresponding to an effective mass $m=0.03m_0$, with $m_0$ being the free electron mass); Rashba spin-orbit coupling coefficient $\alpha=1.25~$meV (i.e., $125~$meV$\cdot$\AA); parent superconducting gap at zero magnetic field $\Delta_0=0.3~$meV; critical Zeeman field associated with the collapse of the parent SC gap $\Gamma_0 = 2~$meV; effective SM-SC coupling $\gamma=0.21~$meV. These parameters are realistic (although slightly optimistic) estimates of the parameters characterizing the currently used experimental SM-SC systems.

\section{Numerical results: Low-energy DOS and LDOS}\label{S3}

In this section we calculate the low-energy properties of a hybrid semiconductor-superconductor wire in the presence of disorder using the modeling tools described above and focusing on the density of states (DOS) and the local DOS (LDOS). We first summarize some key properties of an ideal (i.e., clean and infinitely long) system, to use them as a benchmark. Next, we consider a long disordered Majorana system (corresponding to the parameter values given at the end of the previous section) and investigate the impact of disorder on the low-energy spectral features. These results will help us disentangle the disorder effects from finite size effects, which are ubiquitous (and significant) in short systems. Finally, we discuss the low-energy properties of shorter systems ($L\sim 3~\mu$m) and identify characteristic signatures associated with relevant disorder regimes. Note that disentangling the finite size effects from disorder effects is critical for a full understanding of the spectral properties of the SM-SC systems.

\begin{figure}[t]
\begin{center}
\includegraphics[width=0.35\textwidth]{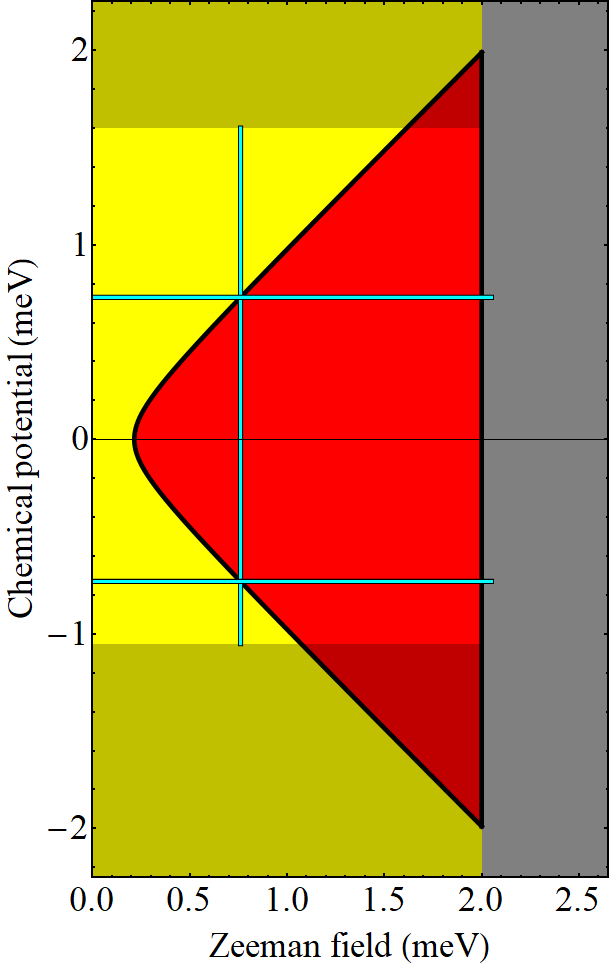}
\end{center}
\vspace{-3mm}
\caption{Topological phase diagram of an ideal (clean and infinitely long) Majorama wire described by the effective Green's function in Eq. (\ref{Eq16}). The  
yellow, red, and dark gray areas represent the trivial SC, topological SC, and gapless phases, respectively. The phase boundary between the trivial and topological SC phases (black line) is given by Eq. (\ref{Eq17}). The low-energy DOS as function of Zeeman field or chemical potential  and energy along the horizontal ($\mu= \pm0.73~$meV) and vertical ($\Gamma=0.75~$meV) blue lines is shown in Fig. \ref{FIG4}. Phase diagrams for disordered Majorana rings corresponding to control parameters within the highlighted area are shown in Fig. \ref{FIG14}.}
\label{FIG3}
\vspace{-1mm}
\end{figure}

\subsection{Ideal Majorana wires}\label{S3_1}

An infinitely long clean system is translation-invariant and the corresponding Fourier transform of the Green's function has the form
\begin{eqnarray}
 G_{SM}\! &=& \!\left[\omega\left(\!1+\frac{\gamma}{\sqrt{\Delta(\Gamma)^2 \!-\! \omega^2}}\right)\!\sigma_0\tau_0 -\xi(k)\sigma_0\tau_z\right. \nonumber \\
 & -&\! \left. (\Gamma \sigma_x \!+ \!\alpha \sin ka ~\sigma_y)\tau_z + \frac{\gamma\Delta(\Gamma)}{\sqrt{\Delta(\Gamma)^2\!-\!\omega^2}} \sigma_y\tau_y\right]^{-1}\!\!\!\!\!\!\!\!,~~~~~~ \label{Eq16}
\end{eqnarray}
with $\xi(k) = 2t(1-\cos ka) -\mu$. 
The low-energy states are given by the poles of the Green's function, i.e., the solutions of the equation $\det[G_{SM}(\omega, k)]=0$. The phase  boundary associated with the topological quantum phase transition  (TQPT) corresponds to the zero-energy solutions for $k=0$, i.e., the solutions of the equation $\det[G_{SM}(0, 0)]=0$. Explicitly, the phase boundary between the trivial and topological SC phases is given by the equation $\Gamma = \Gamma_c(\mu)$, with $-\sqrt{\Gamma_0^2-\gamma^2} \leq \mu\leq \sqrt{\Gamma_0^2-\gamma^2}$ and 
\begin{equation}
\Gamma_c(\mu) =\sqrt{\mu^2+\gamma^2}. \label{Eq17}
\end{equation}
\begin{figure}[t]
\begin{center}
\includegraphics[width=0.48\textwidth]{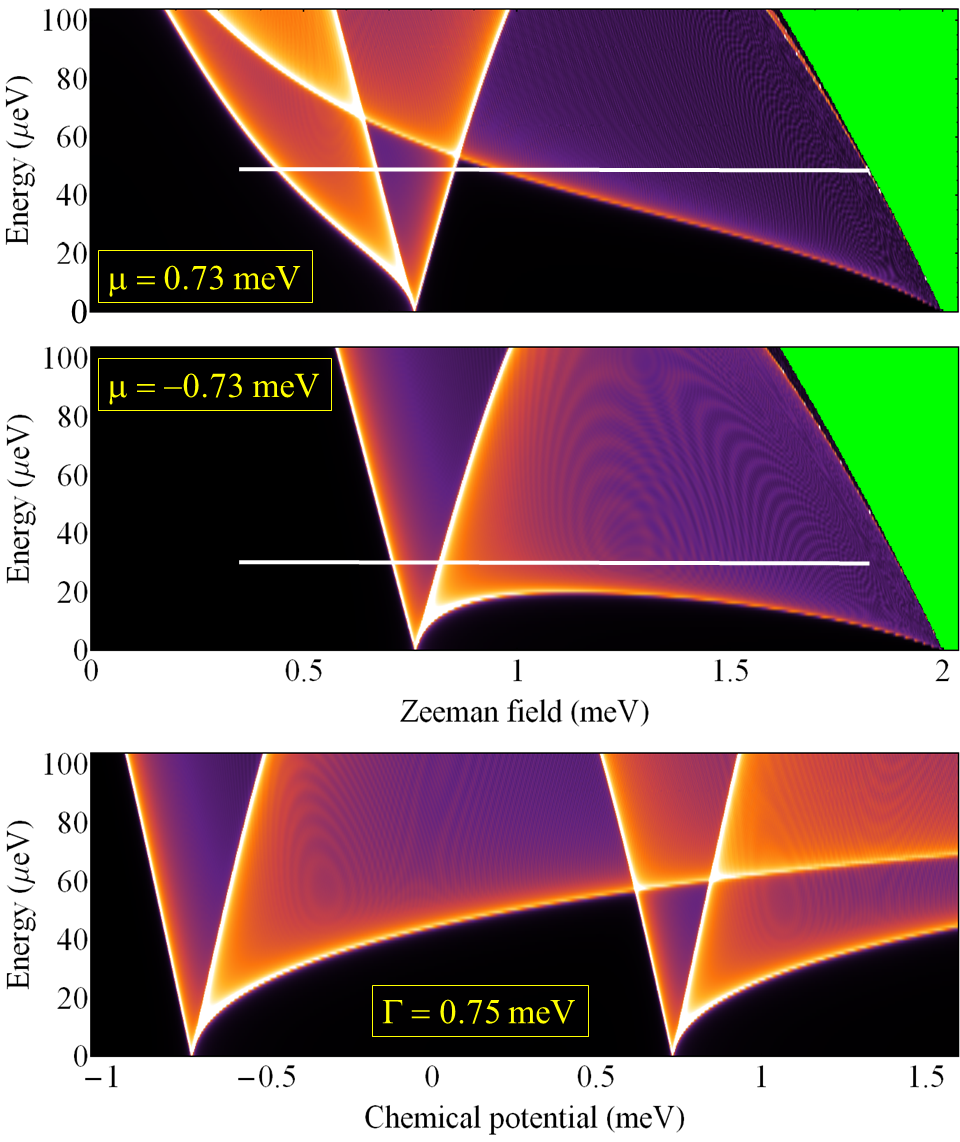}
\end{center}
\vspace{-3mm}
\caption{Density of states as function of energy and Zeeman field (upper two panels) or chemical potential (lower panel) corresponding to the cuts marked by light blue lines in Fig. \ref{FIG3}. White/light yellow shades indicate high DOS. The green region is above the collapsing parent SC gap (see Fig. \ref{FIG1}). Note that the quasiparticle gap vanishes at points corresponding to the topological phase boundary (see Fig. \ref{FIG3}), i.e., $\mu = \pm 0.73~$meV and $\Gamma =0.75~$meV or $\Gamma =\Gamma_0= 2~$meV. The spectral function along the cuts marked by horizontal white lines is shown in Fig. \ref{FIG5}. In all figures (showing DOS or LDOS results) throughout this section brighter (darker) colors indicate higher (lower) DOS values, with white corresponding to the maximum value and black indicating zero DOS.
}
\label{FIG4}
\vspace{-1mm}
\end{figure}
In addition, $\Gamma = \Gamma_0$ defines the boundary between the gapped SC phases and the large-field (trivially) gapless phase.  The corresponding topological phase diagram is shown in Fig. \ref{FIG3}, with the yellow, red, and dark gray areas representing the trivial SC, topological SC, and gapless phases, respectively. Note that, within the approximations used in this work (i.e., single band model, local proximity-induced self-energy, no magnetic field orbital effects) the boundary of the topological SC phase is uniquely determined by two parameters: the effective SM-SC coupling ($\gamma$, which controls the minimum value of the Zeeman field associated with the TQPT) and the critical field associated with the collapse of the parent SC gap ($\Gamma_0$, which controls the upper limit of the gapped phases). In particular, for strongly coupled systems (large $\gamma$) with low values of the SM g-factor or low critical field $B_0$ (implying reduced values of $\Gamma_0$) the topological phase of the ideal system shrinks and eventually disappears for $\Gamma_0\leq \gamma$.  Finally, we note that, in practice, the chemical potential can be controlled using electrostatic gates. In turn, the applied electrostatic potential changes the transverse profile of the wave functions in the SM wire and, implicitly, the effective SM-SC coupling $\gamma$ \cite{Woods2018Effective,Antipov2018Effects,Mikkelsen2018Hybridization}. This effect can be incorporated into the model by assuming $\gamma=\gamma(\mu)$. The phase diagram in Fig. \ref{FIG3} is obtained within the assumption that the variation of the effective SM-SC coupling over the relevant chemical potential range, $|\mu|\lesssim 2~$meV, is negligible. However, we emphasize that within the weak/intermediate SM-SC coupling regime, the shape of the phase boundary is basically controlled by $\gamma(0)$ and is weakly dependent on $\gamma(\mu)-\gamma(0)$. This behavior is a consequence of Eq. (\ref{Eq17}) giving $\Gamma_c(\mu) \approx |\mu| + \gamma(\mu)^2/2|\mu|\approx |\mu|$ for $|\mu| \gtrsim 3\gamma(\mu)$, i.e., almost everywhere, except the ``tip'' of the topological region, where $\gamma(\mu)\approx\gamma(0)$. 

\begin{figure}[t]
\begin{center}
\includegraphics[width=0.48\textwidth]{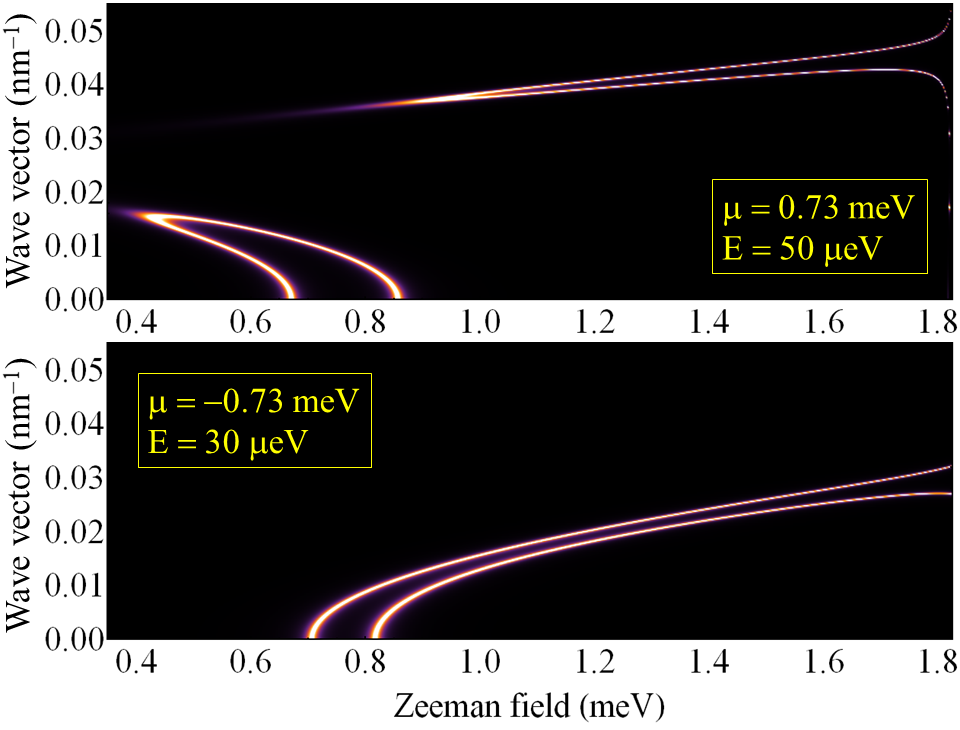}
\end{center}
\vspace{-3mm}
\caption{Dependence of the spectral function $A(\omega, k)$ on Zeeman field and wave vector along the cuts marked by white lines in Fig. \ref{FIG4}. Note that the high DOS V-shaped features in Fig.\ref{FIG4} are associated with long wavelength states ($k \approx 0$). The low-field, low-k branch in the upper panel is associated with the higher energy spin subband (HESS), while the the high-field, high-k branch in the upper panel and the spectral features in the lower panel are associated with the lower energy spin subband (LESS). A schematics of the spin subbands is shown in Fig. \ref{FIG6}.  Note that, upon approaching the gapless regime, $\Delta(\Gamma)\searrow \omega$ (i.e., $\Gamma\rightarrow 1.83~$meV in the upper panel), the spectral features loose weight and diverge either toward large k values (upper branch) or toward $k=0$ (lower branch).}
\label{FIG5}
\vspace{-1mm}
\end{figure}

To illustrate the dependence of the quasiparticle gap on the control parameters, we calculate the DOS as a function of energy (within the low-energy window $0\leq \omega\leq 100~\mu$eV) and Zeeman field or chemical potential (corresponding to the cuts along the light blue lines in Fig. \ref{FIG3}). 
Note that, for a translation-invariant system, we have $\rho(\omega) = \sum_k  A(\omega, k)$, with the spectral function $A(\omega, k) = (-1/\pi) {\rm Im}{\rm Tr}[G_{SM}(\omega+i\eta, k)]$. The results are shown in Fig. \ref{FIG4}. As expected, upon approaching the topological phase boundary $\Gamma = \Gamma_c(\mu)$ by varying the Zeeman field or the chemical potential, the quasiparticle gap closes, then, after crossing into a topologically different SC phase it reopens. The vanishing of the quasiparticle gap at the TQPT is associated with a V-shaped feature characterized by high DOS (white shading in Fig. \ref{FIG4}). In addition, the quasiparticle gap vanishes at the critical field $\Gamma = \Gamma_0$, where the parent SC gap collapses and the entire system becomes gapless.  As general trends, we note that the topological gap (i.e., the quasiparticle gap characterizing the system with control parameters within the red region in Fig. \ref{FIG3}) typically increases with the chemical potential and, for $\mu >0$, it typically decreases with increasing $\Gamma$. For the model parameter values used in this study the maximum topological gap is about $60~\mu$eV, which is consistent with the clean-limit estimates for InAs-Al systems being used currently.

\begin{figure}[t]
\begin{center}
\includegraphics[width=0.4\textwidth]{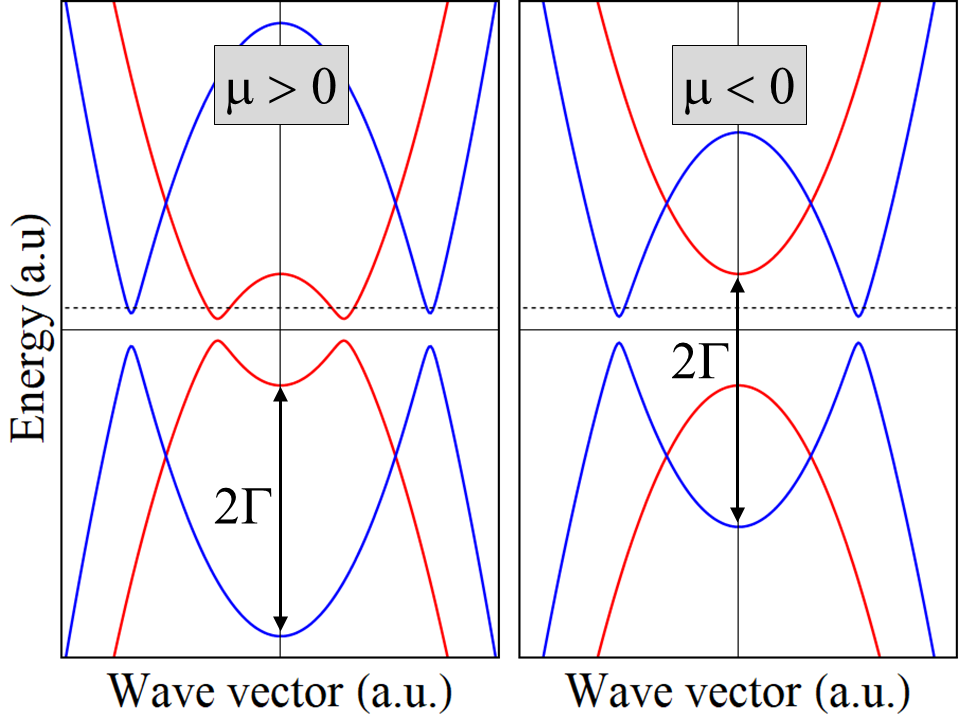}
\end{center}
\vspace{-3mm}
\caption{Schematic representation of the spin subbands for positive (left panel) and negative (right) values of the chemical potential in the presence of a finite Zeeman field. The characteristic k-vectors for quasiparticles having energy $\omega$ correspond to the intersections with the dashed horizontal lines.   For $\mu>0$ there are low-energy quasiparticles associated with both the higher energy (red lines) and the lower energy (blue lines) spin subbands, while for $\mu<0$ the are only contributions from the  lower energy subband. Note that this schematics does not incorporate the effects of proximity-induced energy renormalization.}
\label{FIG6}
\vspace{-1mm}
\end{figure}

To better understand the nature of the states responsible for various low-energy features, we calculate the spectral function $A(\omega, k)$ along the cuts marked by white lines in Fig.  
\ref{FIG4}. The results are shown in Fig. \ref{FIG5}. In addition, we have determined the spin structure of various contributions by calculating the spin-resolved spectral function $A_\sigma(\omega, k)$ for the spin direction parallel to the applied Zeeman field (not shown).  
The key conclusions are the following. First, the high DOS V-shaped features in Fig. \ref{FIG4} responsible for the closing and reopening of the quasiparticle gap at the TQPT are associated with long wavelength states having $k \ll 0.01~$nm$^{-1}$. In the presence of disorder, these long wavelength states are the first to become localized. Second, for negative values of $\mu$ (i.e., chemical potential below the bottom of the SM band at zero magnetic field) the low-energy spectral contributions are associated with the lower-energy spin subband (Fig. \ref{FIG5}, bottom panel), while for $\mu >0$ there are contributions from both spin subbands   (Fig. \ref{FIG5}, top panel). The contribution associated with the higher energy spin subband is characterized by relatively low wave vector values (e.g., $k<0.02~$nm$^{-1}$ in Fig. \ref{FIG5}), while the contribution from the lower energy spin subband has significantly larger characteristic wave vectors (e.g., $k>0.035~$nm$^{-1}$ in Fig. \ref{FIG5}, top panel). This behavior can be easily understood by considering the schematic spin subband structure shown in Fig. \ref{FIG6}. 
We note that for positive (and relatively large) values of the chemical potential the topological gap is (mostly) controlled by states with large characteristic wave vectors associated with the lower energy spin subband. In turn, these states are expected to be more robust against localization by disorder (as compared to the low-k states), i.e., more delocalized. By contrast, the low-k states associated with the higher energy spin subband are expected to become localized even in the presence of relatively weak disorder. This behavior will be investigated in detail in the next section. Our third conclusion concerns the behavior of the spectral function in the vicinity of gapless regime, where 
the parent gap approaches $\omega$ (from above), $\Delta(\Gamma)\searrow \omega$ (see top panels of Figs. \ref{FIG4} and \ref{FIG5}). In this regime, the ratio $\gamma/\sqrt{\Delta(\Gamma)^2 -\omega^2}$ in Eq. (\ref{Eq16}) diverges, which implies strong energy and quasiparticle renormalization \cite{Stanescu2017Proximity}. On the one hand, this leads to a reduced quasiparticle weight. On the other hand, it results in the effective ``flattening'' of the lower energy spin subband (blue lines in Fig. \ref{FIG6}) -- the subband with low-energy contributions at large $\Gamma$ values -- and the ``migration'' of the two branches of characteristic quasiparticle wave vectors toward large $k$ values and toward $k=0$, respectively. Both effects can be clearly observed in the upper panel of Fig. \ref{FIG5} for $\Gamma\rightarrow 1.83~$meV. 

\subsection{Low-energy disorder effects in long Majorana systems}\label{S3_2}

\begin{figure}[t]
\begin{center}
\includegraphics[width=0.46\textwidth]{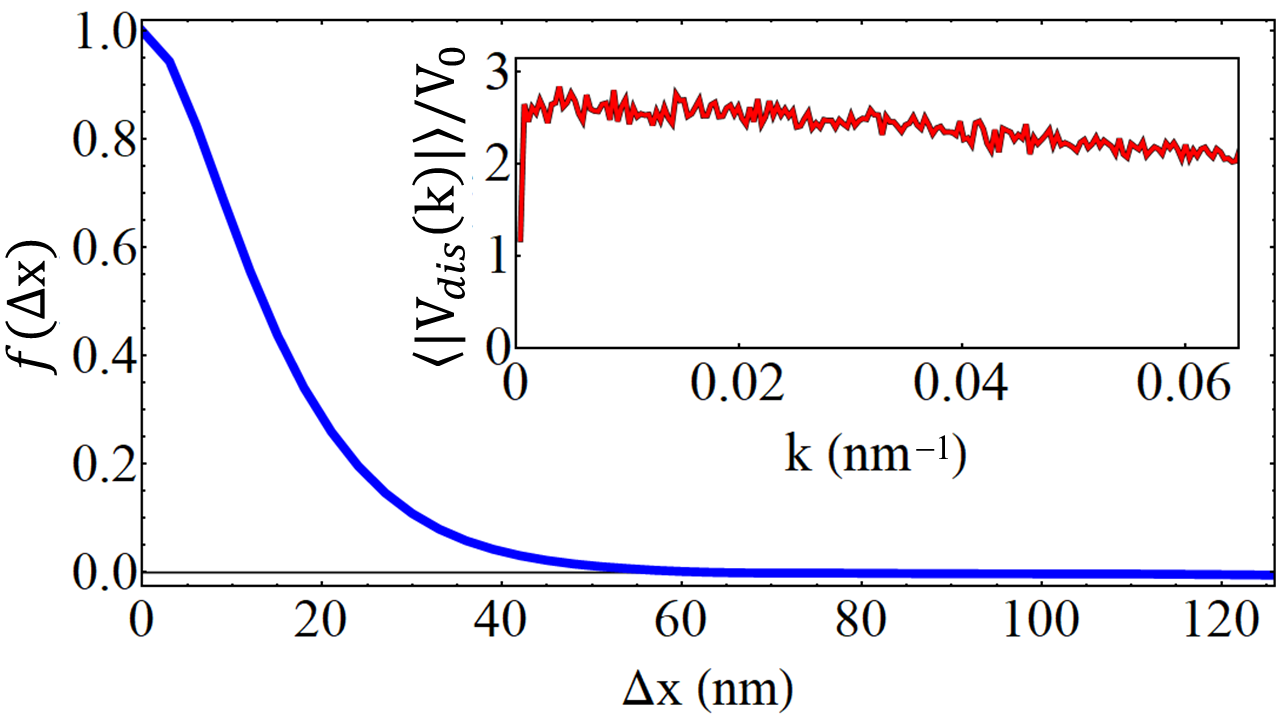}
\end{center}
\vspace{-3mm}
\caption{{\em Main panel}: Dependence of the characteristic correlation function given by Eq. (\ref{Eq19}) on distance. Note that the disorder correlation length is about $15~$nm. {\em Inset}:  Spectral signature of the disorder potential corresponding to the disorder-averaged absolute value of the Fourier transformed $V_{dis}$ given by Eq. (\ref{Eq20}). The results were obtained by averaging over 500 disorder realizations.}
\label{FIG7}
\vspace{-1mm}
\end{figure}

The effects of disorder on the low-energy physics of the hybrid structure are investigated based on a correlated disorder potential of the form 
\begin{equation}
V_{dis}(i) = V_0 \left[\sum_{p=1}^{N_{dis}}\frac{ (-1)^p}{2} \exp\left(-\frac{|i - j_p|}{\lambda}\right) - \overline{\cal V}\right],  \label{Eq18}
\end{equation}
where $V_0$ is the amplitude of the disorder potential and $N_{dis}$ represents the number of ``effective impurities'' having random locations $j_p$ and generating impurity potentials $V_{p}(i) = (-1)^p/2 \exp(-|i - j_p|/\lambda)$. The parameter $\overline{\cal V}$ is the average (over the lattice index $i$) of the first term in the square parentheses, which ensures that the position average of $V_{dis}$ is zero (i.e., there is no overall shift of the chemical potential). In the numerical calculations, for a $40~\mu$m long wire, we have $N_{imp} = 2\cdot 10^4$ (i.e., five effective impurities per micron) and $\lambda = 0.8$ (i.e., $\lambda a = 8~$nm). The disorder potential has a characteristic correlation function 
\begin{equation}
\langle V_{dis}(i)V_{dis}(i^\prime)\rangle = V_0^2 f(\Delta x),  \label {Eq19}
\end{equation}
with $\Delta x = |i - i^\prime| a$. The dependence of the correlation function $f$ on $\Delta x$ is shown in Fig. \ref{FIG7}. To further characterize the disorder potential, we calculate the {\em spectral signature} defined as the disorder-averaged absolute value of the Fourier transformed disorder potential, $\langle|V_{dis}(k)|\rangle$, with
\begin{equation}
V_{dis}(k) = \sqrt{\frac{2}{N+1}} \sum_{i=1}^N V_{dis}(i) \sin(i k a). \label{Eq20}
\end{equation}
Note that each disorder realization corresponds to a different set of (random) impurity positions $\{j_p\}_{p=\overline{1,N_{dis}}}$ in Eq. (\ref{Eq18}). The result is shown in Fig. \ref{FIG7} as an inset. The key feature is that the spectral signature of the disorder potential is almost constant over the relevant range of wave vector values (also see Fig. \ref{FIG5}), which implies that the disorder potential will have comparable impact on all low-energy states. Note that, as a result of the position average of $V_{dis}$ being zero (by construction), we have $V_{dis}(k=0)=0$.  We note that our model disorder is qualitatively consistent with the estimated effective disorder corresponding to experimental SM-SC structures, but we make no effort to ``realistically'' simulate a specific SM-SC structure.

\begin{figure}[t]
\begin{center}
\includegraphics[width=0.49\textwidth]{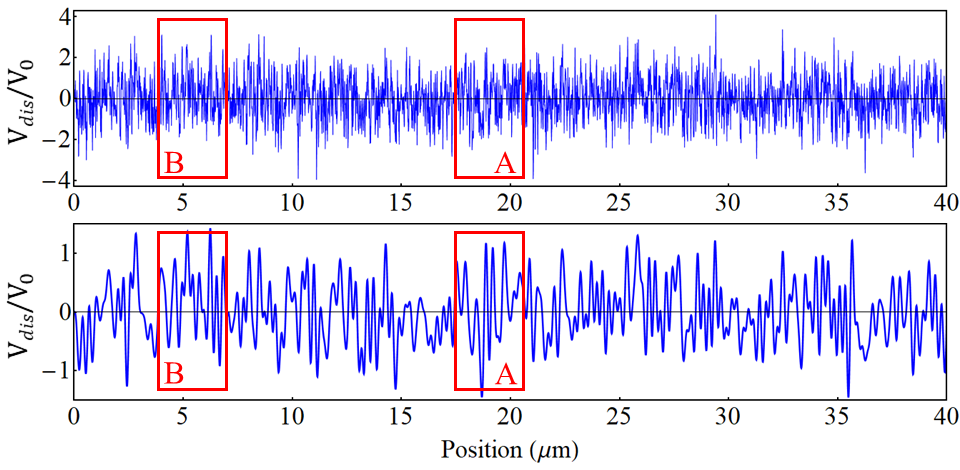}
\end{center}
\vspace{-3mm}
\caption{{\em Top}: Position dependence of the disorder potential used in the calculations discussed in this section. {\em Bottom}: ``Smooth'' disorder potential obtained from the Fourier components of the potential in the upper panel with wave vectors $k \lesssim 0.025~$nm$^{-1}$. The original and ``smooth'' potentials have similar impacts on the long-wavelength (low-$k$) states (see Fig. \ref{FIG5}). The red rectangles mark $3~\mu$m segments that will be studied in Sec. \ref{S3_3} as disorder realizations ``A'' and ``B'' in a short wire.}
\label{FIG8}
\vspace{-1mm}
\end{figure}

Next, we focus on the particular disorder realization that will be used in subsequent calculations. The dependence of the corresponding disorder potential, $V_{dis}$,  on the position along the wire is shown in the top panel of Fig. \ref{FIG8}. We also calculate a ``smooth'' disorder potential obtained by suppressing all Fourier components of the original potential with $k\gtrsim 0.025~$nm$^{-1}$ (see bottom panel of Fig. \ref{FIG8}). We note that the original and the ``smooth'' disorder potentials have similar impacts on the long-wavelength (low-$k$) states (see Fig. \ref{FIG5}), but the effect of the  ``smooth'' potential on the large-$k$ states is minimal. The features of the disorder potential that are effective in localizing long-wavelength states can be identified as features of the ``smooth'' potential, e.g., deep minima in Fig. \ref{FIG5} (lower panel). 

\begin{figure}[t]
\begin{center}
\includegraphics[width=0.49\textwidth]{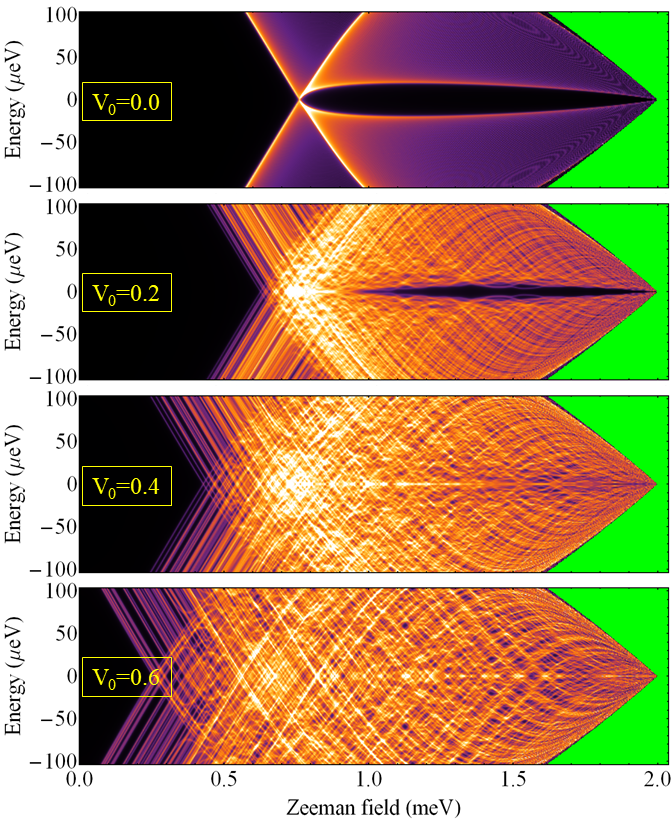}
\end{center}
\vspace{-3mm}
\caption{Dependence of the low-energy DOS on energy and applied Zeeman field for a $40~\mu$m-long Majorana ring with $\mu=-0.73~\mu$eV and disorder potential corresponding to the top panel of Fig. \ref{FIG8} with overall amplitude $V_0$.  The values of $V_0$ are given in meV. The quasiparticle gap is reduced by disorder and collapses above a certain (disorder-dependent) value of the Zeeman field for $V_0\gtrsim 0.4~$meV.}
\label{FIG9}
\vspace{-1mm}
\end{figure}

After this preparatory work, we are finally ready to investigate the impact of disorder on the TQPT. We start with the Zeeman field dependence of the low-energy DOS for a system with $\mu=-0.73~$meV, which corresponds to the middle panel of Fig. \ref{FIG4} in the ``ideal'' case. Specifically, we consider a $40~\mu$m-long Majorana ring in the presence of the disorder potential shown in the top panel of Fig. \ref{FIG8} with overall amplitude $V_0$. The dependence of the low-energy DOS on the Zeeman field for different $V_0$ values  is shown in Fig. \ref{FIG9}. First, we note that the result for the clean system ($V_0=0$) is practically identical to the ``ideal'' case illustrated in Fig. \ref{FIG4}, which demonstrates that the system under investigation is long-enough ($L= 40~\mu$m) within our energy resolution ($\eta = 2~\mu$eV). We point out that, unlike Fig. \ref{FIG4}, the energy window in Fig. \ref{FIG9} includes both positive and negative values,  $-100 \leq \omega \leq 100~\mu$eV.  Second, we note that the quasiparticle gap in the nominally topological phase, which has a maximum value of about $20~\mu$eV in the clean system, gets reduced by disorder and completely collapses for $V_0\gtrsim 0.4~$meV. Hence, for $V_0\gtrsim 0.4~$meV the system has quasiparticles with arbitrarily low energy for all values of the Zeeman field above 
$\Gamma^*(V_0) < \Gamma_c(\mu) \approx 0.76~$meV. We note that, upon increasing the size of the system, these low energy states form a quasi-continuum. We also remind the reader that the system is a Majorana ring, hence these low energy states do not include possible MZMs localized near the ends of an open boundary system. 

\begin{figure}[t]
\begin{center}
\includegraphics[width=0.49\textwidth]{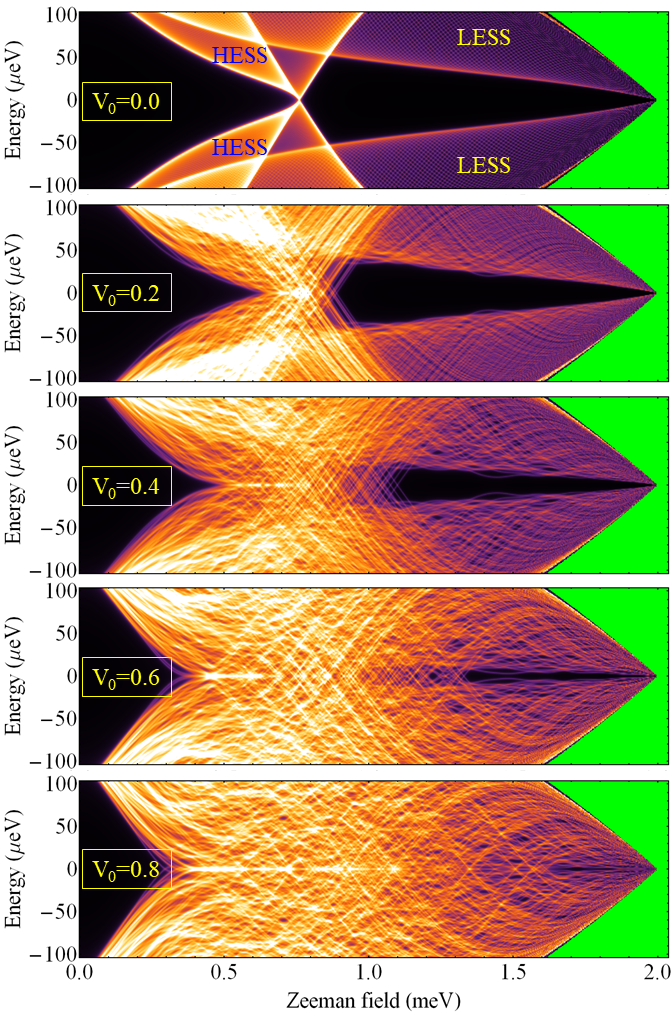}
\end{center}
\vspace{-3mm}
\caption{Dependence of the low-energy DOS on energy and Zeeman field  for a Majorana ring of length $L=40~\mu$m having chemical potential $\mu=0.73~$meV and different values of the disorder strength $V_0$ (with values given in meV). For the clean system (top panel), the contributions associated with the lower/higher energy spin subbands (LESS/HESS) are explicitly marked. The corresponding characteristic wave vectors are given in Fig. \ref{FIG5} (top panel). For $V_0\gtrsim 0.8~$meV, the quasiparticle gap collapses above a Zeeman field of the order $\gamma$ (where $\gamma$ is the effective SM-SC coupling).}
\label{FIG10}
\vspace{-1mm}
\end{figure}

To get further insight, we also consider the positive chemical potential regime.  In Fig. \ref{FIG10} we show the energy and Zeeman field  dependence of the DOS for a system with $\mu=0.73~$meV and different values of the overall disorder potential amplitude, $V_0$. The maximum value of the topological gap (in the clean system) is about $55~\mu$eV. Upon increasing the disorder strength, the system becomes gapless (above a certain Zeeman field value) for $V_0\gtrsim 0.8~$meV. Again, increasing the size of the system results in a quasi-continuous gapless spectrum.
The crucial feature revealed by the sequence shown in Fig. \ref{FIG10} is that two different ``mechanisms'' contribute to the extinction of the quasiparticle gap, as the disorder strength increases: the expansion of the linearly dispersing gapless modes associated (mainly) with HESS states toward higher Zeeman fields and the collapse of the quasiparticle gap associated with LESS states. The first mechanisms is controlled by low-$k$ states within the higher energy spin subband (HESS), while the second mechanism is associated with  high-$k$ states from the lower energy spin subband (LESS). The corresponding characteristic $k$ vectors are given in Fig. \ref{FIG5} (top panel). 

\begin{figure}[t]
\begin{center}
\includegraphics[width=0.49\textwidth]{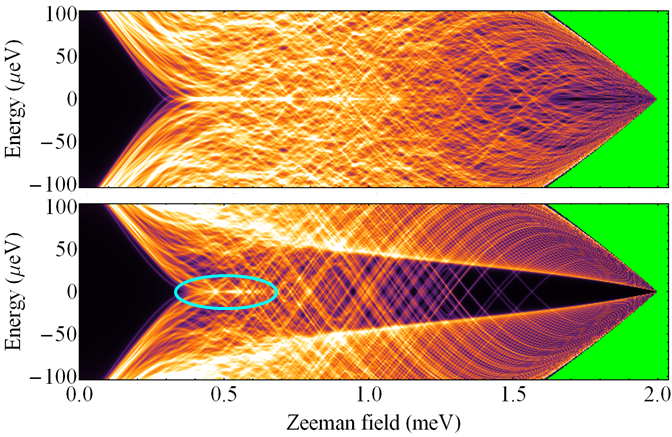}
\end{center}
\vspace{-3mm}
\caption{Comparison between the effects of the ``original'' and ``smooth'' disorder potentials shown in Fig. \ref{FIG8} (with overall amplitude $V_0=0.8~$meV) on a system with $\mu=0.73~$meV. The top panel is identical to the lowest panel in Fig. \ref{FIG10}. Note that the ``smooth'' potential (lower panel) has a minimal effect on the high-$k$ LESS states, but induces gapless HESS states over a large Zeeman field range. The zero energy modes near $\Gamma \approx 0.5~$meV (highlighted region) result from the disorder-induced collapse of the trivial gap associated with HESS states (also see Fig. \ref{FIG10}). In the upper panel, the zero energy DOS is augmented by contributions from LESS states.}
\label{FIG11}
\vspace{-1mm}
\end{figure}

\begin{figure}[t]
\begin{center}
\includegraphics[width=0.49\textwidth]{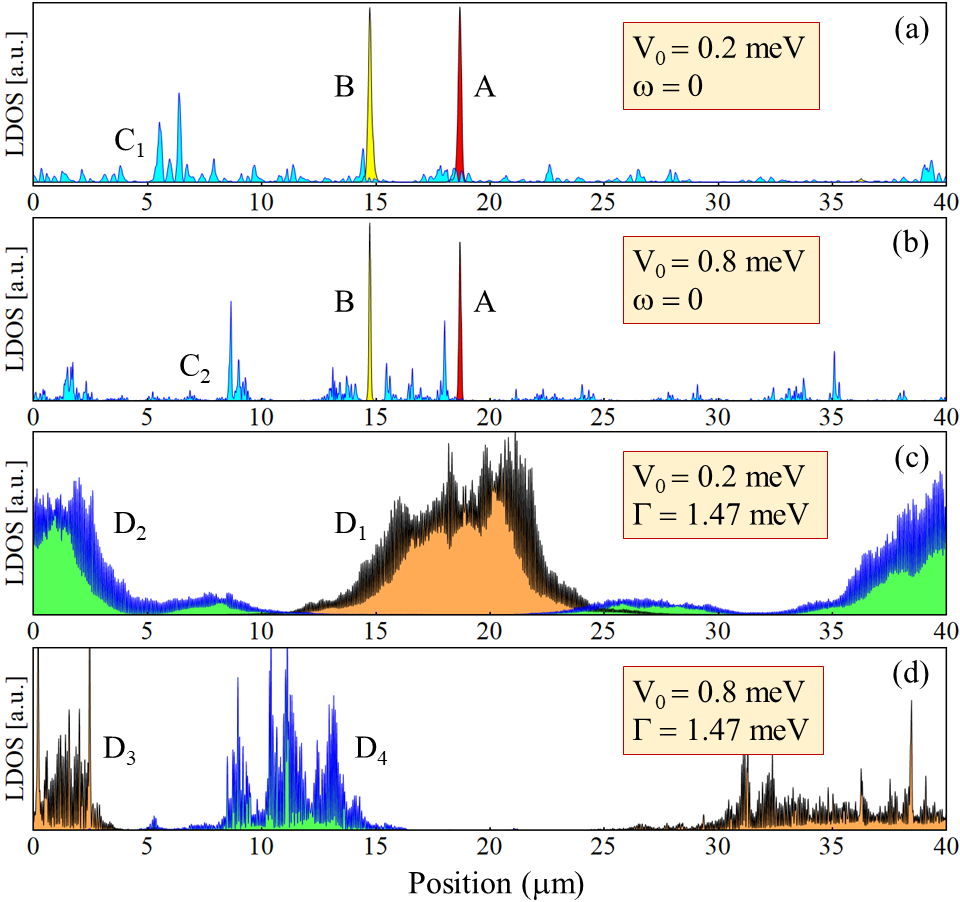}
\end{center}
\vspace{-3mm}
\caption{Position dependence of the LDOS associated with specific low-energy modes in a disordered Majorana ring with $L=40~\mu$m, chemical potential  $\mu=0.73~$meV, and additional parameters corresponding to representative features in Fig. \ref{FIG10}.  Note that $x=0$ and $x = 40~\mu$m represent the same point on the ring. 
Modes A (red filling) and B (yellow filling) are associated with zero energy crossings occurring at Zeeman field values (a) $\Gamma_A = 0.889~$meV and $\Gamma_B = 0.877~$meV (for $V_0=0.2~$meV) and (b) $\Gamma_A = 1.568~$meV and $\Gamma_B = 1.564~$meV (for $V_0=0.8~$meV). These modes represent strongly localized disorder-induced Andreev bound states that cross zero energy at points that can be continuously traced within the $(\mu, \Gamma, V_0)$ parameter space. The zero-energy LDOS contributions $C_1$  (calculated at $\Gamma_{C_1} = 0.703~$meV) and $C_2$  (at $\Gamma_{C_2} =0.573~$meV), which result from the collapse of the quasiparticle gap in the vicinity of the TQPT of the clean system (see Fig. \ref{FIG10}), consist of highly delocalized modes  extending throughout the entire system.  $D_1-D_4$ are finite energy contributions associated with LESS modes near the edge of the ``partial gap''. The corresponding energies are: (c) $\omega_{D_1}=22.2~\mu$eV and  $\omega_{D_2}=25.6~\mu$eV; (d)  $\omega_{D_3}=2.7~\mu$eV and  $\omega_{D_4}=4.6~\mu$eV. Note that for $V_0=0.8~$meV the ``partial gap'' has practically collapsed (see Fig. \ref{FIG10}).}
\label{FIG12}
\vspace{-1mm}
\end{figure}

The most straightforward way to disentangle the two mechanisms is to consider the system in the presence of the ``smooth'' potential shown in the lower panel of Fig. \ref{FIG8} that mainly affects the long wavelength HESS states. A comparison between the effects of the ``original'' and ``smooth''  potentials for a system with chemical potential $\mu=0.73~$meV  is shown in Fig. \ref{FIG11}. First, note that the LESS states are weakly affected by the ``smooth'' potential, being characterized by a quasiparticle gap similar to that of a clean system (for comparison, see Fig. \ref{FIG10}, top panel). By contrast, the HESS spectrum, which in the clean system is gapped everywhere except at the critical point $\Gamma = \Gamma_c \approx 0.76~$meV,  
becomes essentially gapless over a wide Zeeman field range, $0.38 \lesssim \Gamma \lesssim 1.57~$meV.  Taking the limit $L\rightarrow \infty$ and assuming a properly bounded disorder potential  (which essentially limits the maximum Zeeman field associated with zero-energy crossings for given values of the disorder strength  and chemical potential), the system becomes gapless for {\em all} Zeeman field values within a certain range. As shown below (Fig. \ref{FIG12}), these low-energy modes (mainly) associated with low-$k$ HESS states are localized around local minima of the ``smooth'' potential, which are scattered throughout the ring. The bounded disorder practically limits the strength of these local minima. Alternatively, if we consider, for example, Gaussian disorder, rare strong local minima can occur, which implies that, for a given value of the overall disorder amplitude, $V_0$, low-energy modes localized near these minima can  emerge at arbitrarily large Zeeman field values. Nonetheless, the key property of the low-energy spectrum in the presence of the ``smooth'' disorder potential is the existence of a ``partial'' gap associated with the high-$k$ LESS states (see lower panel of Fig. \ref{FIG11}).  
We  note that for $\mu<0$ the low-energy spectral features are associated with a single spin  subband and, consequently, there is a smooth crossover between  low-$k$ and large-$k$ contributions (see Fig. \ref{FIG5}), hence, between the two ``mechanisms''. However, since the topological phase is most 
stable in the positive chemical potential regime (which, consequently, is the most relevant regime) and because various system parameters affect low-$k$ and large-$k$ states differently, it is practically useful to distinguish between the two ``mechanisms''. 

The relative importance of the two ``mechanisms'' that contribute to the collapse of the quasiparticle gap is determined by the system parameters. On the 
one hand, the maximum ``width'' of the topological phase along the Zeeman field axis is controlled by the effective SM-SC coupling $\gamma$ and by the 
critical Zeeman field $\Gamma_0$ associated with the collapse of the parent SC gap. Specifically, for a clean system the topological phase corresponds to  $
\sqrt{\gamma^2+\mu^2} \leq \Gamma\leq \Gamma_0$, with a maximum width equal to $\Gamma_0-\gamma$. Increasing the disorder strength, $V_0$, results in the emergence (within both the trivial and topological phases) of (isolated) gapless modes localized near minima of the (``smooth'') disorder potential. In addition, increasing disorder reduces that quasiparticle gap that characterizes the large-$k$ states associated with the lower energy spin subband; this gap is controlled by the parent gap, $\Delta_0$, the spin-orbit coupling, $\alpha$, and the effective SM-SC coupling, $\gamma$ (see below, Fig. \ref{FIG15}). Finally, the relative impact of disorder on the low- and high-$k$ states depends on the characteristic length scale of the disorder potential, as shown explicitly in Fig. \ref{FIG11}. The main focus of this study is to understand in detail the low-energy properties of the system in the regime characterized by the (near) collapse of the ``partial quasiparticle gap'' associated with large-$k$ states.  

At this point, it is useful to provide more detail on the real space properties of the states responsible for the low-energy features shown in Fig. \ref{FIG10}. More specifically, we calculate the position dependence of the LDOS associated with: (i) modes characterized by a linear low-energy dispersion,   which generate distinct zero-energy crossings at certain (discrete) values of the Zeeman field; (ii) modes associated with the zero-energy DOS in the vicinity of the critical  Zeeman field $\Gamma_c$ corresponding to the TQPT of the clean system; (iii) finite energy LESS modes near the edge of the ``partial gap'' associated with high-$k$ states. Several representative examples are shown in Fig. \ref{FIG12}. First, we consider the linearly dispersing modes (associated with low-$k$ states) that cross zero energy at certain values of the Zeeman field. Modes A and B shown in Fig. \ref{FIG12} (a) and (b) are two specific examples. In general, these modes are strongly localized near minima of the ``smooth'' potential. For example, modes A and B are localized near the minima  at $x\approx 18.4~\mu$m and $x\approx 14.5~\mu$m, respectively (see lower panel of Fig. \ref{FIG8}). We note that the characteristic length scale of these state is (typically) less that one micron and decreases as one increases the amplitude of the disorder potential. We also note that the points at which the energy of these modes vanishes can be continuously traced within the $(\mu, \Gamma, V_0)$ parameter space; panels (a) and (b) of Fig. \ref{FIG12} show these modes at two different sets of parameters.  

\begin{figure}[t]
\begin{center}
\includegraphics[width=0.48\textwidth]{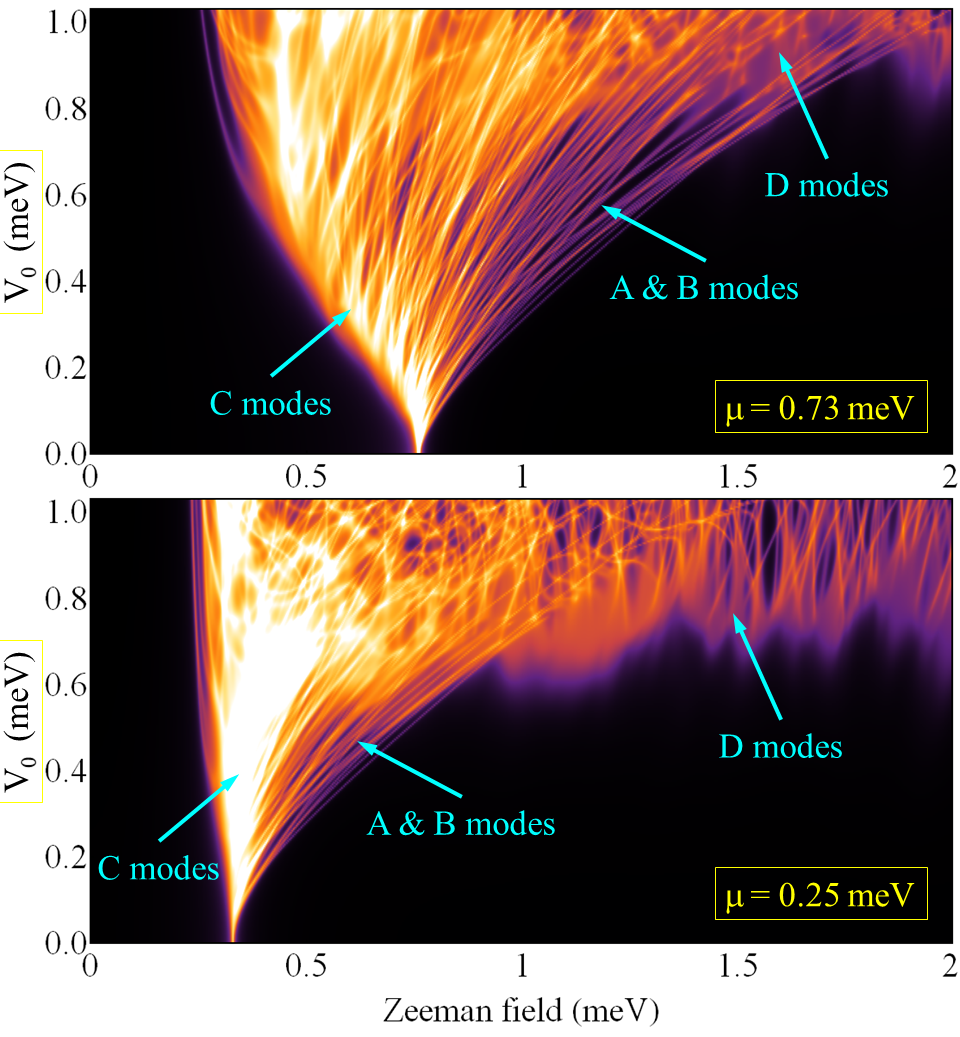}
\end{center}
\vspace{-3mm}
\caption{Map of the zero-energy DOS as function of Zeeman field and disorder strength for a Majorana ring of length $L=40~\mu$m with chemical potential  
$\mu=0.73~$mev  (top) and $\mu=0.25~$mev (bottom). The collapse of the quasiparticle gap associated with large-$k$ states occurs at $V_0 \approx 0.9~$meV in the top panel and $V_0 \approx 0.7~$meV in the lower panel. The modes A/B, C, D denote disorder-induced strongly localized, delocalized, and weakly localized states, respectively, as discussed in the text (also see Fig. \ref{FIG12})}
\label{FIG13}
\vspace{-1mm}
\end{figure}

\begin{figure*}[t]
\begin{center}
\includegraphics[width=\textwidth]{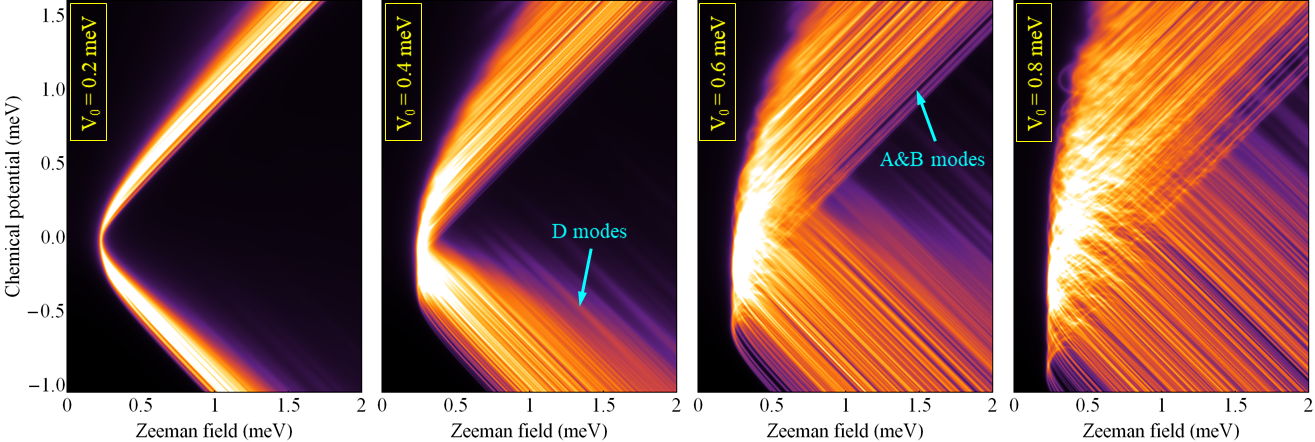}
\end{center}
\vspace{-3mm}
\caption{``Phase diagrams'' obtained by mapping the zero-energy DOS of a Majorana ring ($L=40~\mu$m) as function of Zeeman field and chemical potential for different values of the disorder strength. The gapless large-$k$ D-type modes first emerge at low (negative) values of the chemical potential, then, upon increasing $V_0$, expand into the region characterized by larger chemical values.}
\label{FIG14}
\vspace{-1mm}
\end{figure*}

Next, we consider zero energy modes in the vicinity of the critical  Zeeman field $\Gamma_c$ corresponding to the TQPT of the clean system. Specific examples include the states $C_1$ and $C_2$ (blue shading) shown in panels (a) and (b) of Fig. \ref{FIG12}. Note that these zero energy modes correspond to values of the Zeeman field lower than the critical value for the clean system, $\Gamma_c(\mu) \approx 0.76~$meV. The most relevant characteristic of these modes is that they are highly delocalized, with spectral weight distributed throughout the entire system. This feature is independent of the size of the system, in the sense that upon increasing $L$ one can still find $C$-type delocalized states, although typically within a narrower range of Zeeman fields. Below we investigate the dependence of this type of state on the control parameters (see Fig. \ref{FIG16}).

Finally, we consider $D$-type low-energy modes associated with large-$k$ (LESS) states. Specific examples are shown in panels (c) and (d) of Fig. \ref{FIG12}. There are several critical features associated with this type of states. First, although they are localized in the presence of disorder, their characteristic length scale is relatively large - several microns, up to tens of microns.  Consequently, in short wires (i.e., in systems with lengths of a few microns) these states can be practically delocalized, i.e., they can extend throughout the entire system. Second, the large characteristic momenta associated with these modes are clearly revealed by the highly oscillatory nature of the corresponding  LDOS (see the shape of the black and blue lines corresponding to the $D$  modes in Fig. \ref{FIG12}). Third, the low-energy $D$-type modes are ubiquitous in the regime characterized by the collapse of the gap associated with large-$k$ states. This is in contrast with the highly localized modes associated with low-$k$ states (e.g., the A and B modes discussed above), which disperse (approximately) linearly and become gapped under small variations of the Zeeman field or chemical potential. Most importantly, in a long (but finite) wire the probability of having low-energy $D$-type modes near the ends of the system is significant (unlike the corresponding probability for strongly localized modes, which are relatively rare and isolated). 

Before investigating in more depth the properties of the low-energy LDOS in the regime characterized by gapless large-$k$ modes, let us determine more quantitatively the parameter values associated with this regime. In particular, we want to estimate the disorder strength above which the whole topological phase is characterized by the presence of (arbitrarily) low-energy  large-$k$ modes. First, we consider the dependence of the zero-energy DOS on the Zeeman field and disorder strength ($V_0$) for fixed values of the chemical potential. The results are shown in Fig. \ref{FIG13}. For $V_0=0$ (clean system), the points characterized by a non-vanishing zero-energy DOS (i.e., $\Gamma \approx 0.76~$meV in the top panel and  $\Gamma \approx 0.33~$meV in the bottom panel) mark the TQPT between the low-field trivial phase and the high-field topological phase. Note that the corresponding zero-energy mode is delocalized. In the presence of disorder ($V_0 >0$), the TQPT is expected to occur at values of the critical field different from those characterizing the clean system, but within the range characterized by high values of the zero-energy DOS (light yellow/white in Fig. \ref{FIG13}), where delocalized $C$-type modes can be found. The line-like features  that fan out of the high DOS region are associated with highly localized (low characteristic $k$) states (A \& B-type). The collapse of the quasiparticle gap associated with large-$k$ states is revealed by the presence of zero-energy D-type modes, which occur above a certain ($\mu$-dependent) disorder strength. Typically, this characteristic disorder strength increases with the chemical potential. However, we note that for $V_0 \gtrsim V_0^*\approx 1~$meV the entire parameter region that could host a topological phase becomes gapless (for D-type modes).  

\begin{figure}[t]
\begin{center}
\includegraphics[width=0.48\textwidth]{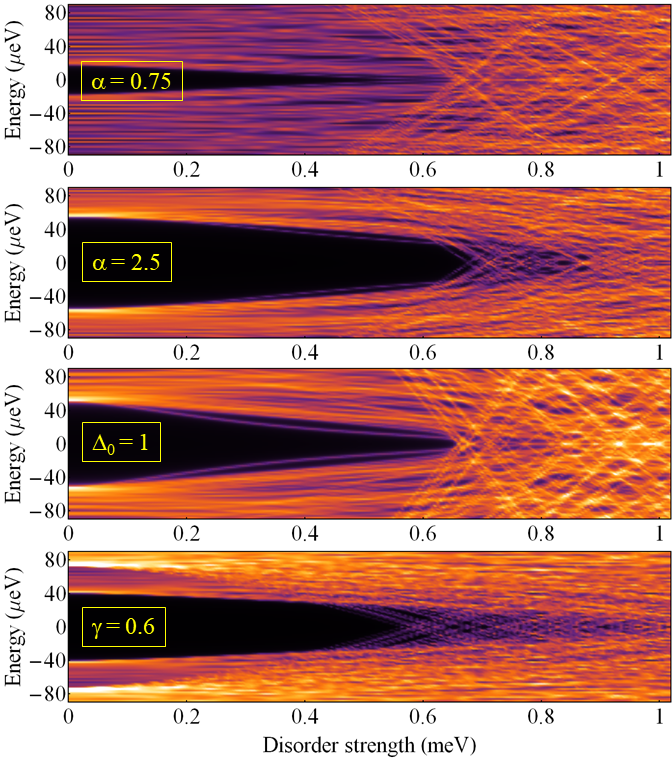}
\end{center}
\vspace{-3mm}
\caption{Dependence of the low-energy DOS on energy and disorder strength for a ring of length $L=40~\mu$m, chemical potential $\mu=0.5~$meV, Zeeman field $\Gamma= 1.4~$meV, and different system parameters. The parameter values that are {\em not} explicitly given in each panel (in units of meV) are $\alpha=1.25~$meV, $\Delta_0=0.3~$meV, and $\gamma=0.21~$meV. In the lowest panel ($\gamma=0.6~$meV), one can distinguish the gap edge associated with both LESS states ($\sim 40~\mu$eV in the clean limit) and HESS states ($\sim 70~\mu$eV in the clean limit).}
\label{FIG15}
\vspace{-1mm}
\end{figure}

To corroborate this picture, we also map the zero-energy DOS as function of the Zeeman field and chemical potential for different values of the disorder strength. The corresponding ``phase diagrams'' are shown in Fig. \ref{FIG14}. We note that, indeed, the (partial)  quasiparticle gap associated with D-type states first collapses at low (negative) values of the chemical potential; with increasing $V_0$, low-energy large-$k$ modes start to emerge at higher $\mu$ values and at $V_0=0.8~$meV they cover almost the entire topological region, consistent with our estimate of $V_0^*$.  In addition, we note that, with increasing disorder, the minimum Zeeman field at which zero-energy modes occur becomes weakly dependent on the chemical potential, with $\Gamma_{min} \approx \gamma$ (where $\gamma= 0.21~$meV is the effective SM-SC coupling) in the strong disorder limit. Finally, we point out that for a disorder strength $V_0=0.4~$meV, which is a factor of only about 2.5 less than $V_0^*$, the system in characterized by a large gapped topological region, with an area comparable to the area of the topological phase of a clean system. 

Before continuing our analysis, we point out that the estimate $V_0^*\approx 1~$meV corresponds to the specific system parameters used in this calculation, in particular Rashba spin-orbit coupling $\alpha_R=\alpha a = 125~$meV$\cdot$\AA, parent SC gap (at zero field) $\Delta_0 = 0.3~$meV, and effective SM-SC coupling $\gamma = 0.21~$meV. A natural question concerns the impact of these parameters on the disorder strength associated with the collapse of the large-$k$ quasiparticle gap. While a detailed quantitative characterization of this impact is beyond the scope of this work, we provide a qualitative characterization in Fig. \ref{FIG15}, which shows the dependence of the low-energy DOS on the disorder strength $V_0$ for a system with $L=40~\mu$m, $\mu=0.5~$meV, $\Gamma=1.4~$meV, and different values of $\alpha$, $\Delta_0$, or $\gamma$. First, by comparing the top two panels, we notice that enhancing the strength of the spin-orbit coupling (SOC) enhances the stability of the LESS quasiparticle gap against disorder. This is consistent with the well-known dependence of the topological gap (in clean systems) on the magnitude of the Rashba SOC coefficient \cite{Sau2010NonAbelian}. Next, enhancing the gap of the parent SC generates an overall enhancement of the quasiparticle gap, but does not affect the zero energy states  (except by increasing their relative weight within the SM wire). This behavior can be easily understood by analyzing the structure of the SM Green's function, e.g., in Eq. (\ref{Eq16}). Indeed, at $\omega=0$ the anomalous contribution becomes $\gamma \sigma_y \tau_y$ (i.e., independent of $\Delta_0$), while the quasiparticle residue, $Z= [1+\gamma/\Delta(\Gamma)]^{-1}$ only affects  the weight of the zero-energy modes, not their dependence on the system parameters. These considerations also hold in the presence of disorder. Finally, the lowest panel in Fig. \ref{FIG15} shows that increasing the effective SM-SC coupling enhances the stability of the LESS quasiparticle gap against disorder (in the SM). Note, however, that strong SM-SC coupling may also enhance the (possible adverse) effects induced by disorder in the parent SC \cite{Stanescu2022Proximity}, which is not included in this calculation. This can be particularly relevant in the vicinity of $\Gamma_0$ (the Zeeman field associated with the collapse of the parent SC gap), where $\gamma \gg \Delta(\Gamma)$ and the system is in the strong SM-SC coupling regime. Note that such a regime exists even for systems that are weakly coupled at zero field ($\gamma < \Delta_0$). In our case, for the parameters used throughout this work (except Fig. \ref{FIG15}), $\gamma > 2\Delta(\Gamma)$ for $\Gamma\gtrsim 1.6~$meV; the results corresponding to this regime should be interpreted with caution, as the stability against disorder may have been overestimated by neglecting disorder inside the parent SC.  
Finally, we point out that (i) for disorder strengths less than $V_0\approx 0.4~$meV the disorder effects are minimal for all parameter values and (ii) the best strategy for protecting the quasiparticle gap (other than reducing disorder) involves enhancing the SOC and SM-SC coupling strengths in combination with using a larger gap parent SC (to minimize the strong SM-SC coupling regime). In addition, this would reduce the characteristic length scales of the low-energy modes, which are large compared to the typical lengths of hybrid systems realized in the laboratory (see Fig. \ref{FIG12}; also Figs. \ref{FIG18}--\ref{FIG20} below). 

Returning to our main analysis, we address the question regarding the location of delocalized (C-type) zero-energy modes (see Fig. \ref{FIG12}) within the control parameter space. For specificity, we focus on a system with disorder strength $V_0=0.6~$meV and identify which of the modes that generate the zero-energy DOS shown in the corresponding panel of Fig. \ref{FIG14} are delocalized. To efficiently characterize a delocalized mode, we introduce following measure, which we dub the ``weakest link'',
\begin{equation}
W\!L = {\rm Min}_{i_0}\left[\sum_{i=i_0}^{i_0+\delta} \rho_L(0,i)\right], \label{WL}
\end{equation} 
where $\rho_L(\omega, i)$ is the local density of states (LDOS) at site $i$ and $\delta = \delta L/a$ defines a segment of the Majorana ring of length $\delta L$.  The site indices  are defined modulus $N$ (consistent with the ring geometry). The quantity $W\!L$ defined by Eq. (\ref{WL}) represents the lowest spectral weight of a zero-energy mode within an arbitrary segment of length $\delta L$. A system containing no zero-energy states or localized zero-energy modes will be characterized by small values of $W\!L$, while delocalized zero-energy modes will be associated with the maxima of $W\!L$, as their spectral weight is distributed throughout the entire system. 

\begin{figure}[t]
\begin{center}
\includegraphics[width=0.48\textwidth]{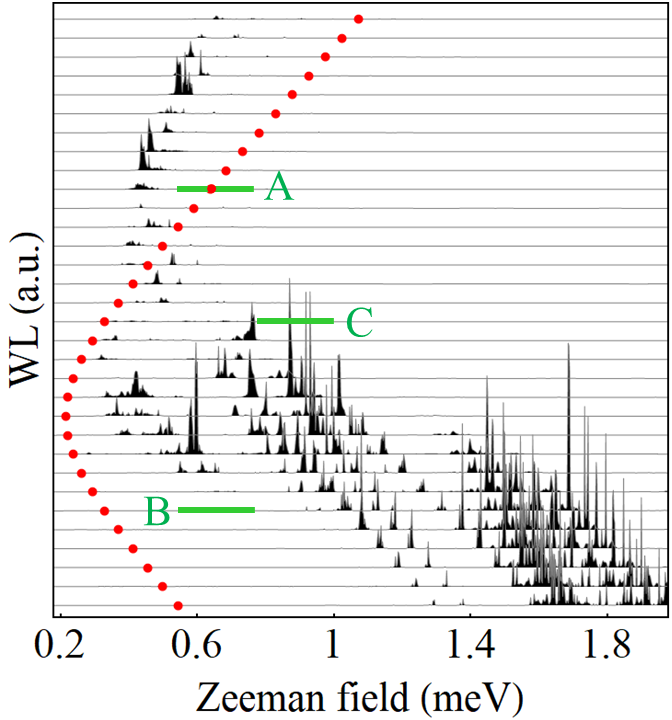}
\end{center}
\vspace{-3mm}
\caption{``Weakest link'' defined by Eq. (\ref{WL}) for a Majorana ring of length $L=40~\mu$m and disorder strength $V_0=0.6~$meV as function of the Zeeman field  for different chemical potential values ranging from $\mu=-0.5~$meV (bottom curve) to $\mu=1.05~$meV (top curve) in steps of $0.05~$meV. The curves are shifted for clarity. The maxima of $W\!L$ correspond to delocalized zero-energy modes. The red circles mark the topological phase boundary of a clean system. The low-energy spectral properties of the system along the cuts marked A, B, and C are investigated below (Figs. \ref{FIG17}--\ref{FIG20}).}
\label{FIG16}
\vspace{-1mm}
\end{figure}

The dependence of $W\!L$ on Zeeman field and chemical potential for a Majorana ring with $L=40~\mu$m and disorder strength $V_0=0.6~$meV is shown in Fig. \ref{FIG16}. For the calculation of $W\!L$ we have considered weak link segments of length $\delta L$ ranging from $500~$nm to $2~\mu$m; the results are essentially the same (up to an irrelevant overall factor). The range of Zeeman fields  associated with the presence of delocalized zero-energy modes (corresponding to maxima of $W\!L$) is relatively narrow for $\gtrsim 0.25~$meV, but becomes significant at lower values of the chemical potential. Nonetheless, one can clearly observe that for $\mu \gtrsim 0.4~$meV the delocalized zero-energy modes emerge outside the topological region associated with the clean system (i.e., to the left of the ``ideal'' topological phase boundary marked by red circles in Fig. \ref{FIG16}), while for low values of the chemical potential the delocalized zero-energy modes emerge well inside the ``ideal'' topological phase. This is consistent with the expected location of the topological phase boundary in the presence of disorder. Furthermore, upon reducing the disorder strength the location of the $W\!L$ maxima approaches the ``ideal'' topological phase boundary, which is expected based on the overall dependence of the parameter space region characterized by non-vanishing zero-energy DOS on disorder (see Fig. \ref{FIG14}). On the other hand, further increasing the disorder strength increases the range of Zeeman fields  associated with the presence of $W\!L$ maxima, which makes this method of estimating the location of the phase boundary unreliable in the strong disorder limit. In this context, we note that large values of $W\!L$ can be generated not only by delocalized C-type modes, but also by (essentially) localized D-type zero modes having characteristic length scales comparable to the size of the system. To eliminate these contributions, one has to increase the size of the system. 

\begin{figure}[t]
\begin{center}
\includegraphics[width=0.5\textwidth]{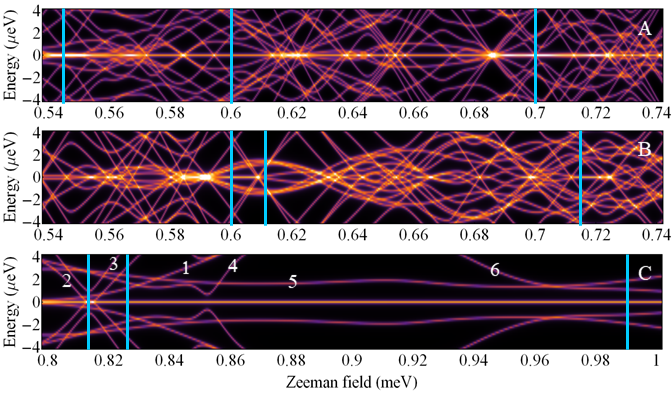}
\end{center}
\vspace{-3mm}
\caption{Low-energy DOS for a Majorana wire (i.e., a system with boundaries) along the cuts marked A, B, and C in Fig. \ref{FIG16}. The (fixed) system parameters are the same as in Fig. \ref{FIG16}. The chemical potential values corresponding to the three cuts are: (A) $\mu=0.6~$meV (top panel); (B) $\mu=-0.25~$meV (middle); (C) $\mu=0.25~$meV (bottom). In panels (A) and (C) one can clearly identify a zero-energy mode extending throughout the entire Zeeman field range associated with the corresponding cuts; the energy of this mode is less than $0.1~\mu$eV, i.e., much smaller than all other relevant energy scales. In panel (B) there are only short segments characterized by zero energy modes. The LDOS corresponding to the vertical (blue) cuts in panels (A), (B), and (C) are shown in Figs. \ref{FIG18}, \ref{FIG19}, and \ref{FIG20}, respectively. The finite energy modes labeled $1, 2, \dots, 6$ in panel (C) are explicitly identified in Fig. \ref{FIG20}.}
\label{FIG17}
\vspace{-1mm}
\end{figure}

Next, we focus on an open system with parameter values in the vicinity of the region characterized by $W\!L$ maxima. More specifically, we calculate the low-energy DOS along the cuts marked  A, B, and C in Fig. \ref{FIG16} for a Majorana wire, i.e., a system with open boundary conditions. We emphasize that all system parameters (including the disorder potential) are the same as in Fig. \ref{FIG16}, but we use open (instead of periodic) boundary conditions. 
The results are shown in Fig. \ref{FIG17}. First, we consider cut A, which extends on both sides of the clean topological boundary, but is positioned to the right of the region characterized by a $W\!L$ maximum (see Fig. \ref{FIG16}). The top panel of Fig. \ref{FIG17} reveals the presence of many low-energy modes (note that the energy range is $\pm 4~\mu$eV), some of them crossing zero energy or even ``sticking'' near zero energy  over some finite (relatively short) Zeeman field interval. However, the most notable feature is a zero energy mode that extends along the entire cut corresponding to a Zeeman field range $\Delta\Gamma = 0.2~$meV. We note that this zero-energy mode can be traced  all the way up to $\Gamma =\Gamma_0$ (not shown). Moreover, for $\Gamma\gtrsim 1.2~$meV this mode sits in the middle of a finite quasiparticle gap (see, e.g., the panel corresponding to $V_0=0.6~$meV in Fig. \ref{FIG14}) and can be clearly identified as a pair of Majorana zero modes (MZMs). On the other hand, in Fig. \ref{FIG17}(A) one can clearly notice variations of the spectral intensity characterizing the zero-energy mode. To unambiguously determine the nature of the zero-energy mode in cut A and identify the source of the spectral intensity variations, we calculate the local density of states (LDOS) as a function of energy and position along the wire for the representative cuts marked by blue lines  in Fig. \ref{FIG17}(A). The corresponding results are shown in Fig. \ref{FIG18}.
\begin{figure}[t]
\begin{center}
\includegraphics[width=0.48\textwidth]{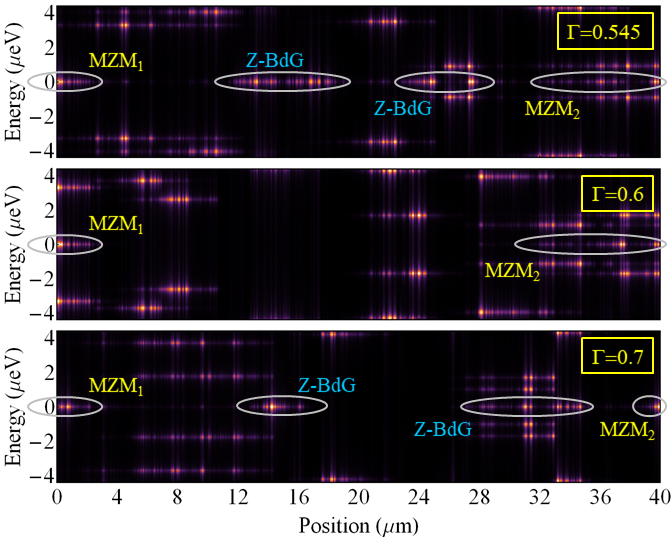}
\end{center}
\vspace{-3mm}
\caption{LDOS as function of position along the wire and energy corresponding to the vertical (blue) cuts in Fig. \ref{FIG17}(A). For all three values of the Zeeman field one can identify a pair of MZMs localized near the ends of the wire (MZM$_1$ and MZM$_2$, respectively). For $\Gamma=0.545~$meV and $\Gamma=0.7~$meV there are additional zero-energy modes (marked Z-BdG) representing ``regular'' BdG states localized inside the wire.  Note that some of the finite energy BdG states are localized near the ends of the wire (see, e.g., the middle panel).}
\label{FIG18}
\vspace{-1mm}
\end{figure}
We note the presence of zero-energy modes localized near the ends of the wire for all three values of the Zeeman field. To determine if these are Majorana modes or regular (fermionic) BdG states [or zero-energy Andreev bound states (ABSs)], we use the following easy-to-check property of the zero-energy LDOS:
\begin{equation}
\pi\eta\sum_{i\in \Delta L_\alpha}\rho_L(0, i) =\left\{
\begin{array}{ll}
1 ~~~~~{\rm if}~~\alpha\equiv{\rm MZM}, \\
2 ~~~~~{\rm if}~~\alpha\equiv{\rm Z\!-\!BdG},
\end{array}\right.  \label{Eq22}
\end{equation}
where $\Delta L_\alpha$ is the segment of the wire that supports the spectral weight associated with mode $\alpha$ and $\eta$ is the broadening used in the calculation of the LDOS. As long as the modes are well separated, applying this criterion is convenient and unambiguous. Using this method we have verified that the robust zero-energy mode in Fig. \ref{FIG17}(A) corresponds to a pair of MZMs localized near the ends of the wire and that the regions with higher (zero-energy) spectral intensity correspond to additional ``regular'' BdG states having (near) zero energy (typically within a narrow Zeeman field range). Specific examples include the modes marked Z-BdG in Fig. \ref{FIG18}. We conclude that cut A is within the topological phase and that the presence of disorder has shifted the topological phase boundary to lower Zeeman field values (as compared to the clean case), as suggested by the position of the $W\!L$ maximum. 

\begin{figure}[t]
\begin{center}
\includegraphics[width=0.48\textwidth]{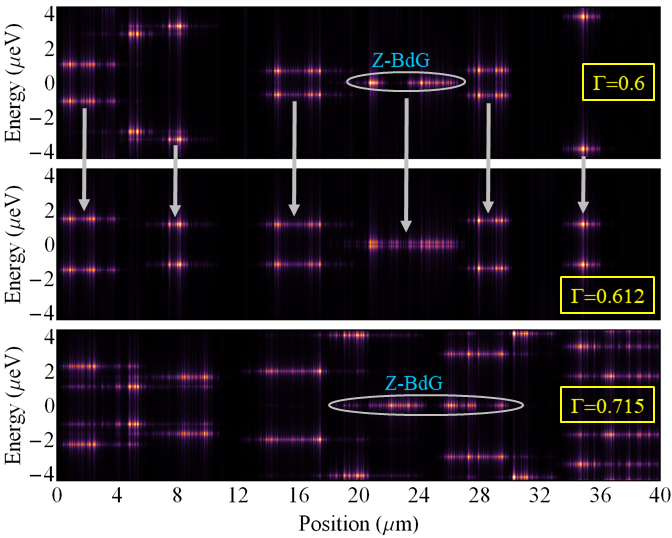}
\end{center}
\vspace{-3mm}
\caption{LDOS as function of energy and position corresponding to the vertical (blue) cuts in Fig. \ref{FIG17}(B). The modes responsible for the zero-energy features in Fig. \ref{FIG17}(B) are (regular) Z-BdG states consisting of a pair of partially separated Majorana modes localized within the wire (see top and bottom panels). Small variations of the Zeeman field result in a gapped lowest energy mode (middle panel). The vertical arrows connect the LDOS features associated with specific low-energy modes at $\Gamma=0.6~$meV and $\Gamma=0.612~$meV. Each mode corresponds to a specific ``line'' in Fig. \ref{FIG17}.}
\label{FIG19}
\vspace{-1mm}
\end{figure}

Next, we consider cut B (see Fig. \ref{FIG16}), which is well within the ``clean'' topological region, but to the left of the $W\!L$ maxima. As shown in the middle panel of Fig. \ref{FIG17}, in this case there is no stable zero-energy mode; only ``accidental'' zero-energy crossings and short, isolated segments that support zero-energy modes. To identify the nature of these modes, we calculate the LDOS for three representative values of the Zeeman field (vertical blue lines). The corresponding results are shown in Fig. \ref{FIG19}. For $\Gamma=0.6~$meV (top panel), one can clearly identify the zero-energy mode as a Z-BdG state localized near the middle of the wire. The relative stability of this mode (over a Zeeman field range of about $0.02~$meV; see Fig. \ref{FIG17}) can be explained  as a result of the Majorana components being partially separated spatially, i.e., forming a so-called partially separated ABS (ps-ABS) \cite{Stanescu2018Robust}, or a pair of quasi-Majoranas \cite{Vuik2019Reproducing}, localized near the middle of the wire. Again, this was explicitly verified using the criterion in Eq. (\ref{Eq22}). Upon slightly increasing the Zeeman field, the ps-ABS becomes gapped, as illustrated in the middle panel of Fig. \ref{FIG19}. 
\begin{figure}[t]
\begin{center}
\includegraphics[width=0.48\textwidth]{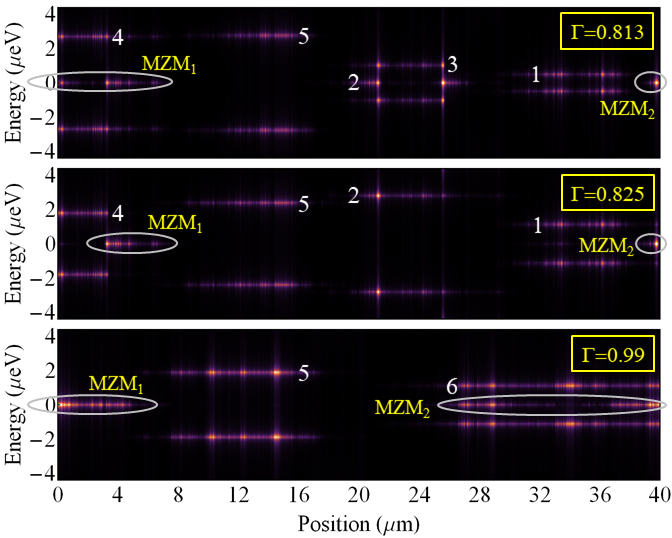}
\end{center}
\vspace{-3mm}
\caption{LDOS as function of energy and position corresponding to the vertical (blue) cuts in Fig. \ref{FIG17}(C). The LDOS features associated with the modes labeled $1,2,\dots,6$ in Fig. \ref{FIG17}(C) are explicitly identified. Note that for $\Gamma = 0.825~$meV (middle panel) MZM$_1$ is located about three microns away from the left end of the wire, while low-energy mode ``4'' is an edge mode. Also note that for $\gamma=0.99~$meV (bottom panel), MZM$_2$ and low-energy edge mode ``6'' have very large characteristic length scales (about $15~\mu$m).}
\label{FIG20}
\vspace{-1mm}
\end{figure}
In this context,  we note that by establishing the correspondence between the low-energy modes at two different values of the Zeeman field (see arrows in Fig. \ref{FIG19}) one can identify the DOS features in Fig. \ref{FIG17} associated with each mode (even when there are accidental degeneracies at a given field value). The lower panel in Fig. \ref{FIG19} confirms that the zero-energy modes along cut B are (trivial) disorder-induced Z-BdGs (typically located inside the wire). We conclude that cut B lies within the trivial phase, again consistent with the estimated phase boundary obtained based on the $W\!L$ maxima. As an additional observation, we point out that the LDOS shown in Fig. \ref{FIG19} reveals the presence of low-energy modes at one or both ends of the wire. A finite resolution differential conductance measurement will generate zero-bias conductance peaks  (ZBCPs) associated with these edge modes. Of course, the height of the ZBCPs will not be quantized, but they may be stable over a finite Zeeman field range and may (accidentally) generate correlated signatures at the two ends. Finally, we point out that when the edge modes are ps-ABSs (which closely mimic the local Majorana phenomenology), the characteristic separation length is determined by disorder, not by the length of the system.

Finally, cut C (see Fig. \ref{FIG16}), which is inside the ``clean'' topological region and to the right of the $W\!L$ maxima, generated a stable zero energy mode that extends throughout the entire Zeeman field range, as shown in the bottom panel of Fig. \ref{FIG17}. Furthermore, we have explicitly verified that this zero-energy mode is associated with the presence of well separated MZMs and  can be traced  all the way up to $\Gamma =\Gamma_0$.  The energy and position dependence of the LDOS corresponding to the Zeeman field values marked by blue lines in Fig. \ref{FIG17}(C) are shown in Fig. \ref{FIG20}. We explicitly identify the contributions associated with the modes  labeled $1,2,\dots,6$ in Fig. \ref{FIG17}(C). One notable feature is the relatively large length scale of MZM$_1$ (several microns) at all Zeeman field values and MZM$_2$ (about $15~\mu$m) for $\Gamma=0.99~$meV.  In short wires, this is expected to result in a strong overlap of the Majorana modes (see Section \ref{S3_3}).  Another significant feature is the presence of low-energy BdG edge modes, consistent with the results for cuts A and B. We emphasize that the energy of these modes can be lower than the experimental resolution corresponding to a tunneling experiment and, consequently, the BdG edge modes can generate zero bias conductance peaks even in the absence of a MZM, or can generate additional contributions when a MZM is present. The last notable feature in Fig. \ref{FIG20} concerns the location of MZM$_1$ in the middle panel ($\Gamma=0.825~$meV). Note that most of the corresponding spectral weight is located more than three microns away from left edge of the system. We also point out that this type of scenario becomes more likely as the disorder strength increases. To verify if the system is in the topological or the trivial phase, we extend the wire and check if the MZM ``migrates'' towards the new edge or gets pinned by disorder.  Note that the additional segments should contain disorder with the same parameters (e.g., overall amplitude, characteristic length scale, etc.) as the ``original'' wire. 

In this section we have investigated the low-energy spectral properties of a long Majorana system ($L=40~\mu$m), within both the ring and wire geometries, in the presence of a disorder potential with a characteristic length scale of about $15~$nm and different values of the overall amplitude, $V_0$, up to $1~$meV. 
The hybrid system has weak effective semiconductor-superconductor coupling ($\gamma = 0.21~$meV) and relatively weak Rashba-type  spin-orbit coupling ($\alpha_R = 125~$meV$\cdot$\AA). Within this parameter regime, we find that the (partial) quasiparticle gap associated with the lower energy spin subband (LESS) decreases with increasing disorder strength and eventually collapses, starting with regions of the parameter space in the vicinity of the topological  phase boundary.  For $V_0 \gtrsim 0.8~$meV the entire topological phase is gapless (or nearly gapless), while it still covers an area of the control parameter space comparable to that corresponding to a clean system (but shifted towards larger chemical potential values). The corresponding low-energy modes are characterized by relatively large characteristic wave vectors and  long characteristic length scales, on the order of $10^0-10^1~\mu$m.  
\begin{figure}[t]
\begin{center}
\includegraphics[width=0.48\textwidth]{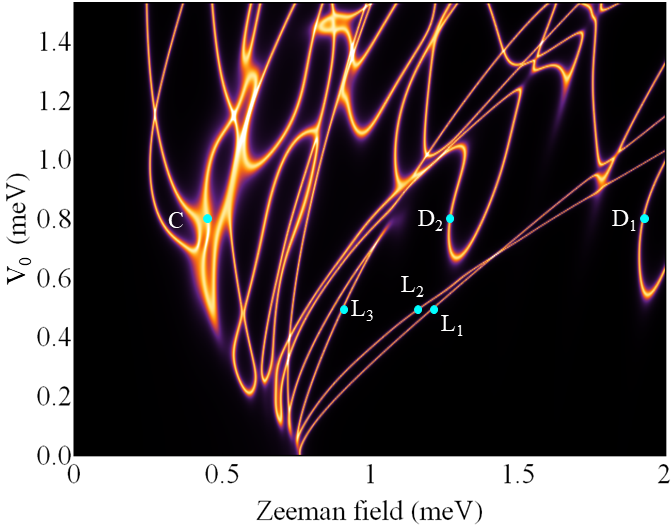}
\end{center}
\vspace{-3mm}
\caption{Zero-energy DOS as function of Zeeman field and disorder strength for a Majorana ring of length $L=3~\mu$m with chemical potential $\mu=0.73~$meV in the presence of a disorder potential corresponding to disorder realization ``A'' in Fig. \ref{FIG8} (top panel). The zero energy features should be compared with the corresponding features for a long system (see top panel of Fig. \ref{FIG13}). The position dependence of the LDOS corresponding to the specific modes marked by blue circles is shown in Fig. \ref{FIG22}.}
\label{FIG21}
\vspace{-1mm}
\end{figure}
By contrast, the  low-energy states associated with the higher energy spin subband (HESS), which are characterized by  lower values of the  characteristic wave vector, are strongly localized by disorder and correspond to ``standard'' low-energy Andreev bound states localized throughout the system. In a system with open boundary conditions (i.e., a Majorana wire), having (non-Majorana) low-energy BdG  edge modes becomes very likely in the regime corresponding to a vanishing LESS quasiparticle gap. The topological phase is characterized by the presence of a pair of well separated MZMs, with a separation length determined by the system size; however, the MZMs are not necessarily located at the very edge of the system, which implies that coupling to them using an end-of-wire probe may be difficult or practically impossible, particularly in the presence of low-energy BdG edge states. Note that a MZM plus a low-energy BdG edge state corresponds to three hybridized, partially overlapping Majorana modes. 
To obtain a robust topological phase characterized by a sizable  LESS quasiparticle gap and relatively short characteristic length scales (on the order of one micron or less in the relevant control parameter range) for the low-energy modes, including the MZMs, one should not only bring the system into the low disorder regime (e.g., below $V_0\approx 0.6~$meV for the system studied in this section), but  enhance the spin-orbit coupling strength, the effective semiconductor-superconductor coupling of the hybrid system, and the parent superconducting gap.  

\begin{figure}[t]
\begin{center}
\includegraphics[width=0.48\textwidth]{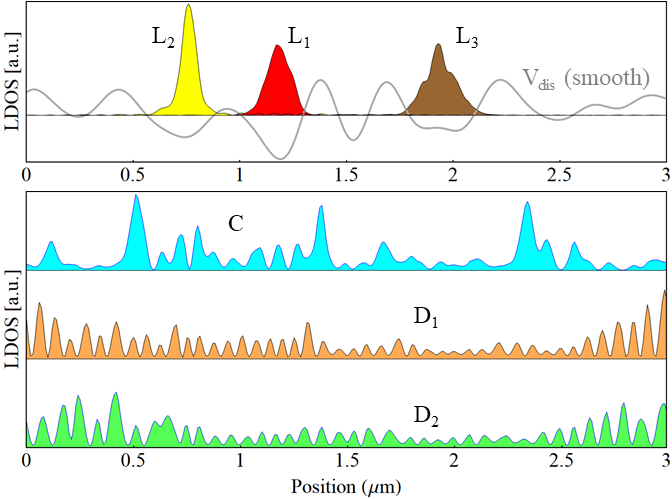}
\end{center}
\vspace{-3mm}
\caption{Position dependence of the LDOS associated with representative low-energy modes (marked by blue circles in Fig. \ref{FIG21}) in a Majorana ring with the same parameters as in Fig. \ref{FIG21}. The modes $L_1$, $L_2$, and $L_3$ correspond to $V_0=0.5~$meV and Zeeman field values $\Gamma =1.21,~ 1.17, ~{\rm and}~ 0.91~$meV, respectively.  These modes are strongly localized near the minima of the (smooth) disorder potential (gray line in the top panel; also see Fig. \ref{FIG8}, bottom panel). Note that $L_1$ is practically identical to mode A in Fig. \ref{FIG12}, being generated by the same local features of the disorder potential. Mode $C$ (with $V_0=0.8~$meV and $\Gamma=0.45~$meV) is reminiscent of the delocalized mode  $C_2$ in Fig. \ref{FIG12}. Modes $D_1$ and $D_2$ (with $V_0=0.8~$meV and $\Gamma=1.92~{\rm and}~1.27~$meV, respectively)  are associated with large-$k$ (LESS) states. Although the corresponding D-type  modes in a long system are localized (see Fig. \ref{FIG12}), the corresponding characteristic length scales are larger than the length of the short wire ($L=3~\mu$m).}
\label{FIG22}
\vspace{-1mm}
\end{figure}

\subsection{Disorder and finite size effects in short Majorana systems}\label{S3_3}

Hybrid Majorana wires realized in the laboratory are much shorter than the system investigated in the previous section, typically ranging between several hundred nanometers and a few microns.  The low-energy properties of a short Majorana system are characterized by an interplay of disorder-induced and finite-size effects.  In this section we investigate these effects by considering a hybrid system of length $L=3~\mu$m, in the ring and wire geometries, with materials-related parameters identical to those characterizing  the long wire discussed above and disorder potential corresponding to segments A or B in Fig. \ref{FIG8}. 

We start with a calculation of the zero-energy DOS as function of the Zeeman field and disorder potential strength, $V_0$, for a 3-micron Majorana ring with disorder potential corresponding to disorder realization ``A''. The result, shown in Fig. \ref{FIG21}, is the short-system equivalent of Fig. \ref{FIG13} (top panel). As expected, for $V_0=0$ (clean system) there is only one  zero-energy mode at the critical field $\Gamma_c\approx 0.76~$meV. In the presence of disorder, multiple zero-energy modes emerge within a Zeeman field range that becomes wider with increasing $V_0$. The manifest quantitative difference with respect to the long system (see Fig. \ref{FIG13}) is the (significantly) lower number of zero energy states within a given Zeeman field range. 

\begin{figure}[t]
\begin{center}
\includegraphics[width=0.48\textwidth]{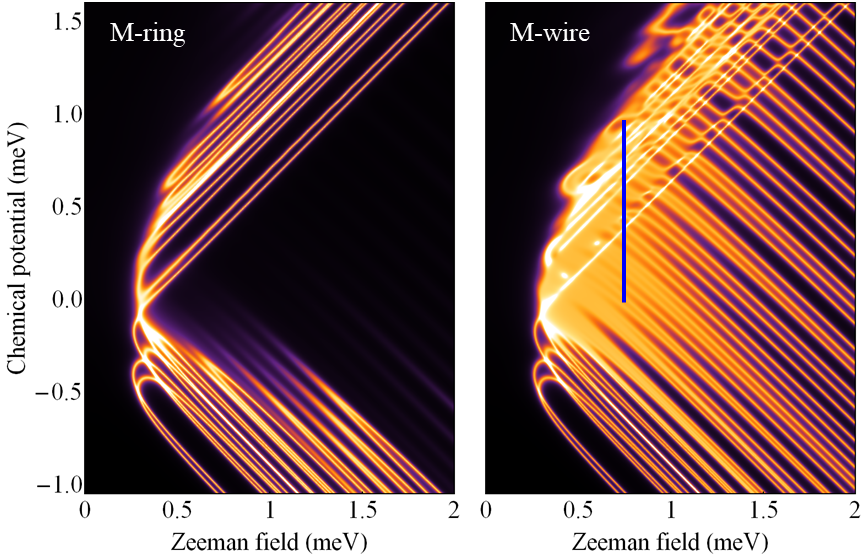}
\end{center}
\vspace{-3mm}
\caption{``Phase diagrams'' obtained by mapping the zero-energy DOS of a Majorana ring of length $L=3~\mu$m (left) and of the corresponding Majorana wire (right) as function of Zeeman field and chemical potential. The disorder potential corresponds to disorder realization ``A'' in Fig. \ref{FIG8}, with an amplitude $V_0=0.4~$meV. For comparison with the long system case, see the corresponding panel in Fig. \ref{FIG14}.  The energy dependence of the DOS along the cut marked by the blue line (right panel) is shown in Fig. \ref{FIG26}. Most of the additional zero-energy features characterizing the wire (right panel) are associated with Majorana modes localized near the ends of the system (see Figs. \ref{FIG26} and \ref{FIG29}).}
\label{FIG23}
\vspace{-1mm}
\end{figure}

To identify the nature of different  zero-energy modes, we calculate the position dependence of the LDOS for a few representative states marked by blue circles in Fig. \ref{FIG21}. The results are shown in Fig. \ref{FIG22}. Similar to the long system case, at low disorder ($V_0 \lesssim 0.6~$meV) and Zeeman field values larger than the critical field one can identify strongly localized modes, e.g., $L_1 - L_3$ (see top panel of Fig. \ref{FIG22}), which are (mainly) associated with low-k HESS states. Note that these states are pinned near the minima of the ``smooth'' disorder potential. All these states can also be identified in the long wire, near the corresponding features of the disorder potential and for similar values of the control parameters. The equivalent of the delocalized (C-type) modes can also be found within a control parameter region approximately corresponding to the location of delocalized modes in Fig. \ref{FIG13}. For example, for $V_0=0.8~$meV the C-type modes emerge at Zeeman fields lower than  $\Gamma_c\approx 0.76~$meV, consistent with the evolution of the topological phase boundary with the disorder strength discussed in the previous section. Finally, for strong-enough disorder we can identify D-type modes, which have long characteristic length scales and are (mainly) associated with high-k LESS states. The crucial difference between the long and short systems is that, while in the long system all D-type states are localized, in the short system they extend throughout the whole ring (i.e., they are practically delocalized; see Fig. \ref{FIG22}).  This is not surprising, considering that the typical D-mode characteristic length scale for the long system was found to be larger than three microns. The key question concerns the effect of these delocalized (D-type) low-energy modes on the stability on the  Majorana bound states (MBSs) that may emerge in systems with open boundary conditions (i.e., wires). On the one hand, since the delocalized mode can couple to a pair of MBSs, the ``topological'' protection of the MBS pair is expected to be affected near the energy minima of the delocalized mode. On the other hand, when these minima occur within a parameter region that is topological in the long wire limit, pairs of (more-or-less stable) MBSs are expected to emerge on both sides of a minimum (along a given direction in parameter space). In other words, in short Majorana wires the minima of the (effectively delocalized) D-type modes are not associated with finite size ``remnant'' topological phase transitions.  This observation is further supported by our analysis below.

\begin{figure}[t]
\begin{center}
\includegraphics[width=0.48\textwidth]{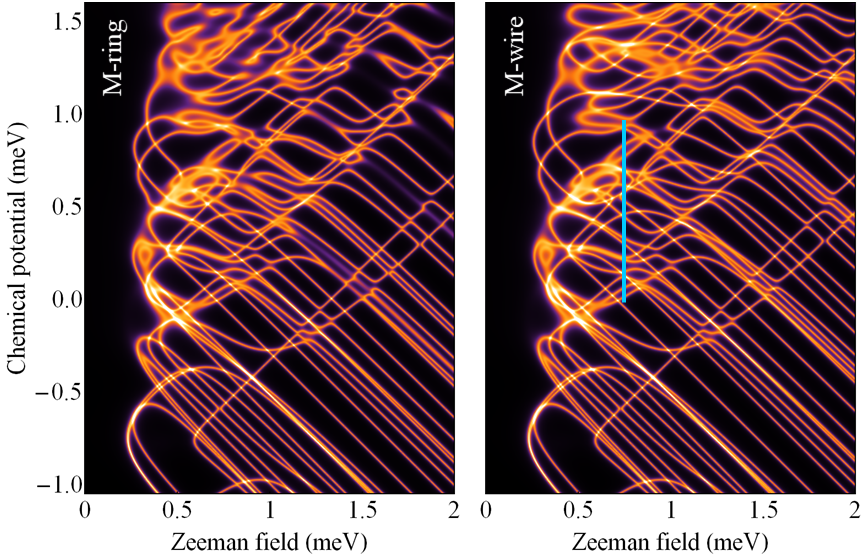}
\end{center}
\vspace{-3mm}
\caption{Same as in Fig. \ref{FIG23}, but for a system with strong disorder, $V_0=1.2~$meV. Note the similarities between the features characterizing the Majorana ring (left) and the Majorana wire (right). Also note that the minimum field associated with the emergence of zero-energy modes is weakly dependent on the chemical potential. The energy dependence of the DOS along the cut marked by the blue line (right panel) is shown in Fig. \ref{FIG26}.}
\label{FIG24}
\vspace{-1mm}
\end{figure}

Next, we calculate ``phase diagrams'' similar to those in Fig. \ref{FIG14} for the short system with disorder realization ``A''. In addition to the zero-energy DOS for a Majorana ring, we also provide the zero-energy DOS for the corresponding Majorana wire, for comparison.   In Fig. \ref{FIG23} we show the results for weak disorder ($V_0 = 0.4~$meV), while Fig. \ref{FIG24} illustrates the strong disorder case ($V_0=1.2~$meV). 
There are two striking differences between the weak and strong disorder cases. The first one concerns the overall shape of the region containing zero energy DOS features. For a weakly disordered Majorana ring (left panel of Fig. \ref{FIG23}), the zero-energy DOS features emerge in the vicinity of the ``clean'' topological phase boundary. Also note the consistency of this diagram with the corresponding panel of Fig. \ref{FIG14} (with the obvious difference that the long wire supports more zero energy modes). On the other hand, for the Majorana ring with strong disorder (left panel of Fig. \ref{FIG24}), the zero energy DOS features emerge within the entire parameter region $\gamma \lesssim \Gamma \leq \Gamma_0$, almost independent of the chemical potential (within the considered range). We point out that an estimate of the chemical potential range over which the lowest field associated with the emergence of zero-energy features is approximately $\mu$-independent provides a measure of the disorder strength. For example, in Figs. \ref{FIG23} and \ref{FIG24} this range is approximately $2V_0$ (see also the diagrams in Fig. \ref{FIG14}, which exhibit a similar property). If, for a device realized in the laboratory, the lever arm associated with the (back or top) gate potential is known (or can be estimated), this method of evaluating the disorder strength can be applied to experimentally measured data.
At this point, we should also emphasize that the minimum Zeeman field at which zero-energy features emerge (which is practically given by the effective semiconductor-superconductor coupling, $\gamma$) is an important energy scale for characterizing the hybrid system. On the one hand, $\gamma$ controls the induced gap  (at zero magnetic field) and can be estimated (at least in the weak/intermediate coupling regime) from the measured value of the gap. On the other hand, combining this with the estimated value of the minimum {\em magnetic field} at which zero-energy features emerge, provides an estimate of the effective g-factor. 

\begin{figure}[t]
\begin{center}
\includegraphics[width=0.48\textwidth]{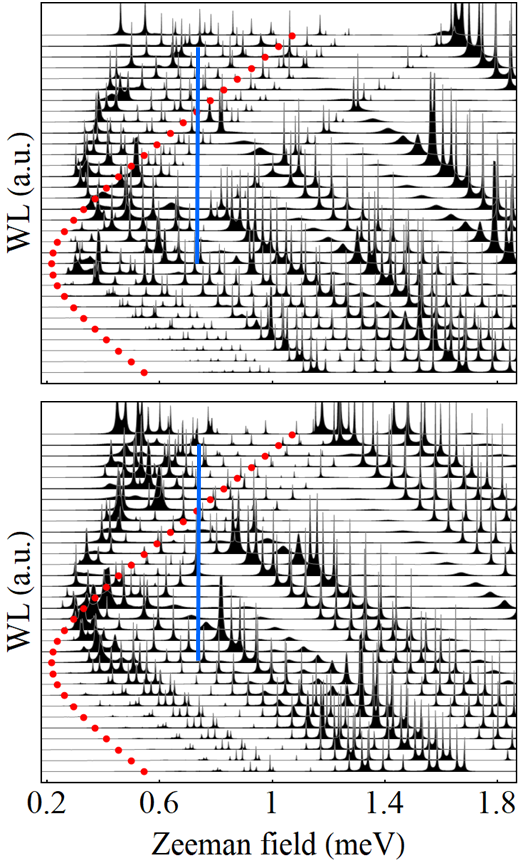}
\end{center}
\vspace{-3mm}
\caption{``Weakest link'' defined by Eq. (\ref{WL}) for a Majorana ring of length $L=3~\mu$m and disorder strength $V_0=0.6~$meV as function of the Zeeman field  for different chemical potential values ranging from $\mu=-0.5~$meV (bottom curve) to $\mu=1.05~$meV (top curve) in steps of $0.05~$meV. The curves are shifted for clarity. The top panel corresponds to disorder realization ``A'', while the bottom panel is for a system with disorder realization ``B''. The maxima of $W\!L$ correspond to extended zero-energy modes. The red circles mark the topological phase boundary of a clean system. The energy dependence of the DOS along the cuts marked by blue lines is given in Figs. \ref{FIG26} and \ref{FIG27}.}
\label{FIG25}
\vspace{-1mm}
\end{figure}

\begin{figure}[t]
\begin{center}
\includegraphics[width=0.48\textwidth]{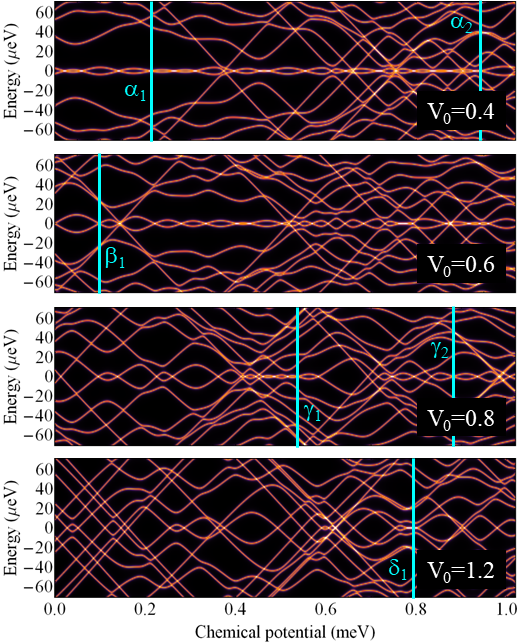}
\end{center}
\vspace{-3mm}
\caption{Low-energy DOS as function of chemical potential and energy for a wire of length $L=3~\mu$m in the presence of a disorder potential corresponding to disorder realization ``A'' and different values of the overall amplitude $V_0$ (explicitly given in meV). The LDOS corresponding to the cuts marked by blue lines are given in Figs. \ref{FIG29}, \ref{FIG31}, and \ref{FIG32}.}
\label{FIG26}
\vspace{-1mm}
\end{figure}

The second important difference between  the two disorder regimes concerns the manifest distinction between the ring and wire results at weak disorder (see Fig. \ref{FIG23}), versus the similarities characterizing the strong disorder results (Fig. \ref{FIG24}). The additional zero-energy features in the right panel of Fig. \ref{FIG23} are associated with the emergence of MBSs in the system with open boundary conditions. These MBSs partially overlap and the resulting mode undergoes energy splitting oscillations (also, see below  Figs. \ref{FIG26} and \ref{FIG29}). The stripy features in the right panel are associated with the nodes of these oscillations, which contribute to the zero energy DOS.  Also  note that features associated with localized states, which are not affected by the boundary conditions, can be clearly identified in both panels of Fig. \ref{FIG23}. Turning now to the strong disorder case (Fig. \ref{FIG24}), we point out that the close similarities between the ring and wire results clearly indicate that the zero-energy DOS features are essentially controlled by disorder, while the boundary conditions have a weak effect. Note that this behavior does not necessarily imply strong localization (although it is definitely consistent with it); more information about the characteristic length scales of the relevant low-energy states are provided below (Fig. \ref{FIG32}). 

Our next goal is to provide a ``global'' characterization of the location within the parameter space of delocalized states, similar to the analysis done for the long system in the context of Fig. \ref{FIG16}. For concreteness, we focus on a short system ($L=3~\mu$m) with the same disorder amplitude as in Fig. \ref{FIG16}, $V_0=0.6~$meV, and two different disorder realization (``A'' and ``B'' in Fig. \ref{FIG8}). The corresponding dependence of the ``weakest link'' defined by Eq. (\ref{WL}), with $\delta L = 50-200~$nm, on Zeeman field and chemical potential is shown in Fig. \ref{FIG25}. The most  notable feature in Fig. \ref{FIG25} is the  presence of zero energy delocalized states throughout most of the relevant parameter space. This is a direct consequence of the system being in a parameter regime characterized by relatively low spin-orbit coupling and low effective semiconductor-superconductor coupling. As discussed above, within this regime (i) the (partial) gap associated with large-$k$ LESS states collapses even  in the presence of  relatively weak disorder and (ii) the corresponding low-energy modes have large characteristic length scales. The combination of these two effects results in the ubiquitous presence of ``delocalized'' zero-energy modes throughout the parameter space. We remind the reader that many of these modes are D-type modes, which are effectively delocalized in a short system, but become localized in long wires.  

\begin{figure}[t]
\begin{center}
\includegraphics[width=0.48\textwidth]{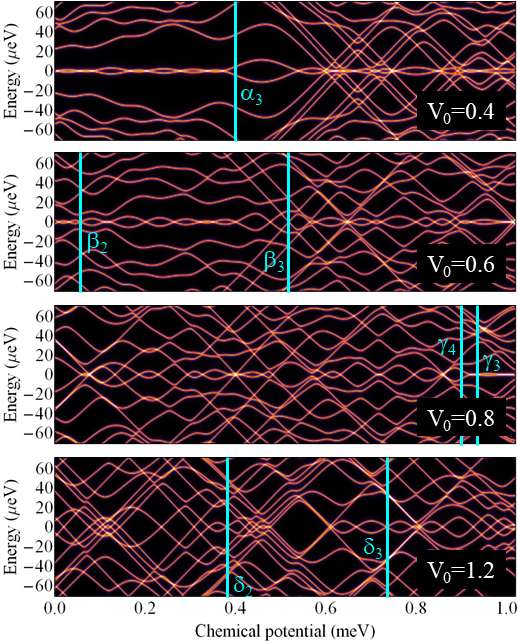}
\end{center}
\vspace{-3mm}
\caption{Same as in Fig. \ref{FIG26} for a disorder potential corresponding to disorder realization ``B''. Note that the main features characterizing the system for a given disorder strength are qualitatively similar to those shown in the corresponding panel of Fig. \ref{FIG26}. The (nearly) linear, x-shaped low-energy features are associated with localized modes. In the top panel (i.e., for relatively low disorder), the lowest energy mode, which consists of a pair of weakly overlapping MBSs, is characterized by energy splitting oscillations with an amplitude (typically) less than $\sim 5~\mu$eV.}
\label{FIG27}
\vspace{-1mm}
\end{figure}

To gain further insight into the low-energy spectral properties of (relatively) short Majorana wires, we calculate the energy and chemical potential dependence of the DOS along a constant field cut corresponding to $\Gamma=0.75~$meV (marked by blue lines in Figs. \ref{FIG23}-\ref{FIG25}) for different values of the disorder strength. The position of the cut was chosen to be well within the topological region (as determined based on long wire calculations) for systems with relatively weak disorder and to avoid the large Zeeman field and large chemical potential regimes, where the finite size effects associated with the presence of large-$k$, long characteristic wavelength modes are expected to be rampant. Note that the size of the cut is $\Delta\mu \approx 1~$meV, which is almost five times larger than $\gamma$ (i.e., the minimum Zeeman field associated with the emergence of zero-energy features). The spectra shown in Fig. \ref{FIG26} are calculated in the presence of a disorder potential corresponding to disorder realization ``A'', while Fig. \ref{FIG27}  shows similar results for disorder realization ``B''. Before discussing in detail the relevant features revealed by these results, let us point out that for an ideal (i.e., clean and long) system the corresponding low-energy spectrum is characterized by a (partial) LESS gap ranging from about $40~\mu$eV to about $60~\mu$eV and a HESS gap that closes and reopens at $\Gamma\approx 0.76~$meV (see Fig. \ref{FIG4}). In the presence of disorder, the HESS states are mostly associated with localized low-energy modes, which can be easily identified (particularly at low/intermediate disorder) through the characteristic, nearly linear x-shaped low-energy features. If we ignore these features, the low-disorder system with $V_0=0.4~$meV (top panels in Figs. \ref{FIG26} and \ref{FIG27}) is characterized by a (partial) quasiparticle gap of magnitude $20-40~\mu$eV over most of the chemical potential range considered here. In the middle of the gap there is a low-energy mode that exhibits energy splitting oscillations with an amplitude smaller than $5~\mu$eV (i.e., about an order of magnitude smaller than the partial LESS gap) over almost the entire $\mu$ range. Increasing the disorder strength to $V_0=0.6~$meV does not modify significantly this qualitative picture, but results in smaller LESS gaps and larger energy splitting amplitudes for the low-energy mode. Further increasing the disorder strength qualitatively changes the low-energy features. 

\begin{figure}[t]
\begin{center}
\includegraphics[width=0.48\textwidth]{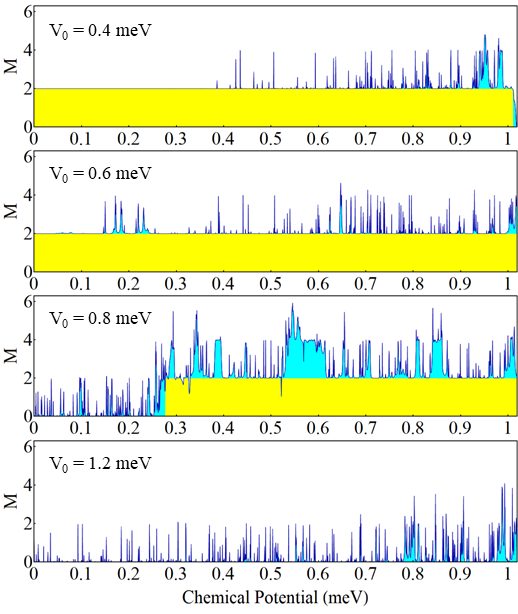}
\end{center}
\vspace{-3mm}
\caption{Dependence of the number $M$ of Majorana modes with energy smaller that $\eta=10^{-5}~$meV on the chemical potential for a long wire with $L=80~\mu$m, $\Gamma=0.75~$meV, and different values of the disorder strengths $V_0$. The yellow regions correspond to the presence of a robust pair of MZMs, $M=2$; additional zero energy states may emerge at specific values of the chemical potential (resulting in $M > 2$). For $V_0=0.4~$meV the system becomes trivial above $\mu\approx 1~$meV, while in the strong disorder limit ($V_0=1.2~$meV) the system is trivial throughout the entire chemical potential range shown in the figure.}
\label{FIG28}
\vspace{-1mm}
\end{figure}

A key question concerns the topological or trivial nature of the lowest energy mode. To provide some perspective, we first address this question in the context of a long, $L=80~\mu$m wire, half of which has a disorder potential as given in Fig. \ref{FIG8} (which includes the segments ``A'' and ``B''), while the other half corresponds to a different disorder realization (with the same parameters).  By analogy with Eq. \ref{Eq22}, we define the (total) number of MZMs within the wire as
\begin{equation}
M=\pi\eta\!\sum_{i=1}^N \rho_L(0,i).            \label{Eq23}
\end{equation}
In the calculation of the LDOS we use a broadening $\eta=10^{-5}~$meV. If the wire supports exactly one pair of (well separated) MZMs having an energy splitting much smaller than $\eta$, we get $M=2$. If there are no states with energy smaller that $\eta$, $M$ will be less than 2, while if zero energy modes in addition to the pair of MZMs are present, we obtain $M>2$. The results corresponding to different values of the disorder strength are shown in Fig. \ref{FIG28}. We note that for weak disorder (top two panels) the system is in the topological phase for $0\leq \mu\leq 1~$meV. At intermediate disorder 
($V_0=0.8~$meV), the wire is in the topological phase above $\mu\approx 0.3~$meV and is topologically trivial below this value. Finally, the strongly disordered system ($V_0=1.2~$meV) is topologically trivial over the entire parameter range considered in the calculation. We also note that there are many zero energy contributions (highlighted in blue) that are not produced by a robust MZM pair. The narrow peaks are typically generated by states localized throughout the wire that cross zero energy at specific values of the chemical potential. The narrow plateaus (e.g., those corresponding to $M\approx 4$ in the $V_0=0.8~$meV panel) are generated by ps-ABSs, i.e., pairs of Majorana modes that are separated by distances smaller than the length of the wire. In general, the presence of these additional contributions is a direct indication of the plethora of low-energy modes that characterize the system in the weak SOC, weak SM-SC coupling parameter regime.   

Using the insight provided by Fig. \ref{FIG28}, we return to the analysis of the low-energy features in Figs. \ref{FIG26} and \ref{FIG27}. In addition to the DOS, we also calculate the position dependence of the LDOS for specific values of the chemical potential marked by blue lines in the two figures. Starting with the weak disorder regime, we calculate the LDOS corresponding to cuts $\alpha_1$, $\alpha_2$ and $\alpha_3$ for systems with $V_0=0.4~$meV and 
$\beta_1$, $\beta_2$ and $\beta_3$ for $V_0=0.6~$meV. The corresponding results are shown in Fig. \ref{FIG29}. Cut $\alpha_1$ illustrates a best case scenario for this type of system. There is a relatively large quasiparticle gap ($> 40~\mu$eV) and two clearly defined Majorana modes localized towards the ends of the wire. However, both the Majorana modes and the finite energy states have characteristic length scales comparable to the size of the system (see top panel of Fig. \ref{FIG29}). As emphasized before, this is a direct consequence of being in a parameter regime characterized by weak spin-orbit and semiconductor-superconductor couplings and has nothing to do with disorder. To further characterize the localization of the Majorana modes, we calculate ${\rm Log}_{10}[\rho_L(\omega,x)/\rho_L^{max}(\omega]$, where the LDOS is normalized (by its maximum value) for convenience. The results shown in Fig. \ref{FIG30} (top panel) clearly indicate that the MBSs are characterized by length scales of order $1~\mu$m and by some exponential decay towards the middle of the wire. For other parameter values, the situation is less favorable, which implies that the low-energy physics is characterized by strong finite size effects. Cut $\alpha_2$ confirms the presence of the low-energy Majorana modes near the high end of the chemical potential range. Again, the characteristic length scale of the Majorana modes is comparable to the size of the system. One can also notice the presence of additional low-energy modes near the ends of the system.  
\begin{figure}[t]
\begin{center}
\includegraphics[width=0.48\textwidth]{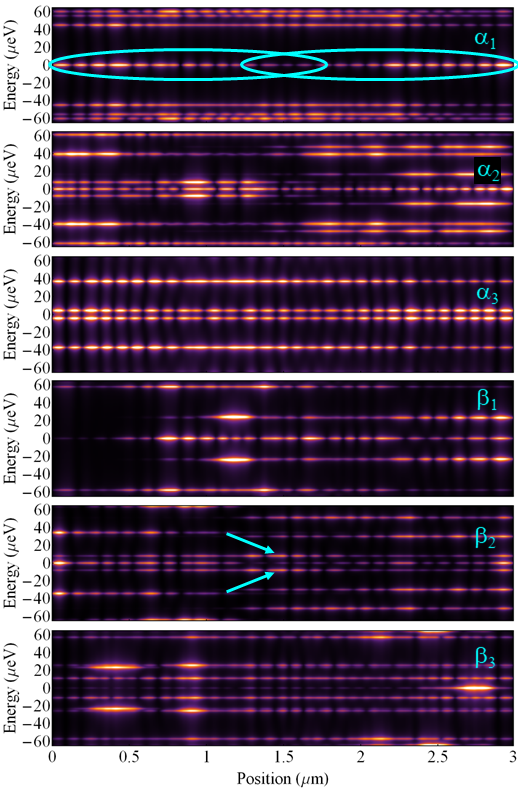}
\end{center}
\vspace{-3mm}
\caption{Position and energy dependence of the LDOS for systems with weak disorder. Different panels correspond to the cuts marked by blue lines in Figs. \ref{FIG26} and \ref{FIG27}. The corresponding values of the chemical potential are: ($\alpha_1$) $\mu=0.21~$meV;  ($\alpha_2$) $\mu=0.94~$meV;  ($\alpha_3$) $\mu=0.4~$meV;  ($\beta_1$) $\mu=0.1~$meV; ($\beta_2$) $\mu=0.06~$meV;  ($\beta_3$) $\mu=0.52~$meV. Typically, the lowest energy mode consists of a pair of MBSs with characteristic length scales smaller than but comparable to the size of the system, which, consequently, have a significant overlap (as suggested in the top panel).}
\label{FIG29}
\vspace{-1mm}
\end{figure}
Cut $\alpha_3$ is near the edge of the largest energy splitting of the lowest energy mode corresponding to $V_0=0.4~$meV (see Fig. \ref{FIG27}). As shown in panel ($\alpha_3$) of Fig. \ref{FIG29} and in the corresponding panel of Fig. \ref{FIG30}, this is a direct consequence of the large overlap of the two MBSs, which basically form a delocalized low-energy mode that extends through the whole system. The next low-energy state is characterized by a relatively large gap ($\sim 40~\mu$eV) and is also delocalized. In this context, we point out that for the parameters characterizing this system the presence of strongly overlapping MBSs forming effectively delocalized low-energy modes is rather generic. In turn, these ``delocalized'' modes may lead to the vanishing of the transport gap.  Of course, this finite size effect does not indicate the presence of a topological quantum phase transition. For example, we may characterize the 
segment $0.4 \lesssim \mu\lesssim 0.5~$meV as ``topologically trivial'', due to the large splitting of the lowest energy mode. However, upon expanding the 
wire, the segment becomes topological, which implies that there is no phase transition. 

\begin{figure}[t]
\begin{center}
\includegraphics[width=0.48\textwidth]{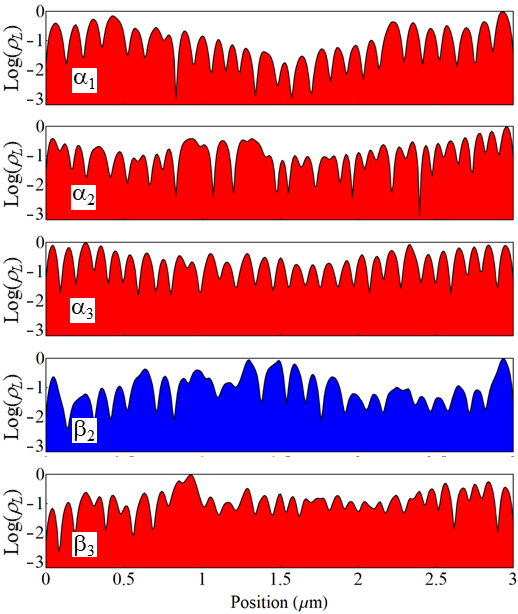}
\end{center}
\vspace{-3mm}
\caption{Position dependence of the logarithm (base 10) of the LDOS (normalized by its maximum value along the wire) for representative low-energy modes. The red shading corresponds to pairs of zero-energy or low-energy Majorana modes from the corresponding panels in Fig. \ref{FIG29}. The blue shaded curve corresponds to the extended state marked by arrows in Fig. \ref{FIG29}.}
\label{FIG30}
\vspace{-1mm}
\end{figure}

Turning now to  the case $V_0=0.6~$meV, we first point out that cut $\beta_1$ is characterized by the presence of a zero energy Majorana mode localized 
away from the left edge of the system (while its counterpart is localized at the right end of the wire). This is a disorder effect that is also present in long 
wires (see, e.g., the middle panel of Fig. \ref{FIG20}). Cut $\beta_2$ is  near the minimum of an extended mode marked by arrows in the corresponding 
panel of Fig. \ref{FIG29}. The logarithm of the LDOS corresponding to the delocalized mode is shown in Fig. \ref{FIG30} (blue shading). We can identify reasonably well separated pairs of Majorana modes in cut $\beta_2$, as well as to the left and right of the corresponding $\mu$ value. Hence, the 
minimum of the delocalized mode is not related to a phase transition. Cut $\beta_3$ corresponds to a large energy splitting of the lowest energy mode. 
Similar to the case $\alpha_3$, this is a consequence of having strongly overlapping MBSs (see also the lower panel in Fig. \ref{FIG30}). We also point out the presence of a zero-energy localized state (near the right end of the wire) corresponding to the x-shaped feature in Fig. \ref{FIG27}. 

\begin{figure}[t]
\begin{center}
\includegraphics[width=0.48\textwidth]{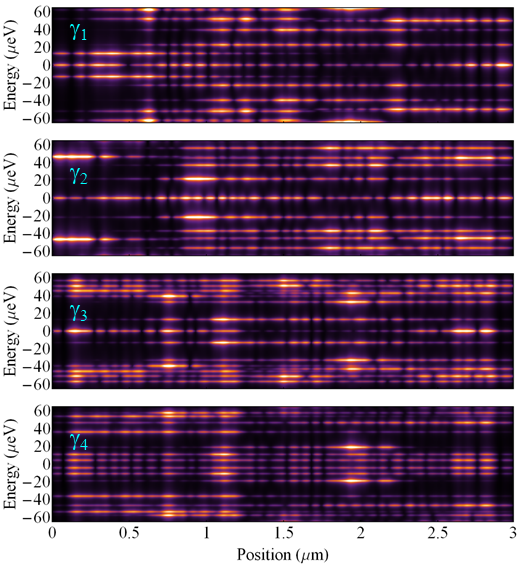}
\end{center}
\vspace{-3mm}
\caption{Position and energy dependence of the LDOS for systems with intermediate disorder. Different panels correspond to the cuts marked by blue lines in Figs. \ref{FIG26} and \ref{FIG27}. The chemical potential values are: ($\gamma_1$) $\mu=0.54~$meV;  ($\gamma_2$) $\mu=0.88~$meV;  ($\gamma_3$) $\mu=0.94~$meV;  ($\gamma_4$) $\mu=0.9~$meV.}
\label{FIG31}
\vspace{-1mm}
\end{figure}

The main conclusions of our analysis of the weak disorder regime are twofold. First, as a consequence of the specific parameter regime characterizing the system under investigation, the characteristic length scales of the Majorana modes are comparable to the size of the system. In turn, the finite size effects in wires of length $L=3~\mu$m are strong and the Majorana modes are weakly protected, despite the disorder being relatively weak.  Second, one can identify reasonably well separated Majorana modes characterized by an energy splitting  that is small compared to the topological gap of the clean system  over large segments of chemical potential, two to five times larger than the energy scale $\gamma$ (associated with the minimum Zeeman field at which zero energy features emerge). In the $\Gamma-\mu$ plane, these ``topological'' regions form large ``patches'' (or ``islands'') with characteristic linear sizes larger than $\gamma$. Upon expanding the wire length, these large patches merge into a single topological region. 

\begin{figure}[t]
\begin{center}
\includegraphics[width=0.48\textwidth]{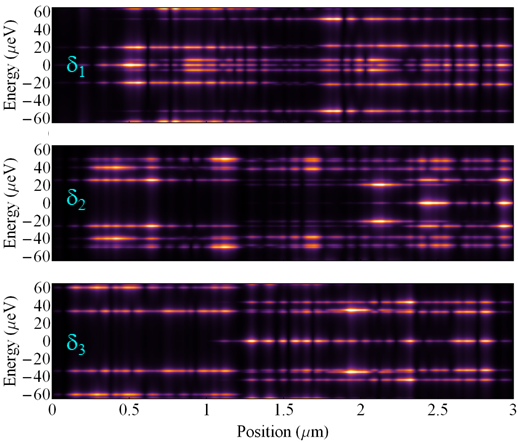}
\end{center}
\vspace{-3mm}
\caption{Position and energy dependence of the LDOS for systems with strong disorder and different chemical potential values corresponding to the cuts  marked by blue lines in Figs. \ref{FIG26} and \ref{FIG27}. The chemical potential values are: ($\delta_1$) $\mu=0.79~$meV;  ($\delta_2$) $\mu=0.38~$meV;  ($\delta_3$) $\mu=0.74~$meV.}
\label{FIG32}
\vspace{-1mm}
\end{figure}

For the system with  intermediate disorder strength (corresponding to $V_0=0.8~$meV),  the DOS exhibits low-energy modes with small energy splitting only within relatively short $\mu$ segments with lengths of order $\gamma$ (see Figs. \ref{FIG26} and \ref{FIG27}). To verify the Majorana nature of the low-energy modes, we calculate the LDOS corresponding to representative values of the chemical potential (marked by blue lines labeled $\gamma_1$-$\gamma_4$ in Figs. \ref{FIG26} and \ref{FIG27}). The results shown in Fig. \ref{FIG31} confirm the presence of  reasonably well separated Majorana modes (particularly in panels $\gamma_1$ and $\gamma_3$).  Again, we emphasize the presence of low-energy states characterized by large length scales (comparable to the size of the system). The presence of these states is the combined effect of having stronger disorder and being in a parameter regime characterized by weak SOC and weak SM-SC coupling. Some of these low-energy states hybridize with both Majorana modes, affecting their stability. A specific example is provided in panel $\gamma_3$ of Fig. \ref{FIG31}, which shows a pair of well separated MBSs hybridized with a low-energy mode with energy of about $15~\mu$eV. Note that the low-energy mode has negligible amplitude near the left end of the wire. A relatively small change in the chemical potential enhances the hybridization, generating two low-energy hybrid modes that extend throughout  the whole system. The presence of these modes may lead to a small (practically vanishing) transport gap, but, again, this is not associated with a phase transition. Upon increasing the size of the system this finite size effect disappears, as the separation of the two MBSs increases and the hybridization with the (localized) finite energy mode becomes negligible. 

The strong disorder case is illustrated in the bottom panels of Figs. \ref{FIG26} and \ref{FIG27}, with the corresponding LDOS shown in Fig \ref{FIG32}. First, we note that there is no stable low-energy mode. The near-zero-energy features that emerge within narrow chemical potential ranges and appear to be consistent with the presence of MBSs are investigated by calculating the LDOS corresponding to $\mu$ values labeled $\delta_1$-$\delta_3$. As shown in Fig. \ref{FIG32}, the lowest energy mode in panel $\delta_2$ is strongly  localized near the right end of the wire, while that in panel  $\delta_3$ is roughly contained within the right half of the system. The scenario in panel $\delta_1$ appears to be similar to the weak disorder case $\beta_1$: a pair of partially separated Majorana modes with one MBS (the left one) localized away from the corresponding edge. However, there is a fundamental difference between the two cases: increasing the length of the weakly disordered system increases the separation of the Majorana modes (and their stability), while extending the size of the strongly disorder system has no effect on the Majorana modes (which become a disorder-induced ps-ABS localized within the wire). 

One main conclusion regarding the intermediate and strong disorder regimes is that the characteristic size of the (connected) parameter space regions that support reasonably well separated Majorana modes decreases with increasing disorder strength. For $V_0=0.8~$meV (i.e., intermediate disorder) we have evaluated the characteristic size of these ``topological islands'' (or ``patches'') to be of order $\gamma$. We also note that the emergence of these ``islands'' is, essentially, a finite size effect. Upon increasing the size of the system, the ``islands'' corresponding to the topological regime grow and eventually merge into a single region, while the ``islands'' (``patches'') that emerge within parameter ranges corresponding to the trivial phase shrink and disappear.  This suggests that investigating the dependence of the size of the ``topological islands'' on the wire length (using wires of different lengths fabricated under the same conditions) could corroborate (or disprove) the topological nature of these ``patches''.  
Another important point is that a disordered system characterized by weak spin-orbit and SM-SC couplings generically supports low-energy modes with large characteristic length scales (comparable to the size of the system). While the energy minima of such ``effectively delocalized'' states may generate the apparent closing (and reopening) of the transport gap, they  are not (typically) associated with topological quantum phase transitions.   On the other hand, in (relatively) short systems the low-energy extended states can overlap significantly with the Majorana end modes, even when they are reasonably well separated, potentially affecting their stability. Hence, the energy gap characterising the extended states is an important {\em operational measure} of the stability of the Majorana modes. We emphasize, however, that the extended modes can overlap strongly with the Majorana modes (which also have relatively large characteristic length scales in this regime) even if they do not extend throughout the whole wire, thus remaining ``invisible'' in a transport experiment.

In the next section, we consider the issue of ``topology'' in finite size disordered SM-SC nanowires based on their transport properties, by calculating the local and non-local conductance, as well as the topological invariant defined in terms of the zero-energy reflection matrix. We discuss the relevance and the consequences of our findings in the context of the ongoing experimental investigations of SM-SC hybrid systems.

\section{Conductance characterization of topological patches}\label{S4}

\subsection{Topological invariant}
One way to determine whether a one dimensional superconductor is in a topological phase is to use the topological charge $Q$ defined from the zero-energy reflection matrix $r$ according to the relation~\cite{Akhmerov2011Quantized}
\begin{align}
&Q=Det(r).\label{eq:Q}
\end{align}
This quantity $Q$ has also been referred to as the topological visibility in the literature~\cite{DasSarmaPhysRevB.94.035143}. 
Particle-hole symmetry constrains $Q$ to be real. For a system with a single lead $Q=1$ for any finite system.
Therefore, the reflection matrix $r$ in the above equation is defined from one of the normal leads of a three-terminal superconducting device where $-1<Q<1$. Points in parameter 
space where $Q=0$ for a system longer than the mean-free path describes a critical point of the system with quantized thermal conductance~\cite{Akhmerov2011Quantized}. This is the TQPT in the SM-SC nanowires, with $Q < (>) 0$ being topological (trivial) by definition.
This motivates the definition of the scattering matrix topological invariant as $\mathcal{Q}=\textrm{sign}(Q)$~\cite{Fulga2011Scattering}.
The topological superconducting phase, defined here conventionally as $\mathcal{Q}=-1$, is expected to be characterized by the presence of a MZMs with a localization length comparable to the coherence length (inverse Lyapunov exponent) $\xi\ll L$, where $L$ is the length of the superconducting wire. One can estimate the Lyapunov exponent from the scattering matrix using the exponential asymptotic scaling of the transmission $trans\sim e^{-2L/\xi}$ as 
\begin{align}
&L/\xi_{trans}=-(1/2)\textrm{Log}(trans),\label{eq:xitrans}
\end{align}
where $trans$ is the total quasiparticle transmission probability through the device. 
Note that we have now separated the definition of 'topological' $(Q <0)$ from the   existence of end MZMs $(L > \xi_{trans})$, on a calculational level, and although for infinite pristine systems, isolated MZMs must appear whenever $Q<0$, there is no such mathematical guarantee for disordered finite systems.  Indeed, we find that there are situations where the topological patches have $Q<0$, but there are no end MZMs in the system because of the proliferation of low energy Andreev bound states throughout the bulk of the wire completely destroying any meaningful distinction between the bulk and the boundary (as emphasized in sec. IV in describing the spectral properties).  Such topological patches are 'topological' with an apparent gap, but do not carry any MZMs at the ends 
even when apparent ZBCPs may appear from both ends at finite temperatures because of the presence of numerous disorder-induced low-energy Andreev bound states.  

\begin{figure}[t]
\centering
{\includegraphics[width = 1.4\linewidth]{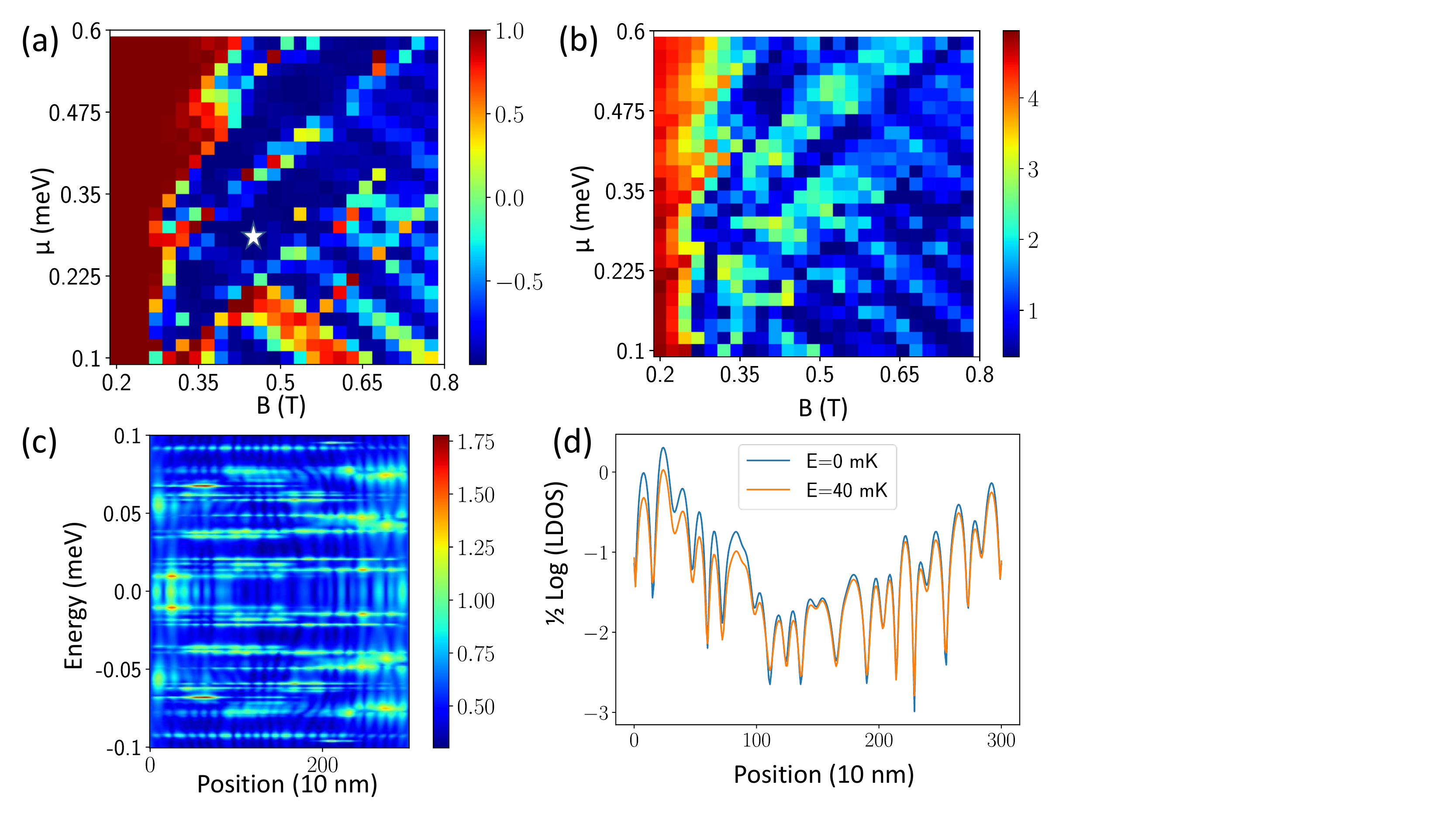}}
\caption{(a) Topological charge $(Q=Det r)$ (see Eq.~\ref{eq:Q})as a function of magnetic field $B$ and chemical potential $\mu$ in the superconductor for a $L=3\,\mu m$ for parameters matched to InAs quantum wire devices~\cite{aghaee2022inas}  and disorder amplitude $V_{dis}=0.8$ meV  shows a regime of topological charge. (b) $(L/\xi)_{trans}=1/2\textrm{Log}(trans)$ (see Eq.~\ref{eq:xitrans}) corresponding to the topological charge shows the regimes where the transmission is low enough (i.e. $(L/\xi)_{trans}\gtrsim 2$) to support a well-defined topological phase. (c) LDOS at $\mu=0.29$ meV and $B=0.42$ T (with $Q=-0.995$,marked by star in panel (a)) suggests MZM at zero energy localized at ends. (d) Plot of $1/2\textrm{Log}(LDOS)$ i.e. cross-sections of LDOS from (c) 
at two energies $E=0,40$ mK shows linear 
decay of wave-function amplitude into the bulk with a length-scale consistent with $(L/\xi)_{trans}=2.6$ at that point.
}
\label{fig:Topinv}
\end{figure}

\begin{figure}[t]
\centering
{\includegraphics[width = 1.0\linewidth]{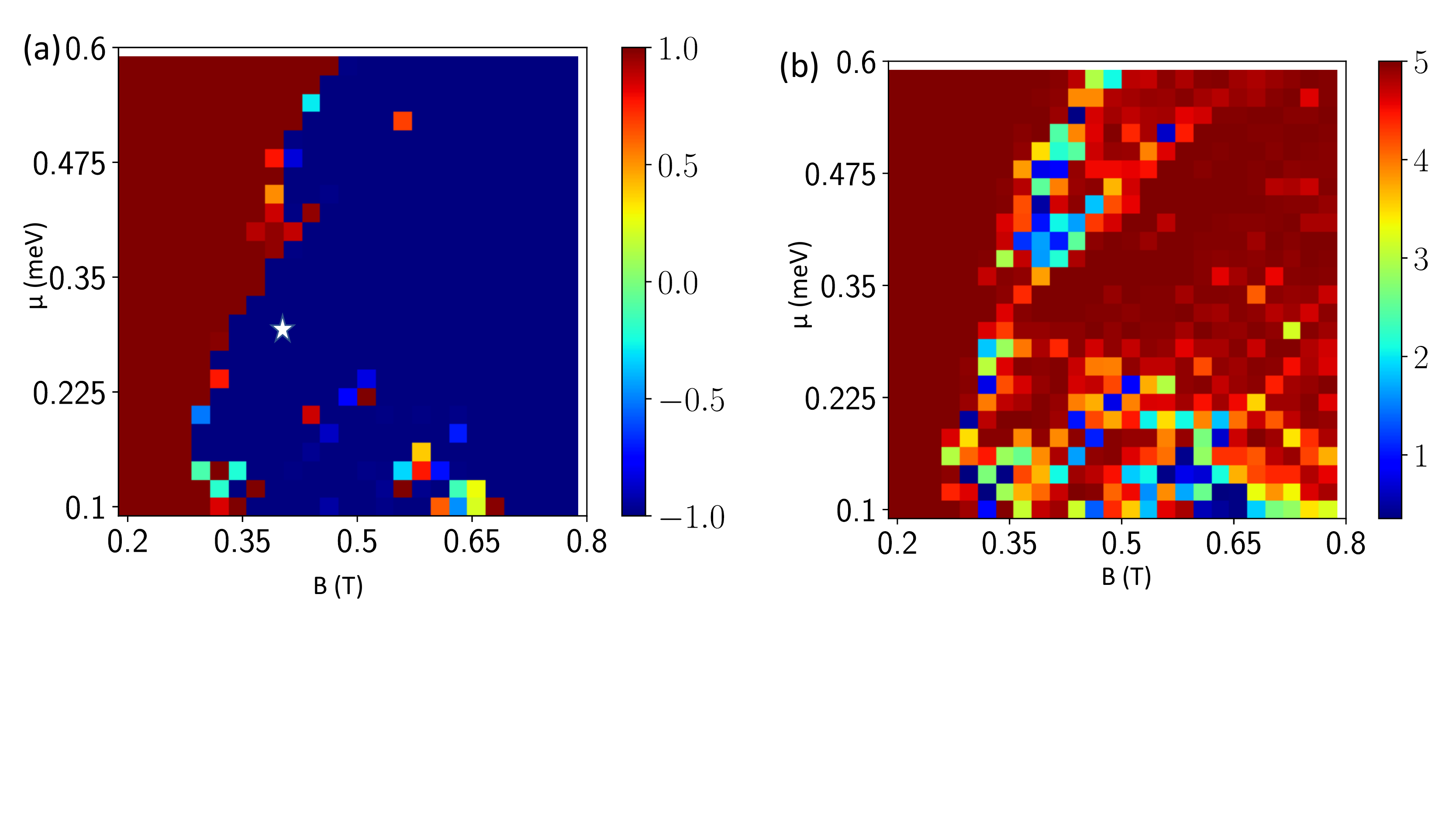}}
\caption{(a,b) Similar to Fig.~\ref{fig:Topinv} except for an $L=24\,\mu m$ long wire. This figure shows that the reference point marked by a star in Fig.~\ref{fig:Topinv}(a) is in the topological phase.
}
\label{fig:Topinvlong}
\end{figure}

\subsection{Topological superconducting phase}
Fig.~\ref{fig:Topinv}(a) shows the topological charge $Q$ for a $L=3\,\mu m$ long Majorana nanowire~\cite{Lutchyn2010Majorana,Oreg2010Helical} with parameters chosen to match InAs wires from the recent Microsoft 
experiment~\cite{aghaee2022inas}, except that we ignore the magnetic field suppression of the bulk superconducting order parameter. 
All calculations in this section are done using the KWANT transport package~\cite{kwant}.
Following the estimates from the recent experiment we choose Rashba coupling $\alpha=8 meV-nm$, bulk superconducting order parameter $\Delta_0= 0.12$meV and the superconducting coupling is $\delta=0.15 meV$. The topological charge is plotted in this figure versus the chemical potential $\mu$ of the Majorana nanowire (controlled by the plunger gate) and the applied magnetic field $B$. The disorder potential for this calculation is chosen to have a correlation length of $\xi_{dis}\sim 30 nm$ with an RMS amplitude $V_{dis}=0.8$ meV so as to support an unambiguous topological phase in long wires with large $L/\xi_{trans}\gg 2$ (see Fig.~\ref{fig:Topinvlong}). The putative topological phase boundary $Q=0$ seen from the figure shows substantial mesoscopic fluctuations consistent with previous results on short wires~\cite{Adagideli2014Effects}. We mention that these mesoscopic fluctuations are what lead to topological 'patches (or islands)' where $Q<0$ regions over small parameter (e.g. magnetic field) regimes coexist with nearby nontopological $(Q>0)$ regions in the parameter space as can be seen in both Figs.~\ref{fig:Topinv} and (less so)~\ref{fig:Topinvlong}. These fluctuations are seen to be substantially reduced in the longer wire results shown in Fig.~\ref{fig:Topinvlong}. Some of the topological phase boundary can be discerned through the vanishing of $L/\xi_{trans}$ seen in Fig.~\ref{fig:Topinv}(b), which would correspond 
to a quantized thermal conductance at the $Q=0$ boundary~\cite{Akhmerov2011Quantized}. The topological phase $Q<0$ (seen in blue in panel a) is seen in Fig.~\ref{fig:Topinv}(b) to potentially support localized MZMs based on the low transmission that can be inferred from $L/\xi_{trans}>2$. 

\begin{figure*}[tbp]
\centering
{\includegraphics[width = \textwidth, height=6cm]{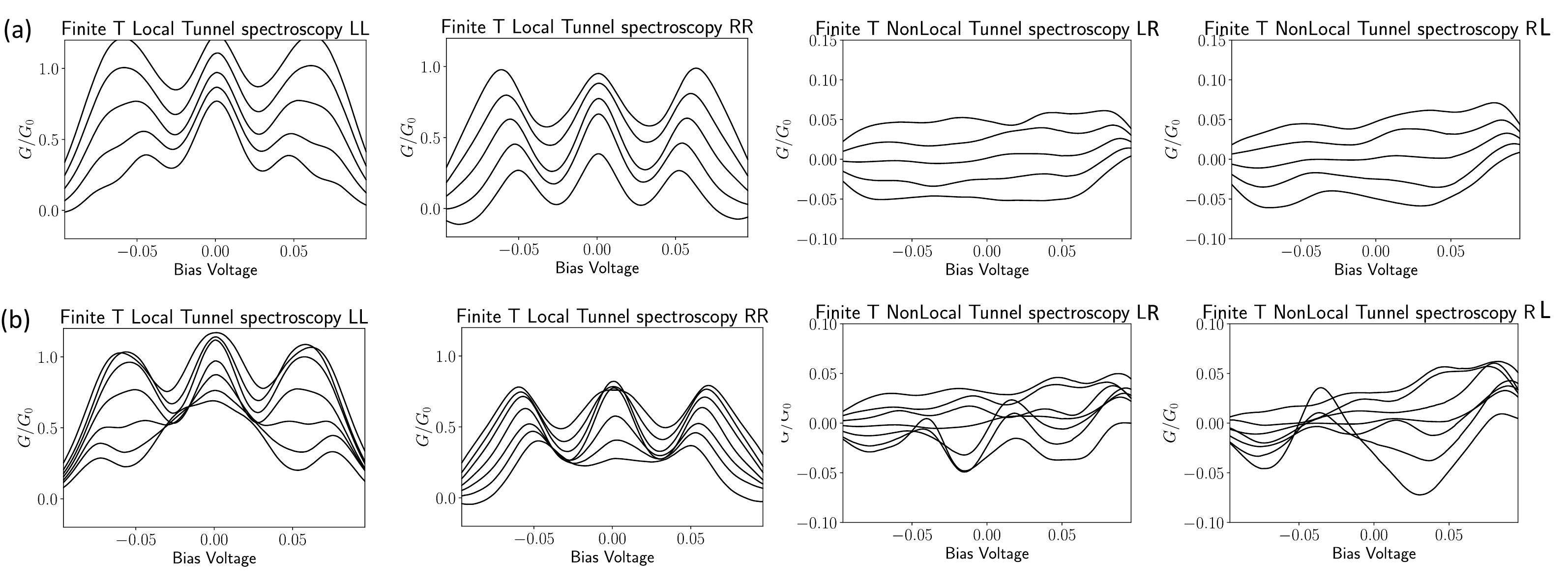}}
\caption{Conductance tensor $G_{ab}$ of a three-terminal Majorana device where $a,b$ label leads $L,R$ at a temperature of $T=40\, mK$. The local conductances $G_{LL}$ and $G_{RR}$ show zero-bias conductance peak signatures of the topological phase. The non-local conductance $G_{LR},G_{RL}$ can indicate a topological gap closure. The conductance $G_{ab}$ are plotted as magnetic field $B$ and chemical potential $\mu$ is varied around the reference point $\mu=0.29$ meV and $B=0.42$ T in Fig.~\ref{fig:Topinv}.
(a) Shows a waterfall of $G_{ab}$ where $B$ varies from $B=0.378$ T to $B=0.462$ T. (b) Shows a waterfall of $G_{ab}$ where $\mu$ varies from $\mu=0.24$ meV to $\mu=0.34$ meV. The
conductance curves in both panels are shifted vertically symmetrically about the reference point to make the curves visible.
}
\label{fig:Conductance1}
\end{figure*}

\begin{figure*}[tbp]
\centering
{\includegraphics[width = 1.0 \linewidth]{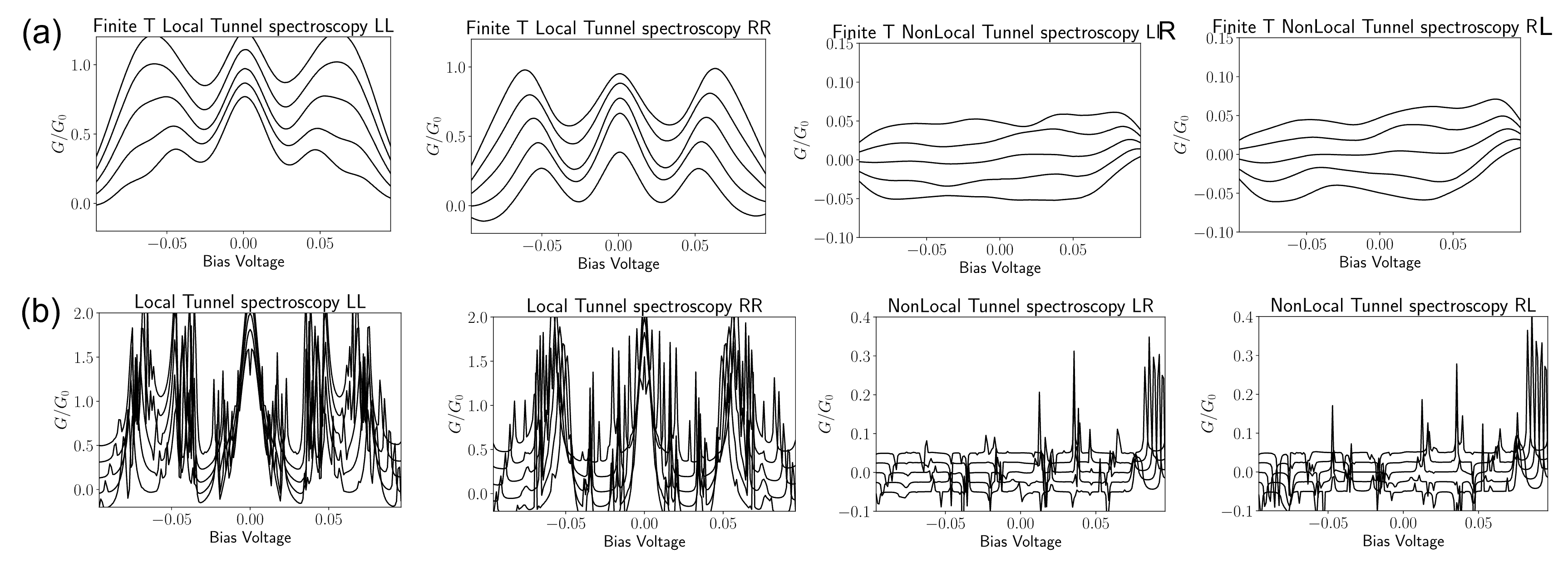}}
\caption{(a) Same as panel Fig.~\ref{fig:Conductance1}(a) repeated here for reference. (b) Presents the zero-temperature results from which panel (a) is generated at $T=40$ mK. We see that 
small gaps at $T=0$ mK often disappear at $T=40$ mK. While this is not apparent because of the vertical shift of the conductances in the wter-fall plots, we note that the $T=0$ local conductance peak heights near zero bias are nearly (i.e. within $5\%$) quantized with $G\sim 2 G_0$ both from the left (i.e. $G_{LL}$)and right (i.e. $G_{RR}$).
}
\label{fig:Conductance1T0}
\end{figure*}

To verify this conjecture, we plot the LDOS as a function of energy $E$ and position $x$ at a reference point $\mu=0.29$ meV and $B=0.42$ T in 
Fig.~\ref{fig:Topinv}(c). This reference point, which is marked by a star in Fig.~\ref{fig:Topinv}(a), is at a parameter value where the topological charge is $Q=-0.995$. Furthermore,
the reference point is deep in the topological region, away from the fluctuating phase boundary for the long wire phase diagram shown in Fig.~\ref{fig:Topinvlong} and is in this sense unambiguously in the topological superconducting phase. This, however, is not obvious just by looking at the finite size effective phase diagram in Fig.~\ref{fig:Topinv}. Note that to enhance visibility in panel (c), the LDOS plot (in this and following figures) is raised to power $1/3$ to prevent localized states from dominating the color scale. This has no impact on the qualitative interpretation of the figure and only implies that the color bar cannot be interpreted quantitatively for these figures. We see that the reference point LDOS at $E=0$ shows strong intensity that is localized near the ends of the wire. Thus, the system has end MZMs here. The quantitative form of this localization becomes clear in the sectional 
plot of the logarithm of the LDOS at $E=0,40$ mK shown in Fig.~\ref{fig:Topinv}(d). The peaks of the $\textrm{Log}(LDOS)$ shows a linear decrease from the ends of the wire to the middle 
consistent with a the transmission estimate $L/\xi_{trans}>2$ from transmission.

\begin{figure}
\centering
{\includegraphics[width = 1.3\linewidth]{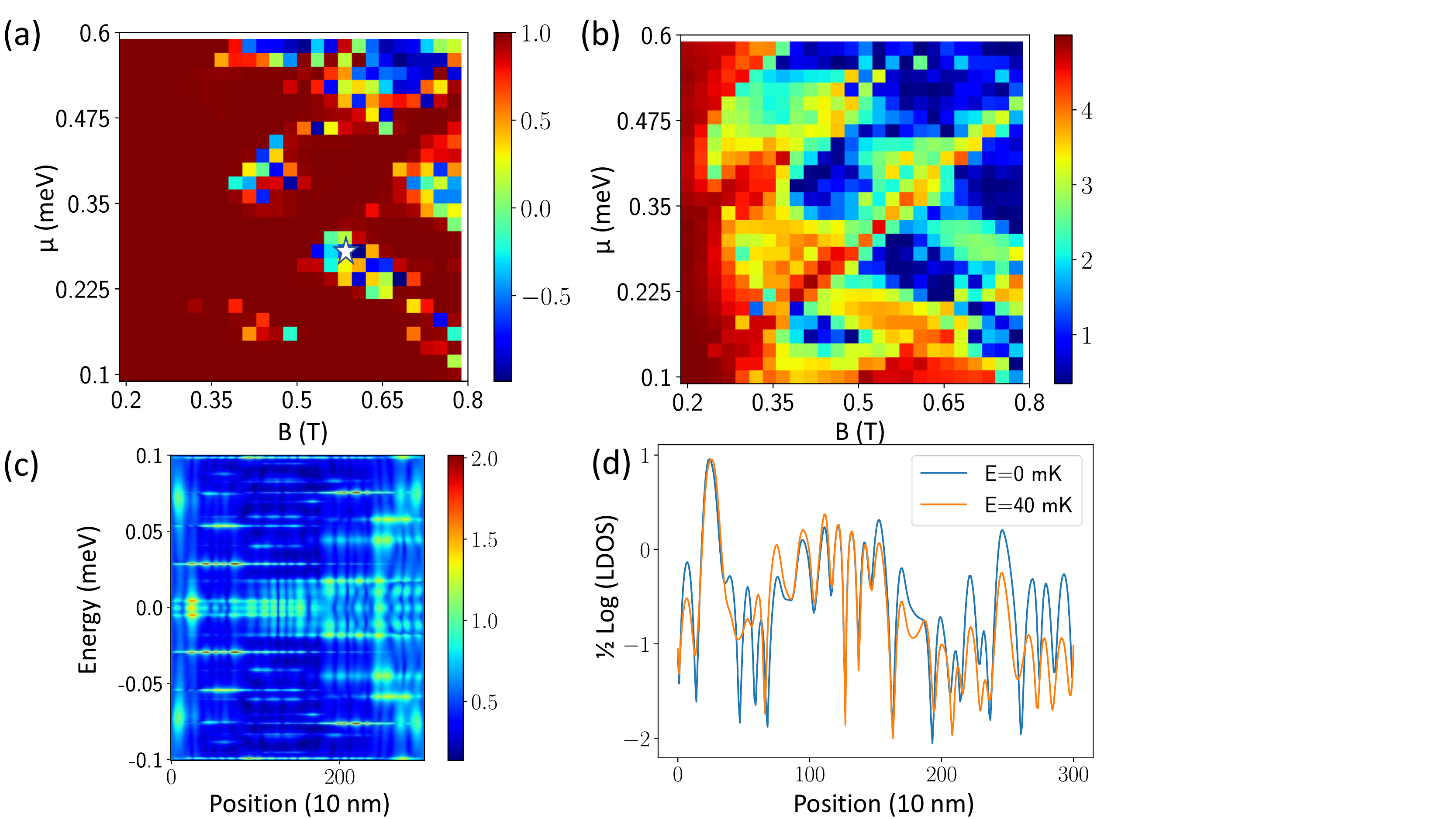}}
\caption{(a) Topological charge $(Q=Det r)$ as a function of magnetic field $B$ and chemical potential $\mu$ in the superconductor for a $L=3\,\mu m$ for same parameters as Fig.~\ref{fig:Topinv} with disorder amplitude increased $V_{dis}=1.5$ meV.  shows topological patches (b) $(L/\xi)_{trans}=1/2\textrm{Log}(trans)$ corresponding to the topological charge shows the regimes where the transmission is low enough (i.e. $(L/\xi)_{trans}\gtrsim 2$) to support a well-defined topological phase. (c) LDOS at $\mu=0.28$ meV and $B=0.56$ T (with $Q=-0.846$, marked by star in panel (a)) shows delocalized zero-energy state (d) Plot of $1/2\textrm{Log}(LDOS)$ i.e. cross-sections of LDOS from (c) 
at two energies $E=0,40$ mK shows delocalized state consistent with $(L/\xi)_{trans}=0.8$ at that point.
}
\label{fig:Topinvpatch}
\end{figure}

\begin{figure}[t]
\centering
{\includegraphics[width = 1.1\linewidth]{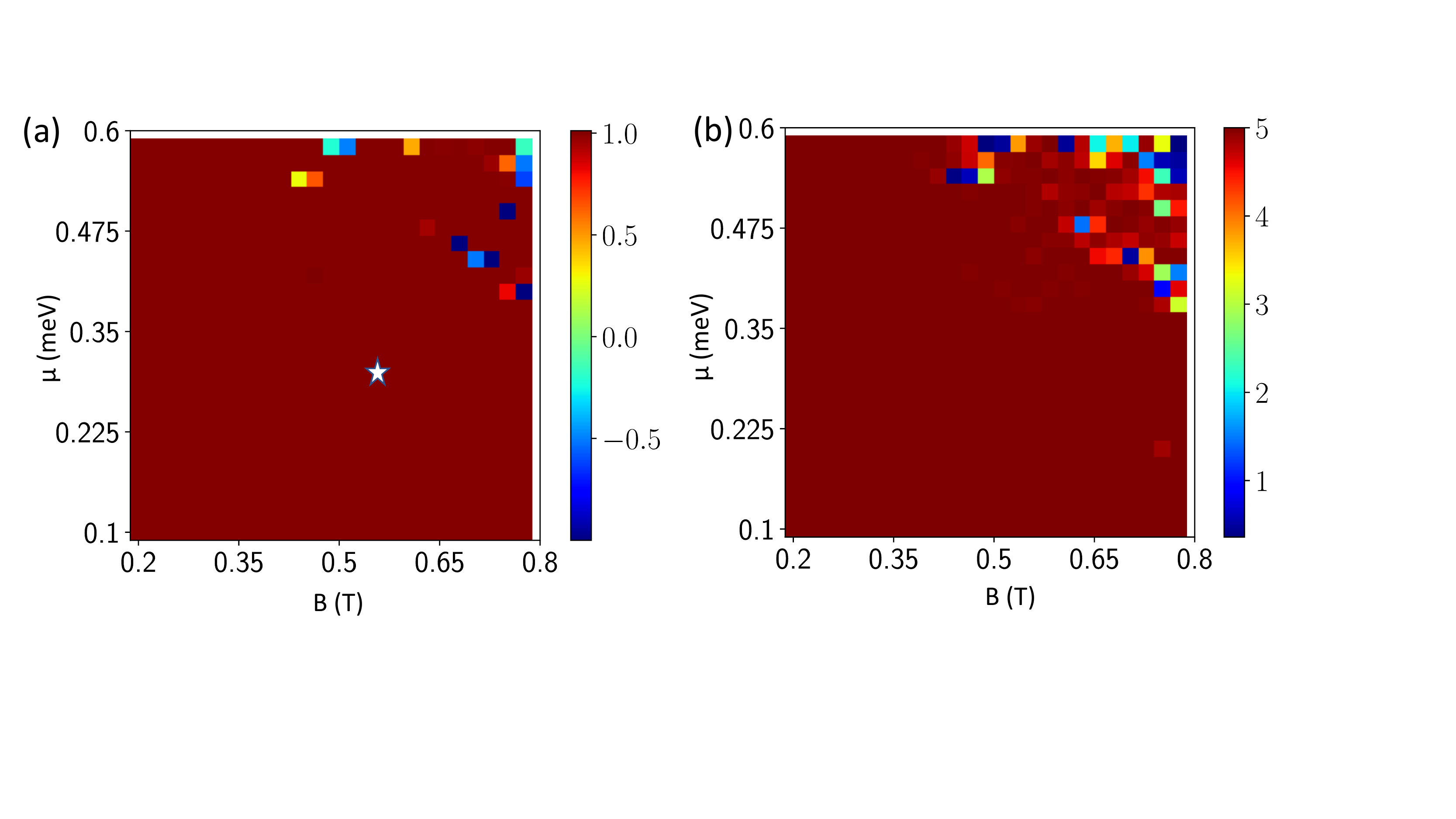}}
\caption{(a,b) Similar to Fig.~\ref{fig:Topinvpatch} except for an $L=40\,\mu m$ long wire. This figure shows that the topological patch reference point marked by a star in Fig.~\ref{fig:Topinvpatch}(a) when extended to a longer wire ends up in the trivial phase.
}
\label{fig:Topinvpatchlong}
\end{figure}

The topological charge $Q$ is not measurable and even the thermal conductance which experimentally marks the boundary $Q=0$~\cite{Akhmerov2011Quantized} of the topological phase is difficult to measure (and we know of no attempt to measure the thermal conductance experimentally in this context). However, the topological phases have been proposed to be electrically detectable based on zero-bias peaks in local conductance~\cite{Sengupta2001Midgap} as well as gaps in non-local conductance~\cite{rosdahl2018andreev,PanThreeterminal2021}.
\begin{figure*}[tbp]
\centering
{\includegraphics[width = \textwidth]{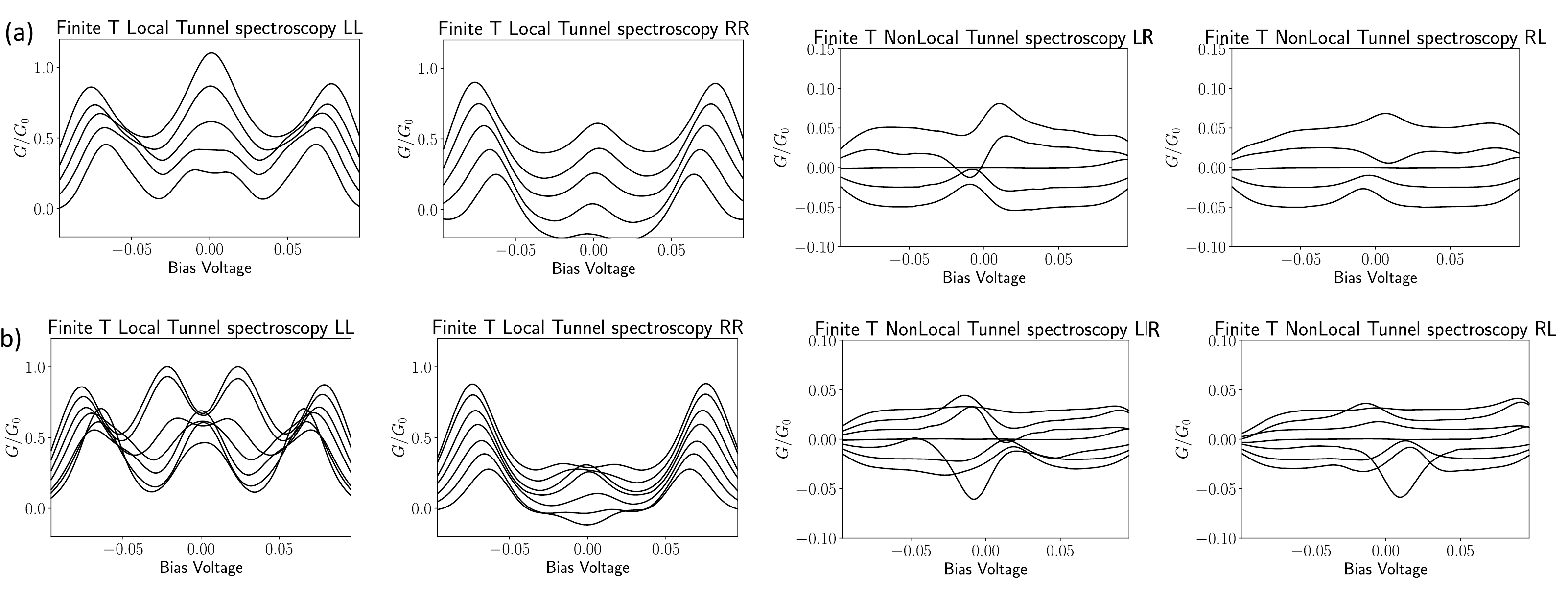}}
\caption{Conductance tensor $G_{ab}$ of a three-terminal Majorana device where $a,b$ label leads $L,R$ at a temperature of $T=40\, mK$. The local conductances $G_{LL}$ and $G_{RR}$ show zero-bias conductance peak signatures of the topological phase. The non-local conductance $G_{LR},G_{RL}$ can indicate a topological gap closure. The conductance $G_{ab}$ are plotted as magnetic field $B$ and chemical potential $\mu$ is varied around the reference point $\mu=0.29$ meV and $B=0.42$ T in Fig.~\ref{fig:Topinv}.
(a) Shows a waterfall of $G_{ab}$ where $B$ varies from $B=0.378$ T to $B=0.462$ T. (b) Shows a waterfall of $G_{ab}$ where $\mu$ varies from $\mu=0.24$ meV to $\mu=0.34$ meV. The
conductance curves in both panels are shifted vertically symmetrically about the reference point to make the curves visible.
}
\label{fig:Conductance2}
\end{figure*}
\begin{figure*}[tbp]
\centering
{\includegraphics[width = \textwidth]{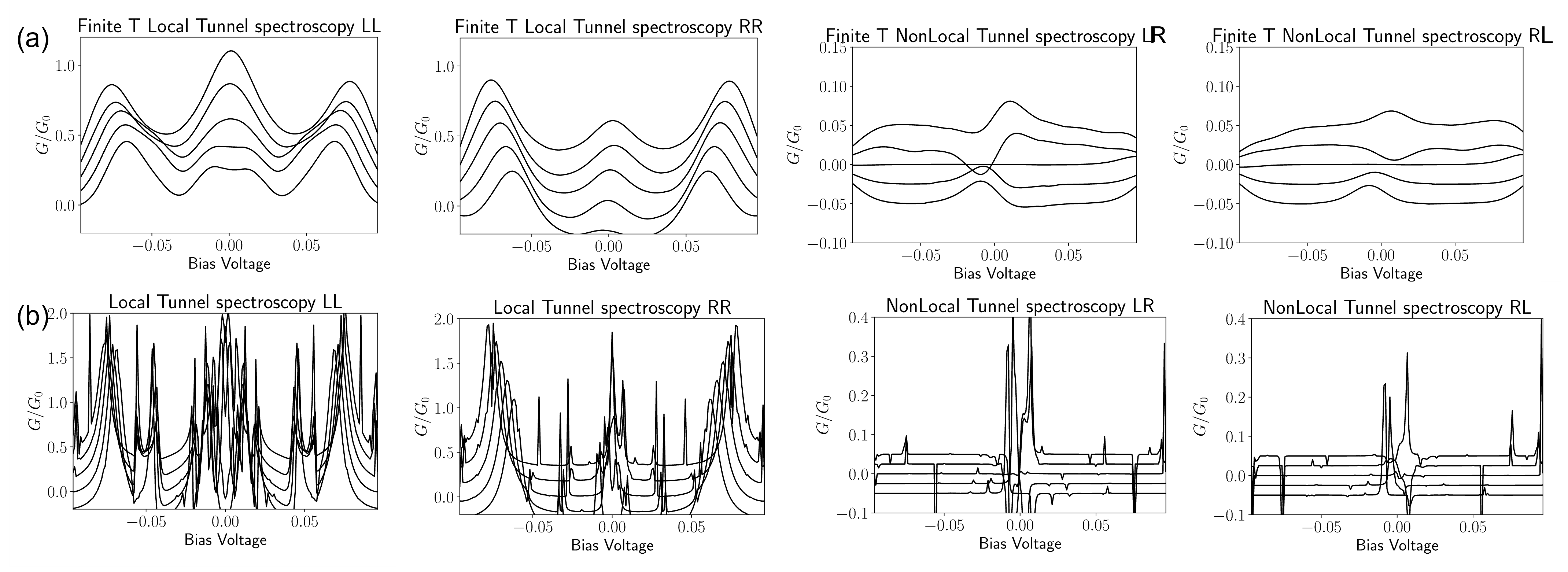}}
\caption{(a) Same as panel Fig.~\ref{fig:Conductance1}(a) repeated here for reference. (b) Presents the zero-temperature results from which panel (a) is generated at $T=40$ mK. We see that 
small gaps at $T=0$ mK often disappear at $T=40$ mK. The local zero-bias conductance peaks on the right are seen to be strongly split and substantially below quantization. This is also true to a lesser extent for most of the ZBCPs on the left. This can be thought to be a result of strong hybridization of the zero energy state with other states.
}
\label{fig:Conductance2T0}
\end{figure*}
\begin{figure}
\centering
{\includegraphics[width = 1.3\linewidth]{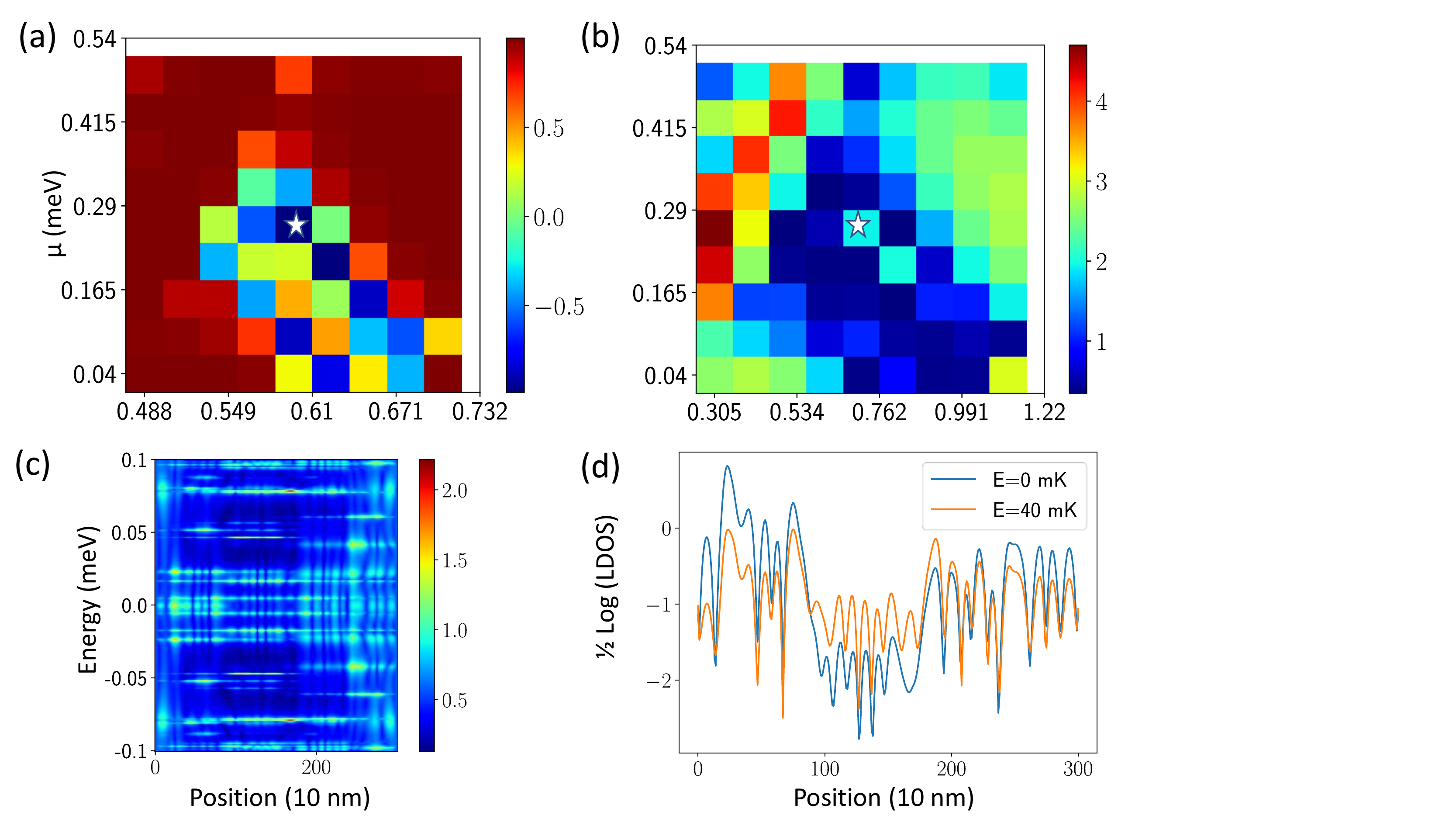}}
\caption{(a) Zoom in of Fig.~\ref{fig:Topinvpatch}(a) shows a reference point $\mu=0.29$ meV and $B=0.61$ T (with $Q=-0.983$) marked by star. (b)
Shows that this point has weak transmission i.e. $(L/\xi)_{trans}\sim 1.9$) and may support a well-defined topological phase. (c) LDOS at  shows a somewhat localized state where the DOS dips significantly in the middle (d) This is confirmed by the section of the LDOS plot on a log scale. 
}
\label{fig:Topinvpatchzoom}
\end{figure}
Fig.~\ref{fig:Conductance1} shows the three-terminal conductance tensor $G_{ab}$ (with $a,b$ labeling leads $L,R$) at temperature $T=40\,mK$, which is consistent with the apparent electron temperature in the recent experiments~\cite{aghaee2022inas}. The middle curve in each of these panels shows the conductance of the reference topological point $\mu=0.29$ meV and $B=0.42$ T marked by a star in Fig.~\ref{fig:Topinv}(a).  The other curves in Fig.~\ref{fig:Conductance1}(a) shows the variation of the conductance with changing magnetic field $B$ while Fig.~\ref{fig:Conductance1}(b) shows variation with varying chemical potential $\mu$. All parameters plotted in this figure are in the topological phase according to the topological invariant $Q<0$ as is suggested by the presence of a robust ZBCP in all local conductance curves $G_{LL}$ and $G_{RR}$. The non-local conductances $G_{LR},G_{RL}$ are more complicated to interpret. While the reference point appears gapped with a nearly flat non-local conductance, changing the magnetic field $B$ or chemical potential $\mu$ appears to rapidly close the gap suggesting a 
very small patch in for the topological phase. This contradicts the phase diagram Fig.~\ref{fig:Topinv}(a), which shows a rather contiguous region of topological phase with a relatively low transmission according to Fig.~\ref{fig:Topinv}(b). The apparent contradiction is resolved by comparing the above plots with the zero-temperature plot shown in Fig.~\ref{fig:Conductance1T0}, which shows a small gap in the non-local conductance. This suggests that while a robust topological phase might be possible for a $L=3\,\mu m$ wire, non-local conductance at $T=40\,mK$ might not 
be able to distinguish the relatively well-defined topological phase discussed in this section with the regime of small topological patches in the next subsection. Thus, not observing a finite temperature gap in the nonlocal conductance does not necessarily rule out the existence of a topological phase!

\subsection{Topological patch regime}
\begin{figure}[t]
\centering
{\includegraphics[width = 1.1\linewidth]{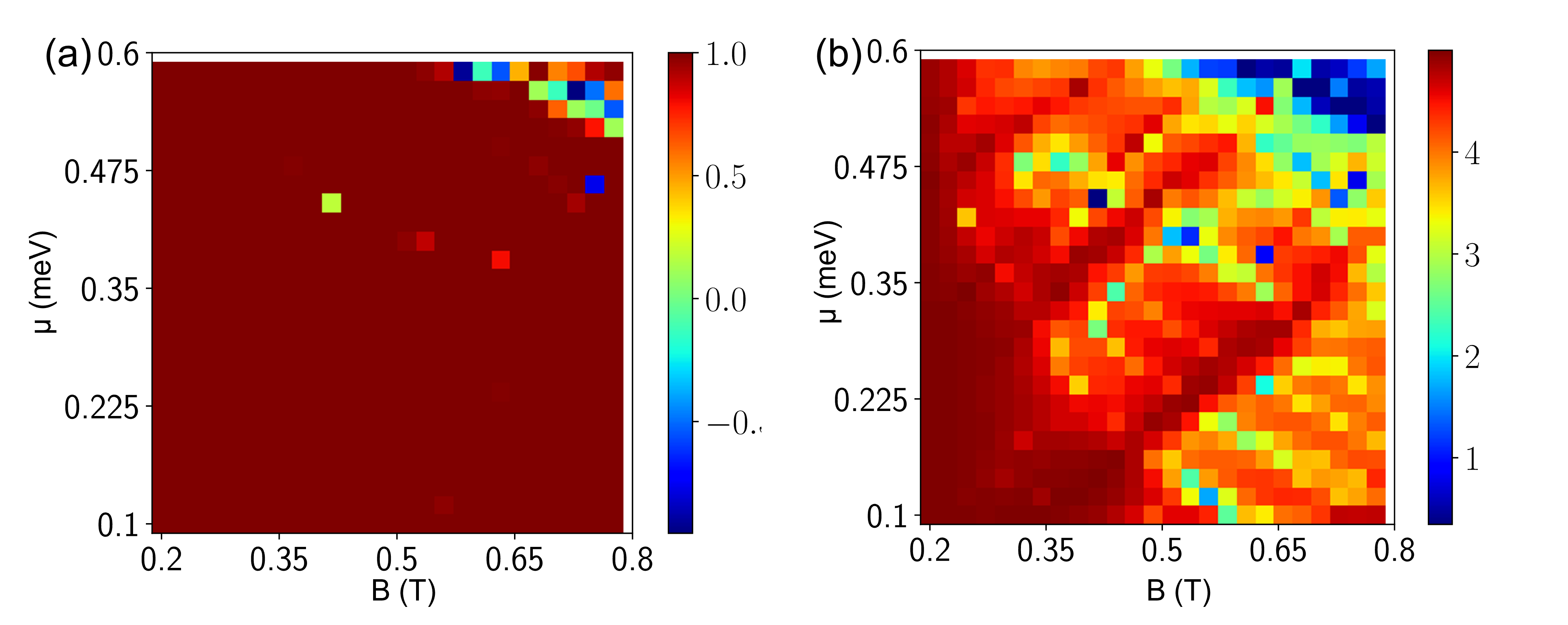}}
\caption{(a,b) Similar to Fig.~\ref{fig:Topinvpatch} except for increased disorder $V_{dis}=2.0$ meV. This figure shows that the topological patches in Fig.~\ref{fig:Topinvpatch} are eliminated with increasing disorder. 
}
\label{fig:Topinvstrong}
\end{figure}

Fig.~\ref{fig:Topinvpatch}(a) shows the topological charge $Q$ for a $L=3\,\mu m$ long Majorana nanowire model~\cite{Lutchyn2010Majorana,Oreg2010Helical} with parameters chosen to match InAs wires from the recent Microsoft 
experiment~\cite{aghaee2022inas}. The topological charge is plotted in this figure versus the chemical potential $\mu$ of the Majorana nanowire (controlled by the plunger gate) and the applied magnetic field $B$. The disorder potential for this calculation is chosen to have an RMS amplitude $V_{dis}=1.5$ meV (larger than the value $0.8 \mu eV$ used in the examples of Figs.~\ref{fig:Topinv}-~\ref{fig:Conductance1T0}), which supports only small 
topological patches. The putative topological phase boundary $Q=0$ seen from the figure shows substantial mesoscopic fluctuations, including islands of topological phase in the non-topological region, which is consistent with previous results on short wires~\cite{Adagideli2014Effects}. These fluctuations are seen to be substantially reduced in the longer wire results shown in Fig.~\ref{fig:Topinvpatchlong} and reveal an unambiguously trivial superconducting phase 
for this system.  These topological patches are also eliminated by 
increasing the strength of the disorder by $30\%$ as seen in Fig.~\ref{fig:Topinvstrong}. The topological patch $Q<0$ (seen in blue in panel a) is seen in Fig.~\ref{fig:Topinvpatchlong}(b) might be expected to support MZMs based on the non-trivial topological invariant. 
Thus, finding topological patches in finite wires with small gaps may not necessarily imply the robust existence of topology even when these patches satisfy the $Q<0$ criterion, as emphasized already in sec. IV.

Despite the suggestion of MZMs, the LDOS as a function of energy $E$ and position $x$ at the reference point $\mu=0.28$ meV and $B=0.56$ T (star in Fig.~\ref{fig:Topinvpatch}(a)) shown in 
Fig.~\ref{fig:Topinvpatch}(c) indicates a zero-energy state that is delocalized across the wire.  This delocalization is confirmed by the sectional 
plot of the logarithm of the LDOS at $E=0,40$ mK shown in Fig.~\ref{fig:Topinvpatch}(d). The peaks of the $\textrm{Log}(LDOS)$ appear to be relatively uniform from the ends of the wire to the middle of the wire, which is consistent with the relatively high transmission ($(L/\xi)_{trans}=0.8$). Even in the absence of localized zero-bias peaks, we find that the end conductance 
 at the reference point, shown as the  middle curve in each of these panels of Fig.~\ref{fig:Conductance2} displays a zero-bias conductance peak at both ends.  The other curves in Fig.~\ref{fig:Conductance2}(a) shows the variation of the conductance with changing magnetic field $B$ while Fig.~\ref{fig:Conductance1}(b) shows variation with varying chemical potential $\mu$. Variation of the magnetic field suggests a robust ZBCP in all local conductance curves $G_{LL}$ and $G_{RR}$ despite the fact that only a few of these curves are in the topological phase. This is quite different from the variation of the chemical potential, which seems to split the ZBCP rather rapdily. The non-local conductances $G_{LR},G_{RL}$ shows that reference point is gapped with a nearly flat non-local conductance. However, changing the magnetic field $B$ or chemical potential $\mu$ appears to rapidly close this gap, which would be interpreted as a very small topological patch according to the criteria used to analyze the experimental data~\cite{aghaee2022inas}. While this is qualitatively consistent with the phase diagram Fig.~\ref{fig:Topinvpatch}(a) surrounding the reference point, the patch from the topological gap appears rather smaller than the theoretical one for this example. The absence of end localized MZMs could be diagnosed from the zero temperature local conductance (Fig.~\ref{fig:Conductance2T0}) which 
 show ZBCPs to be strongly split and substantially below quantization. 
 
The delocalized zero-energy state in the LDOS in Fig.~\ref{fig:Topinvpatch}(c,d) was accompanied by a low transmission. However, a closer examination of the same phase diagram in 
the vicinity of the topological patch, shown in Fig.~\ref{fig:Topinvpatchzoom}(a) shows that a slightly different reference point ($\mu=0.29$ meV $B=0.61$ T) shows a much stronger 
topological charge $Q=-0.983$. The corresponding transmission shown in Fig.~\ref{fig:Topinvpatchzoom}(b) is also found to be much smaller corresponding to an effective length 
$(L/\xi)_{trans}\sim 1.9$. Indeed, this less that $10\%$ change of the reference point along either direction relative to Fig.~\ref{fig:Topinvpatch} introduces a dip in the the middle of the 
profile of the LDOS shown in Fig.~\ref{fig:Topinvpatchzoom}(c). An examination of the log plot Fig.~\ref{fig:Topinvpatchzoom}(d), shows that this dip is consistent with the transmission coherence length $\xi_{trans}$. Thus, topology in this more disordered case for the finite system has no end MZMs even in the patches where the $Q<0$ condition is unambiguously satisfied.  Going to long wires in this case establish decisively that the short wire topological patches were simply finite size fluctuations and cannot be construed to be 'topological' in any practical sense since there are no end MZMs in the system.  We emphasize that transport experiments cannot detect the existence or not of the end MZMs unless the local conductance is robustly quantized at $2e^2/h$ -- just seeing a ZBCP does not ensure the existence  of MZMs.~\cite{PanPhysRevResearch.2.013377}

\subsection{Summary of analysis of topological patches}
To summarize the findings of this section, an intermediate strength of disorder ($\sim 1.5\,meV$) for a finite length nanowire leads to a topological "phase diagram" with small topological patches (Fig.~\ref{fig:Topinvpatch}). This 
is consistent with previous analysis of topological superconductors~\cite{Adagideli2014Effects,aghaee2022inas}. Stronger disorder ($\sim 2.0\,meV$) eliminates even the patches at $L=3\,\mu m$ (Fig.~\ref{fig:Topinvstrong}). Here we use quotation marks to refer to phase diagram because a finite 
size system does not technically represent a phase unless it can be shown to survive extension to the so-called thermodynamic limit. Indeed, the parameters that demonstrate such topological 
patches at $L=3\,\mu m$ are found to be topologically trivial over the broad vicinity of the patches for longer wires (see Fig.~\ref{fig:Topinvpatchlong}), making it clear that the wires are 
non-topological in a thermodynamic sense. However, the finite wire still shows somewhat robust ZBCPs and a gap at a reference point for the $L=3\,\mu m$ long wire (see Fig.~\ref{fig:Conductance2}), which are characteristics that could be associated with MZMs~\cite{aghaee2022inas}. However an examination of the LDOS shown in Fig.~\ref{fig:Topinvpatch} shows 
something essentially delocalized making this interpretation difficult to justify. On the other hand, the wire with slightly lower disorder (Fig.~\ref{fig:Topinv}) is found to show exponentially localized MZMs in the LDOS as well as a relatively large topological regime. In this case the topological regime falls within the range of the topological phase of a longer wire (Fig.~\ref{fig:Topinvlong}) that is essentially in the thermodynamic limit with weak fluctuations. However, even the intermediate disorder topological patch regime does show a singular point (Fig.~\ref{fig:Topinvpatchzoom}) with a potentially localized MZM and low transmission. The challenge is that this point cannot be separated from the delocalized MZM state by transport measurements at $T=40$ mK, making the intermediate disorder topological patch regime problematic to interpret. Unless the topological patch is stable over a reasonably large parameter regime (e.g. magnetic field) for the finite wire, we cannot be sure that the system is topological in any operational sense even if the nonlocal conductance manifests a small gap in this patch.

\section{Conclusion} \label{S6}

Motivated by the recent Majorana nanowire experiment from Microsoft Quantum~\cite{aghaee2022inas}, we have carried out an extensive analysis of the ``topological'' properties of finite length disordered semiconductor nanowires (e.g., InAs) proximitized by a superconductor (e.g., Al) in the presence of spin-orbit coupling and Zeeman 
splitting~\cite{Sau2010Generic,Sau2010NonAbelian,Lutchyn2010Majorana,Oreg2010Helical,Lutchyn2011Search,Stanescu2011Majorana}. The key issues are the extent to which 
topology can be inferred based on simultaneous local and nonlocal tunneling conductance spectroscopy, and the precise meaning of such an inference.  In particular, we investigate the density of states, the local density of states, the Majorana localization length, the relevant topological index (in the scattering 
geometry), and the conductance matrix (i.e. both local and nonlocal components) to conclude that disordered finite length nanowires may produce mixed signals with respect to the existence or not of topological properties depending on system parameters, disorder strength, and the wire length.  For example, just seeing a gap opening in the nonlocal conductance over some small regions of parameter 
values, which superficially implies the existence of a topological gap, may be misleading because the corresponding local density of states may not manifest any Majorana zero modes at the wire ends.  Such superficial signatures of an apparent topological gap may simply reflect the existence of disorder induced low-energy states with localization lengths comparable to the wire length.  The only way to ensure ``topology'' is to find a topological gap region which is stable over a large 
range of parameter values (e.g., magnetic field and gate voltage).  The typical magnetic field range over which the topological signatures should remain stable is expected to be comparable to the typical magnetic field where the zero bias conductance peaks show up.  
The key problem we identify is that the existence of gap closing and opening signatures in nonlocal conductance along with the existence of zero bias conductance peaks in the local conductance, by themselves, are not definitive in ascertaining ``topology'' in finite disordered wires, unless such features persist over a large variation in the tuning parameters, such as the Zeeman splitting and chemical potential. To make things even more complicated, there is the possibility that occasionally real MZMs may be localized somewhat away from the ends of the system, thus producing no signatures in the local tunneling and leading to an erroneous conclusion that the system is trivial (since no ZBCP shows up from tunneling at the end).  A similar false negative may also arise in the nonlocal tunneling where the gap reopening signal may occasionally be  far too weak for it to effectively manifest itself in the experiment even when a real topological gap actually exists.

One key aspect of the current SM-SC nanowire samples playing a crucial role in the Majorana physics is  the fact that the relative magnitudes of the three important length scales in the problem are comparable:  mean free path (or localization length), SC coherence length, and the wire length.  Of these three length scales, only the wire length is approximately ($\sim$ 3 microns) known in the Microsoft experiment.  Our best estimates of the other two length scales indicate that the conditions necessary for the realization of isolated end MZMs, namely, that the wire length is much longer than the coherence length ($\sim$ 1-3 microns)  which in turn is much less than the mean free path  ($\sim$ 1 micron),  may not be decisively satisfied because of the relatively small disorder to gap ratios in the system.  These problems are soluble by using longer wires and larger gap to disorder ratios in the future experiments. 
This approximate one micron estimate for the mean free path is also consistent with the interface random charged impurity density of $\sim 3.10^{12} /{\rm cm}^2$, as estimated for the samples used in the Microsoft experiment.~\cite{aghaee2022inas} We note that this experiment is already a huge improvement over all earlier nanowire experiments in this context, since the earlier experiments used much shorter  wires (of lengths $\sim 1$ micron) having larger disorder.

A key finding of ours, which was not realized in earlier works on Majorana zero modes (including our own work), is that it is insufficient to observe simultaneous unquantized ZBCPs from both ends in the local tunneling conductance along with weak gap closing/opening features in the nonlocal conductance over small variations of the system parameters (most particularly, the Zeeman field) in order for a decisive claim of topology.  To have stable Majorana zero modes, the ``topological phase'' should persist over a reasonably large parameter region, instead of consisting of small patches surrounded by trivial regions, with repeated field-tuned transitions between topological and trivial regimes indicating the presence of considerable disorder-induced low energy 'Griffiths' states in the bulk of the wire.  The problem is that such low energy states throughout the bulk, if present, completely suppress the end MZM properties, destroying the bulk-boundary correspondence, and in such situations, having a small bulk gap may not necessarily imply the existence of anything resembling a pair of isolated non-Abelian end MZMs, since the bulk of the wire is filled with ``effective'' Majorana modes created by the random disorder.  Our theory explicitly shows that such a disturbing situation is generic in the intermediate disorder regime, and experimental observations of fragile gap closing/opening along with weak ZBCPs is insufficient for the manifestation of  any topology with non-Abelian properties. A pair of Majorana modes (strongly) overlapping with disorder-induced low-energy states may exhibit some stability within a small patch in the control parameter space, but this does not imply that it possesses non-Abelian properties and does not guarantee that its stability will be enhanced by increasing the system size.
Of course, it is possible, by sheer luck, that some of these fragile ``topological'' patches inferred on the basis of nonlocal gap closing/opening are in fact truly topological, with non-Abelian end MZMs, but this is unlikely if the stability of the effective ``topology'' persists only over small magnetic field regions.  In any case, local and nonlocal transport is incapable of decisively settling this point  in finite disordered wires, unless the effective topology (as obtained by the existence of gap closing/opening along with ZBCPs) is stable over a large parameter space.  Disorder-induced low energy bulk states may very well lead to artificial ZBCPs at finite temperatures, mimicking MZMs, but these ZBCPs are neither quantized nor conclusive signatures for isolated end MZMs.  In the same vein, some of these bulk localized  states may have localization lengths comparable to or larger than the wire length, thus leading to artificial gap opening features which are not the topological bulk gaps.  Such disorder-induced gap opening features in the nonlocal conductance are misleading and would disappear in longer wires.  Since the current experiments cannot directly measure any topological invariants, there are limited options in deciding about the topological nature of the system from tunneling transport spectroscopy, including (1) efforts to measure the thermal conductance which should be quantized at the TQPT; (2) trying to find quantized ZBCPs stable to large variations in system parameters, including tunnel barrier conductance~\cite{lai2022quality}; (3) observing stable topological features in transport over large regimes of the Zeeman field and the gate voltage; (4) trying experiments to look for non-Abelian properties (e.g. braiding). 
We point out that although the measurement of nonlocal conductance is necessary to establish the closing and the opening of the bulk gap at the TQPT, such nonlocal measurements will be unsuccessful in long wires if the disorder is not sufficiently reduced so that the mean free path is longer than the wire length, and therefore, both longer wires and lower disorder are necessary to demonstrate the exponential protection characterizing Majorana zero modes which underlie topological quantum computation.
We believe, and our current work establishes that the gap to disorder ratio in the SM-SC hybrid systems must increase considerably 
before one can be certain that decisive topology has been established. 
In this context, it may be worthwhile to consider some additional diagnostic measurements for Majorana signatures such as  the fractional Josephson effect,~\cite{Kwon2004Fractional} noise correlation spectroscopies~\cite{bolech2007observing,Akhmerov2011Quantized}, finite -frequency spectroscopy around the TQPT~\cite{tewari2012probing}, and ZBCP fidelity studies~\cite{lai2022quality} which could meaningfully complement the local and nonlocal transport spectroscopies of Ref. \onlinecite{aghaee2022inas}.  It may also be worthwhile to carry out noninvasive local tunneling measurements along the wire to ensure that the bulk is free of abundant low energy bound states suppressing Majorana physics.
The breakthrough Microsoft experiment provides clear guidance for future efforts and tells us precisely how the experiments should improve in our search for Majorana. Although some of the current Microsoft samples are likely to have manifested topological patches with small gaps, longer wires  and more robust stability to parameter variations as well as larger values of gap to disorder ratios seem to be necessary  for developing  Majorana-based topological qubits in future experiments.  The current work indicates that a factor of 2-4 increase  both in the gap to disorder ratio and the wire length  compared with the current Microsoft samples in Ref.~\onlinecite{aghaee2022inas} may be adequate to reach the decisively topological regime with stable Majorana zero modes at the wire ends.  An encouraging recent materials development in this context is the just-appeared report of  the fabrication of InAs-Al hybrid SM-SC structures with mobilities exceeding $10^5 cm^2/V.s$~\cite{zhang2023mobility}.

We believe that the Microsoft experiment is an essential breakthrough in the search for topological superconducting phase, but unless the topological gap regime is stable over a large parameter regime, it is unclear that the system would manifest non-Abelian topological properties.  Our best guess based on our work is that the current systems, although much improved from the earlier nanowires used in the older experiments, may still be in the intermediate disorder regime with considerable low energy subgap Andreev bound states, which may produce some signatures (e.g., gap opening/closing over small patches of parameter values) of topological gap. However, these signatures may disappear if longer wires are used, as shown in our concrete examples.  We cannot rule out that some of the studied nanowires are indeed topological with small gaps and weak stability, but any definitive conclusion would necessitate wires with much reduced disorder to gap ratios, which can be achieved either by reducing the disorder or by increasing the induced gap (or both).  We remain optimistic that future experiments on cleaner samples would provide definitive evidence for a topological superconducting phase with localized non-Abelian Majorana zero modes. 

\begin{acknowledgements}
We are grateful to Roman Lutchyn, Chetan Nayak, Haining Pan and Anton Akhmerov for valuable discussions and advice. JS acknowledges using KWANT code from Bas Nijoholt's gitbuh (https://github.com/basnijholt/majorana-nanowire-conductance) for Sec.~\ref{S4}. This work is supported by the Laboratory for Physical Sciences and by NSF-2014156.
\end{acknowledgements}

\bibliography{references.bib}

\end{document}